\documentclass[a4paper,fleqn,usenatbib]{mnras}

\usepackage[T1]{fontenc}
\usepackage{ae,aecompl}

\usepackage{graphicx}	
\usepackage{amsmath}	
\usepackage{amssymb}	

\usepackage{ulem}
\usepackage{amsbsy}
\usepackage{mathrsfs,bm}
\newcommand{\bcdot}{\ensuremath{%
  \mathchoice%
   {\mskip\thinmuskip\lower0.2ex\hbox{\scalebox{1.5}{$\cdot$}}\mskip\thinmuskip}}%
   {\mskip\thinmuskip\lower0.2ex\hbox{\scalebox{1.5}{$\cdot$}}\mskip\thinmuskip}%
   {\lower0.3ex\hbox{\scalebox{1.2}{$\cdot$}}}%
   {\lower0.3ex\hbox{\scalebox{1.2}{$\cdot$}}}%
}

\newcommand{\bnabla}{\ensuremath{\boldsymbol{\nabla}}}
\newcommand{\vect}[1]{\boldsymbol{#1}}
\newcommand{\vecbf}[1]{\mathbfit{#1}}
\newcommand{\highresfac}{1}
\newcommand{\matrixfac}{1}
\newcommand{\evolutionfac}{1}

\usepackage                             {hyperref}
\usepackage[dvipsnames]{xcolor}
\usepackage{array}

\title[Magnetised jets and cosmic rays in galaxy clusters]
      {Simulations of the dynamics of magnetised jets and cosmic rays in galaxy clusters}

\author[K. Ehlert et al.]{
K. Ehlert$^{1}$\thanks{E-mail: kehlert@aip.de},
R. Weinberger$^{2}$,
C. Pfrommer$^{1}$,
R. Pakmor$^{2}$,
V. Springel$^{2,3,4}$
\\
$^{1}$Leibniz Institute for Astrophysics, An der Sternwarte 16, D-14482 Potsdam, Germany\\
$^{2}$Heidelberg Institute for Theoretical Studies, Schloss-Wolfsbrunnenweg 35, 69118 Heidelberg, Germany\\
$^{3}$Zentrum f\"ur Astronomie der Universit\"at Heidelberg, Astronomisches Recheninstitut, M\"onchhofstr. 12-14, 69120 Heidelberg, Germany\\
$^{4}$Max-Planck-Institut f\"ur Astrophysik, Karl-Schwarzschild-Str. 1, D-85741 Garching, Germany\\
}
\date{Accepted XXX. Received YYY; in original form ZZZ}

\pubyear{2018}

\begin{document}
\label{firstpage}
\pagerange{\pageref{firstpage}--\pageref{lastpage}}
\maketitle

\begin{abstract}
Feedback processes by active galactic nuclei in the centres of galaxy clusters
appear to prevent large-scale cooling flows and impede star formation. However,
the detailed heating mechanism remains uncertain. One promising heating scenario
invokes the dissipation of Alfv\'en waves that are generated by streaming cosmic
rays (CRs). In order to study this idea, we use three-dimensional
magneto-hydrodynamical simulations with the \textsc{arepo} code that follow the
evolution of jet-inflated bubbles that are filled with CRs in a turbulent
cluster atmosphere. We find that a single injection event produces the CR
distribution and heating rate required for a successful CR heating model. As a
bubble rises buoyantly, cluster magnetic fields drape around the leading
interface and are amplified to strengths that balance the ram pressure. Together
with helical magnetic fields in the bubble, this initially confines the CRs and
suppresses the formation of interface instabilities. But as the bubble continues
to rise, bubble-scale eddies significantly amplify radial magnetic filaments in
its wake and enable CR transport from the bubble to the cooling intracluster
medium. By varying the jet parameters, we obtain a rich and diverse set of jet
and bubble morphologies ranging from Fanaroff-Riley type I-like (FRI) to
FRII-like jets. We identify jet energy as the leading order parameter (keeping
the ambient density profiles fixed), whereas jet luminosity is primarily
responsible for setting the Mach numbers of shocks around FRII-like sources. Our
simulations also produce FRI-like jets that inflate bubbles without detectable
shocks and show morphologies consistent with cluster observations.
\end{abstract}




\section{Introduction}

The hot, X-ray emitting gas in cool-core (CC) clusters is expected to cool on
time scales $\lesssim 1\ \mathrm{Gyr}$. The absence of observed large-scale
cooling flows and low star formation rates suggests an efficient heating process
that balances cooling \citep{Peterson2006}. The observed X-ray cavities in the
centres of clusters, which correspond to low density, hot bubbles inflated by
jets from active galactic nuclei (AGN), contain sufficient energy to heat the
intracluster medium \citep[ICM,][]{Birzan2004,Birzan2008}. The correlation
between jet power and cluster cooling rate supports the now established idea of
heating through an AGN that is powered by a supermassive black hole (SMBH) found
in the centre of every cluster \citep{McNamara2007,McNamara2012}. The exact
mechanisms by which AGNs heat clusters in a volume-filling fashion remains
however unclear.

Proposed models for heating clusters include AGN-initiated weak shocks
\citep{Fabian2003b,Li2016a,Martizzi2018}, sound waves \citep{Sanders2008,Fabian2017}, gravity
waves \citep{Reynolds2015,Bambic2018}, or mixing of hot bubble material with
the ambient medium \citep{Hillel2015}. Thermal conduction is likely relevant in
the outskirts of the cluster core but is locally unstable to thermal
perturbations \citep{Kim2003,Soker2003} and thus unable to provide the global
solution. However, anisotropic thermal conduction renders the ICM unstable to
thermal buoyancy instabilities \citep[in particular the heat-flux driven
  instability in the core region of CC clusters,][]{Quataert2008}, facilitating
mixing of the AGN energy input and thereby increasing the coupling efficiency of
feedback energy \citep{Yang2016a,Kannan2017}. Earlier models that suggest energy
dissipation of strong shocks and turbulence volumetrically are challenged by
velocity measurements of the X-ray satellite {\it Hitomi}. {\it Hitomi} inferred
velocities in the ICM of the Perseus cluster of
$\approx150\ \mathrm{km\ s^{-1}}$ \citep{HitomiCollaboration2016}. Kinetic
energy is quickly dissipated before it could reach cooling regions that are
distant from the bubbles \citep{Fabian2017}.

As lobes of FRI-type jets \citep{Fanaroff1974} rise buoyantly in the CC cluster
atmosphere, their pressure content reaches equilibrium with the ambient
ICM. These lobes contain at most a small admixture of thermal pressure, implying
a dominant non-thermal pressure component with CR protons being the likely
candidate \citep[e.g.,][]{Morganti1988,Croston2008a,Croston2018}. After CRs have
diffused from the lobes into the ICM, these CRs can excite Alfv\'en waves via
the streaming instability \citep{Kulsrud1969,Zweibel2013}. The process of
scattering on the waves confines the CR population to move macroscopically close
to the Alfv\'en speed in the ICM \citep{Wiener2013}. The dissipation of the
Alfv\'en waves through damping processes such as non-linear Landau or turbulent
damping effectively transfers CR energy to thermal heat
\citep{Wentzel1971,Guo2008}. This provides a promising alternative heating
mechanism in CC clusters \citep{Loewenstein1991,Guo2008,Pfrommer2013}.  Assuming
steady state, a combination of central CR heating and thermal conduction at
larger radii can balance radiative cooling in a large sample of CC clusters,
suggesting a stable, self-regulated cycle of CR heating and radiative cooling
\citep{Jacob2016a,Jacob2016b}. Idealised three-dimensional (3D) simulations show
that self-regulated CR feedback can smoothly heat the centres of clusters
\citep{Ruszkowski2017a}. The model depends crucially on the interplay between CR
transport and AGN bubble dynamics.

Simulations of AGN bubbles are able to reproduce the general morphology of
observed X-ray cavities \citep[e.g.,][]{Churazov2001,Reynolds2001,Bruggen2001}
and explain the absence of radio synchrotron emission of so called ghost
cavities \citep{Ensslin2002,Bruggen2002a}. In these earlier simulations, the
bubbles are modelled as low-density cavities. Subsequent hydrodynamical (HD)
simulations started inflating bubbles self-consistently via a subgrid jet model
\citep{Sternberg2008a}. On these scales, the jet is assumed to be sufficiently
slow such that non-relativistic HD can be used \citep[but see][]{Perucho2017}. A
propagating jet introduces significant heating through the dissipation of the
accompanying bow shock \citep[e.g.,][]{Reynolds2002,Bruggen2002}.

However, first simulations showed a discrepancy between short disruption times
of HD bubbles \citep{Churazov2001,Bruggen2002a} and observed long-lived bubbles
in the outskirts of cluster cores \citep[e.g.,][]{Fabian2000,Fabian2011}. The
issue can be alleviated with the addition of viscosity
\citep{Reynolds2005,Sijacki2006}, magnetic fields
\citep{Ruszkowski2007,Bambic2018}, or modelling the stage of bubble
  inflation \citep{Sternberg2008a}. Jets dominated by kinetic energy form
radially elongated cavities at large radii. This is in contrast to observed,
light jets that are energetically dominated by CRs and lose momentum more
quickly because of the lower jet inertia. The CR pressure causes the jet to
expand laterally and to displace more ICM at smaller cluster-centric radii,
naturally producing wider cavities near cluster centres in agreement with X-ray
observations \citep{Sijacki2008,Guo2011}. This holds when CR diffusion is added \citep{Ruszkowski2008}. Lobes inflated by jets in cosmological cluster
simulations show deviations from the initial jet axis due to bulk motions of the
ICM and substructure \citep{Heinz2006,Morsony2010,Mendygral2012}.

To explore the feasibility of Alfv\'en-wave heating in CC clusters, we simulate
a single AGN jet event leading to the formation and evolution of CR-filled
bubbles in a turbulent, magnetised ICM. We focus on the resulting CR
distribution due to anisotropic diffusion as well as on the consequences for
cluster magnetic fields. By varying jet power and lifetime, we study general trends
of the CR distribution, lobe morphology and mixing efficiency.

The outline of our paper is as follows: in Section \ref{sec:methods}, we
describe our initial conditions, numerical modelling, and detail the different
types of simulations. The general evolution of the jet and subsequent formation
of the bubble is analysed in Section \ref{sec:evolution}. In Section
\ref{sec:magneticfield}, we discuss the stabilisation of the bubble due to
magnetic fields and the influence of the bubble on the external magnetic fields
and mixing efficiency. The CR distribution and relevance of Alfv\'en heating is
the topic of Section \ref{sec:cosmicrays}. In Section \ref{sec:jetparameters},
we focus on the influence of jet power and jet lifetime on bubble morphology,
magnetic field structure and CR distribution, in particular in light of the
FRI/FRII dichotomy. We briefly summarise our results in Section
\ref{sec:conclusions}. In Appendix~\ref{sec:magneticfield_generation}, we detail
our procedure for generating turbulent magnetic fields.  Finally, in
Appendices~\ref{sec:resolution_study} and \ref{sec:BubbleCooling}, we perform a
resolution study and assess how varying parameters of our subgrid CR cooling
models impact our results.

\begin{figure*}
\centering
\includegraphics[trim=0.25cm 1.3cm 4.75cm .45cm,clip=true, width=0.48\textwidth]{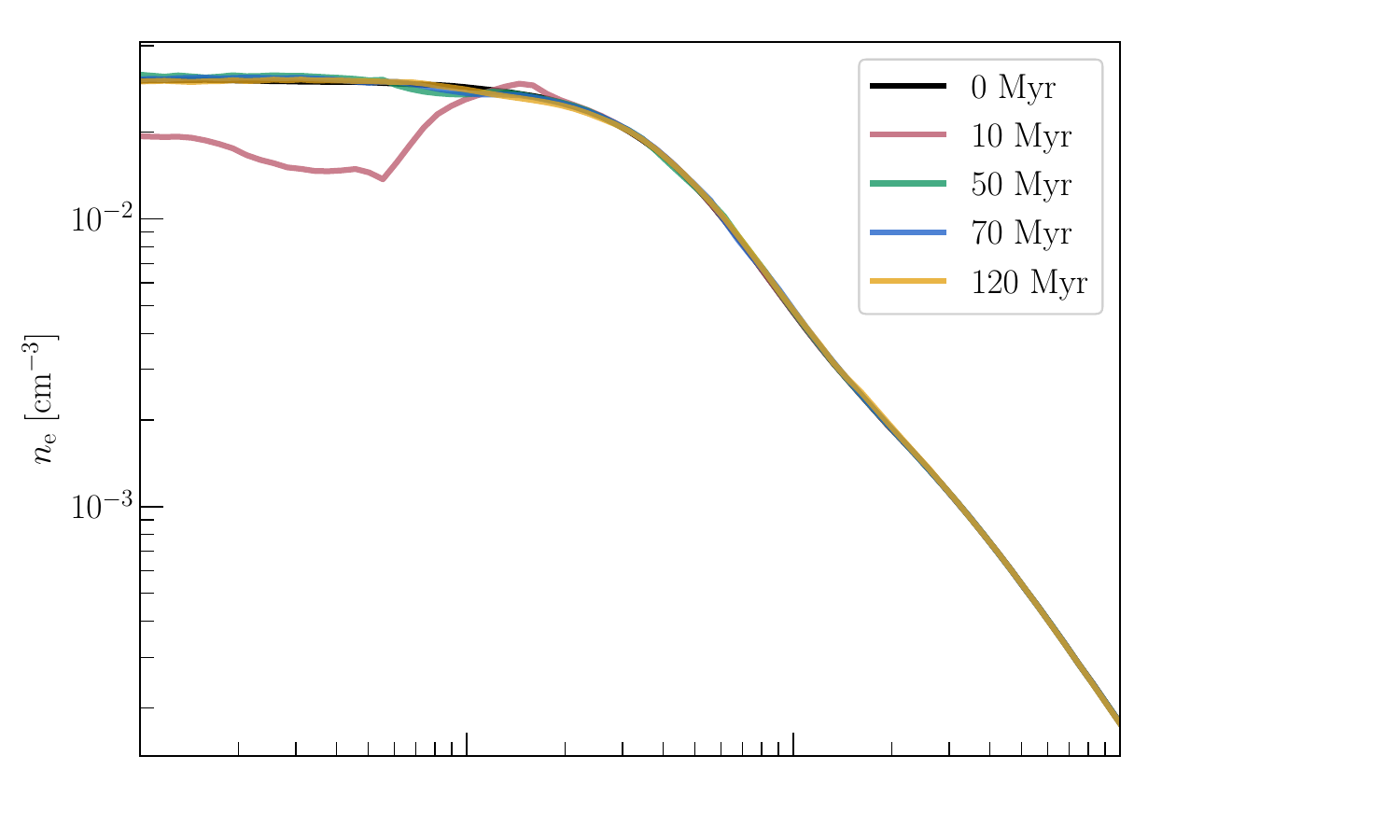}
\hspace{0.01\textwidth}
\includegraphics[trim=0.25cm 1.3cm 4.75cm .45cm,clip=true, width=0.48\textwidth]{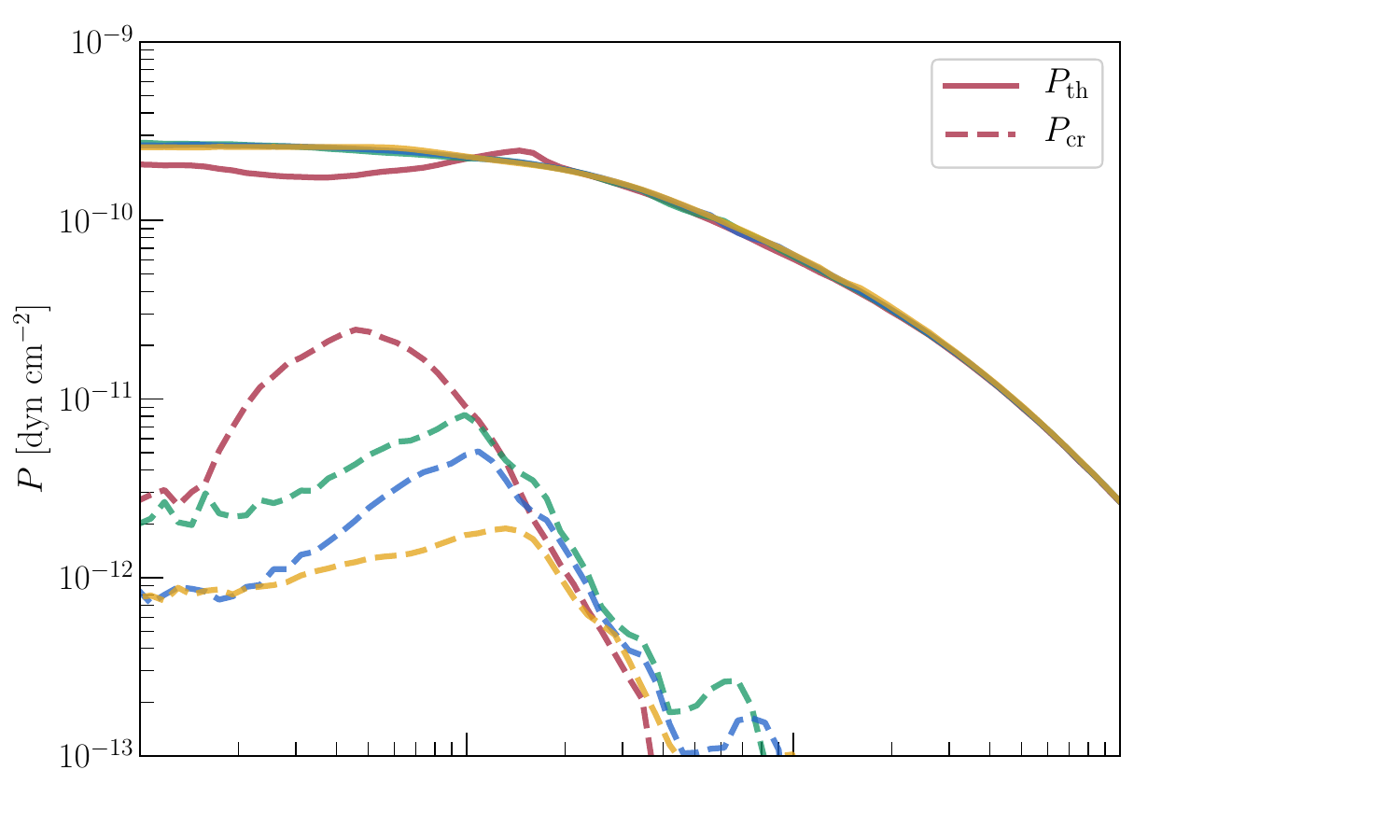}
\includegraphics[trim=0.25cm 0.1cm 4.75cm .45cm,clip=true, width=0.48\textwidth]{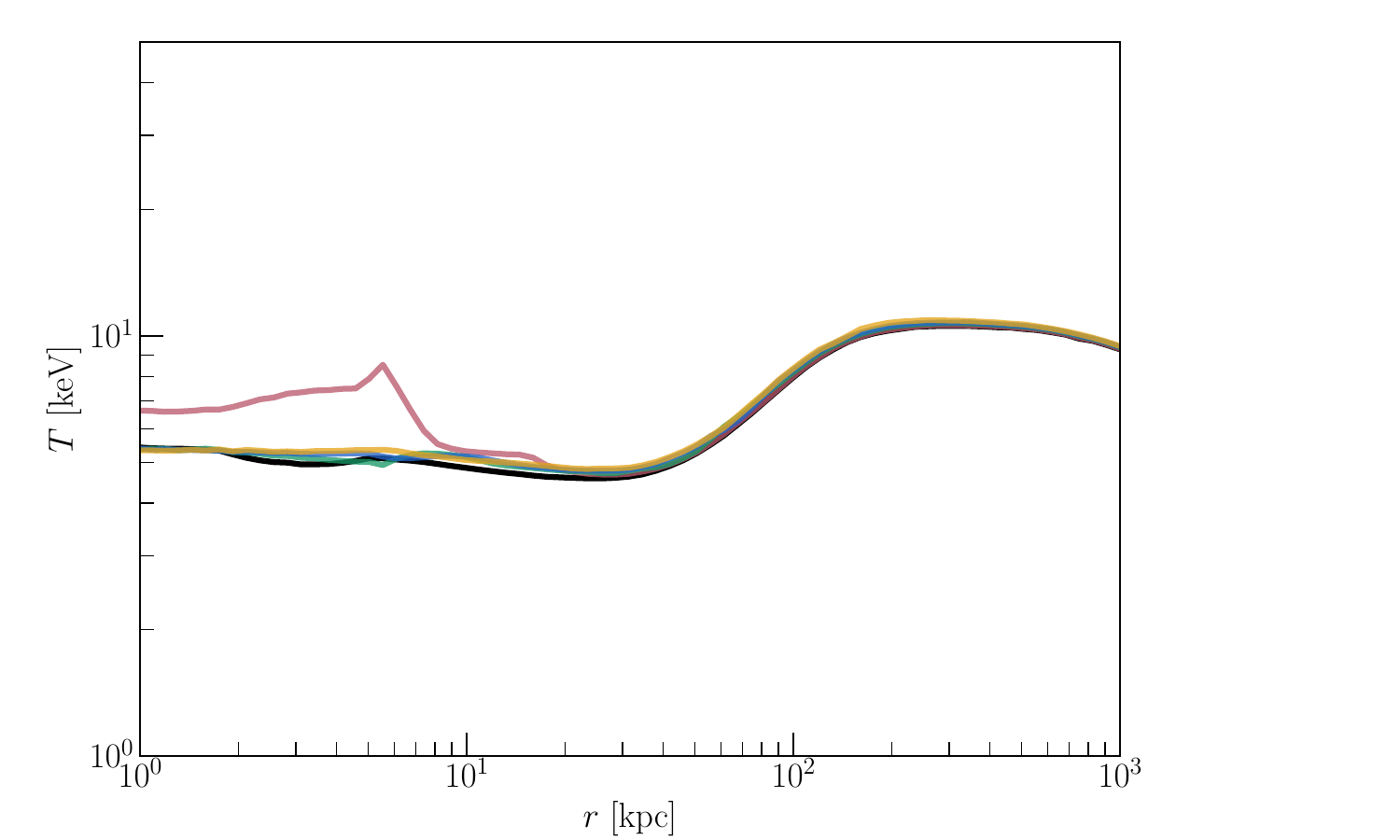}
\hspace{0.01\textwidth}
\includegraphics[trim=0.25cm 0.1cm 4.75cm .45cm,clip=true, width=0.48\textwidth]{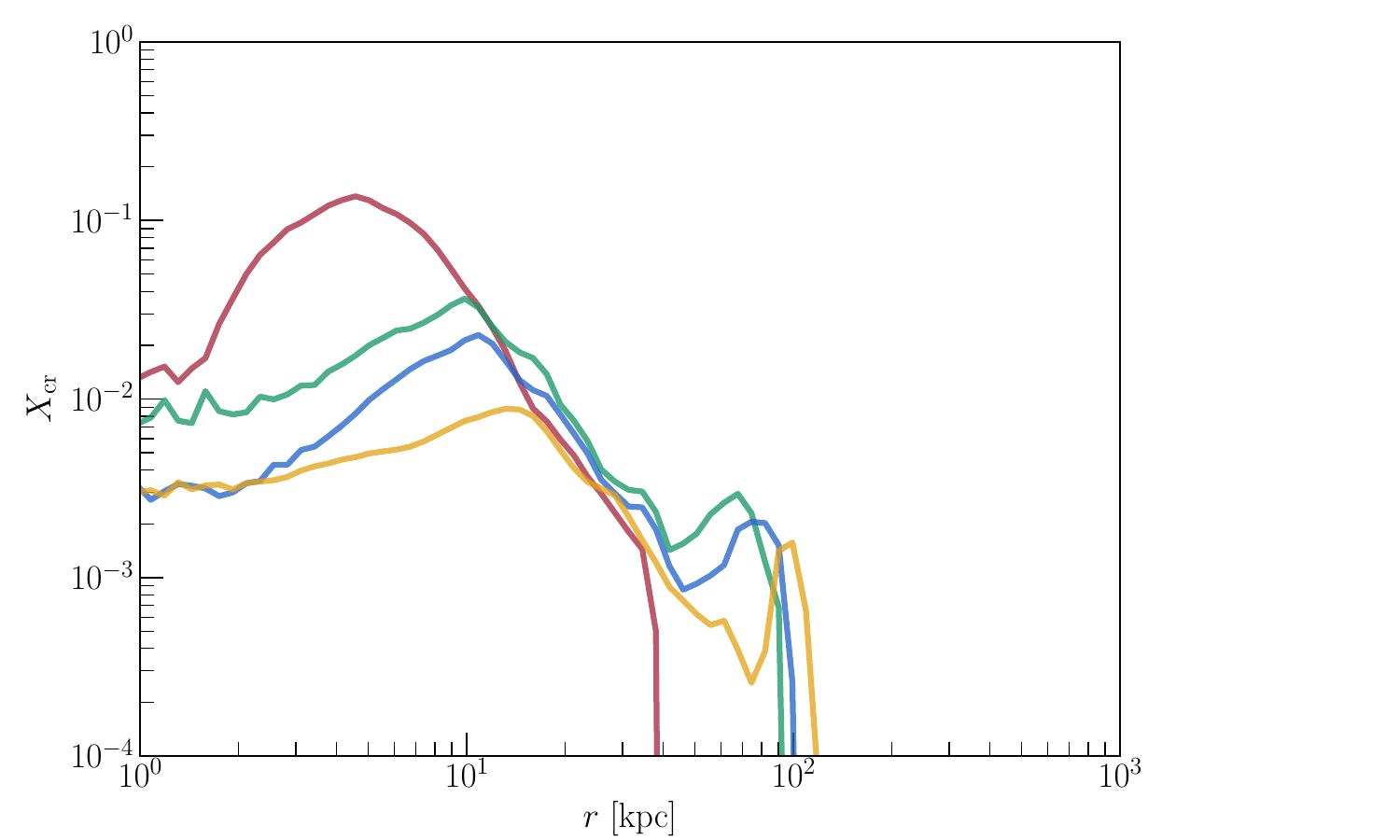}
\caption{Radial profiles of the gas for the initial conditions at
  $0\ \mathrm{Myr}$ (black) and the fiducial model at stated times. The number
  density $n_e$ is volume-weighted, the temperature $T$ is mass-weighted and the
  CR pressure $P_\mathrm{cr}$ as well as the thermal pressure $P_\mathrm{th}$
  are both volume-weighted. The CR-to-thermal pressure ratio $X_\mathrm{cr}$ is
  obtained by dividing both volume-weighted quantities. The propagating jet is
  visible as a low-density and hot feature close to the cluster centre at
  10~Myr. The CRs are distributed within the inner $100\ \mathrm{kpc}$ and their
  profile drops off steeply at larger radii.}
    \label{fig:radialProfile_general}
\end{figure*}

\section{Methods and simulation models}
\label{sec:methods}

Here, we introduce our numerical set-up of the cluster model and the adopted
magnetic structure of the ICM.  We employ magneto-hydrodynamical (MHD)
simulations with the moving-mesh code \textsc{arepo} \citep{Springel2010}
that evolves thermodynamic quantities of the gas on an unstructured moving mesh
  defined by the Voronoi tessellation of a set of discrete points that initially
  obey a quasi-Lagrangian distribution. Of particular importance for this work
are our models for launching AGN jets and CR transport, which we describe in
detail here. An overview of our simulation models closes this section.

\subsection{Initial conditions}
\label{sec:ICs}

By analogy with \cite{Weinberger2017}, we model the cluster density distribution
with a Navarro-Frenk-White (NFW) profile \citep{Navarro1996,Navarro1997} with
virial radius\footnote{We define the virial cluster radius as the radius at
  which the mean interior density equals $200$ times the critical density of the
  universe today.}  $R_{200,\mathrm{c}}=2.12\ \mathrm{Mpc}$, virial mass
$M_{200,\mathrm{c}}=10^{15}\ \mathrm{M}_\odot$, and concentration parameter
$c_\mathrm{NFW}=5$. The electron number density $n_\mathrm{e}$ is modelled after
the Perseus galaxy cluster and based on a double-beta profile fit to the
de-projected density distribution from X-ray observations
\citep{Churazov2003}. In addition, the number density is scaled in order to
retain a gas fraction of $16\%$ within $R_{200,\mathrm{c}}$:
\begin{equation}
\begin{split}
n_\mathrm{e} &= 26.9\times10^{-3}\left[1+\left(\frac{r}{57\ \text{kpc}}\right)^2\right]^{-1.8}\text{cm}^{-3}\\
 &\quad +2.8\times10^{-3}\left[1+\left(\frac{r}{200\ \text{kpc}}\right)^2\right]^{-0.87}\text{cm}^{-3}.
\end{split}
\end{equation}
The pressure in each cell is obtained from the assumption of hydrostatic
equilibrium and the restriction that the pressure vanishes at a radius of
$3\ \mathrm{Mpc}$. Initially, we adopt a Cartesian mesh with a box size of
$1.5\ \mathrm{Mpc}$ that we iteratively refine to ensure similar mass content
per cell. This yields an adaptive mesh with smoothly increasing cell sizes
towards larger radii.

Gravity is treated as a static background provided by a NFW dark matter
distribution, neglecting the effects of self-gravity and the gravitational
interactions with the SMBH. We show radial profiles of the initial conditions of
the electron number density $n_\rmn{e}$, temperature $T$, and thermal pressure
$P_\mathrm{th}$ in Fig.~\ref{fig:radialProfile_general} (note that there are
no CRs in the initial conditions).

To mimic a realistic ICM, we generate a Gaussian-distributed, turbulent magnetic
field $\vecbf{B}$ (see Appendix~\ref{sec:magneticfield_generation}, for
details). First, we lay down our initial magnetic field on a Cartesian mesh,
which needs to have a resolution that is smaller than the smallest (Lagrangian)
cell size of our high-resolution initial conditions, $\Delta x
=700\ \mathrm{pc}$.\footnote{Note that the cells in the AGN lobe have a
  super-Lagrangian resolution with typical grid-size $V^{1/3} = \Delta x =
  188\ \mathrm{pc}$ in our highest resolution simulations.} Fourier transforming
such a large Cartesian grid is numerically not feasible. Thus, for our
high-resolution simulation we set up three nested meshes, with resolutions decreasing from the cluster centre. Generally, the magnetic field meets the
following requirements:
\begin{enumerate}
\item $\vecbf{B}$ is divergence-free: $\bnabla\bcdot\vecbf{B}=0$;
\item $\vecbf{B}$ follows a Kolmogorov spectrum in the inertial range for
  wave numbers larger than the injection scale $k_\mathrm{inj}$, and a
  random white noise power spectrum for $k<k_\mathrm{inj}$;
\item $\vecbf{B}^2$ is scaled at each radius to obtain a constant average
  magnetic-to-thermal pressure ratio of $X_{B,\mathrm{ICM}}$;
\item $\vecbf{B}$ fields on different nested meshes do not interconnect.
\end{enumerate}

The resulting cluster magnetic field is then interpolated from this nested mesh
onto our adaptive, smoothly varying mesh. To ensure pressure equilibrium in the
initial conditions, we adopt temperature fluctuations of the form $n
k_\rmn{B}\delta T=-\delta \vecbf{B}^2/(8\pi)$. We then relax our mesh to obtain
a computationally more efficient non-degenerate tessellation structure for
\textsc{arepo}. However, this setup does not balance the magnetic tension. The resulting turbulent motions initiate the decay of magnetic power. To reinitialize the prescribed magnetic-to-thermal pressure ratio, we rescale the temperature as
well as the magnetic field to the desired initial magnetic-to-thermal pressure
ratio $X_{B,\mathrm{ICM}}$ while maintaining hydrostatic equilibrium.

The described procedure generates our initial conditions for simulations with a
turbulent magnetic field. The initiated turbulence during the relaxation of the mesh initiates bulk velocities, which remain part of the initial conditions. For our comparison simulations without a magnetic
field, we proceed as before, but set $\vecbf{B}={\bf 0}$. Thus, the atmosphere
remains turbulent through the presence of initial flows and pressure
irregularities. However, these runs show a lower degree of turbulence than the
runs with finite magnetic field as the magnetic tension can drive and sustain
turbulence on longer time scales.

As explained above, the stresses of the tangled magnetic field induce turbulent
gas motions. In the absence of a driver of turbulence (as in our simulations)
the turbulence gradually dissipates and the magnetic field strength and gas
velocities decrease as a function of time.

\subsection{Jet model}
\label{sec:jet}

To study the influence of a SMBH-driven and CR-filled jet in a turbulent cluster
environment, we employ the jet model of \citet{Weinberger2017}. Here, we provide
a brief summary of the implementation and describe modifications related to CR
acceleration, CR cooling and the magnetic isolation of the injection region.

The jet becomes active at time $t$ after the last injection event at
$t_\mathrm{inj}$ when the available energy to the jet,
$\dot{E}_\mathrm{jet}(t-t_\mathrm{inj})$, exceeds the required energy to
redistribute the gas considering adiabatic changes and inject a predefined amount of energy (thermal and
  non-thermal). When the jet is
active, the model identifies two opposing jet regions close to the black hole. A
low density state ($\rho_\mathrm{jet}/\rho_\mathrm{ICM}\sim 10^{-4}$) is set up
in pressure equilibrium with the surrounding medium. Mass and thermal energy are
redistributed to and from neighbouring cells to ensure mass conservation. If
desired, a magnetic field is included in a purely toroidal
configuration. Due to gas flows in the bubble, the magnetic field is
  reshaped, but maintains its helical morphology. After accounting for
adiabatic losses, the remaining jet energy is injected as kinetic energy to
launch a bipolar outflow. To identify the bubble, an advective scalar
$X_\mathrm{jet}$ is used, which corresponds to the mass fraction of the jet
material in the cell. In the remainder of this work we refer to
  $X_\mathrm{jet}$ as jet tracer. The tracer is initialised in the jet
injection region with a value of $X_\mathrm{jet}=1$. The strong density contrast
between bubble and cluster is maintained through refinement criteria based on
the density gradient and cell volume \cite[for details, see][]{Weinberger2017}. Throughout this work, we define lobe material to
exhibit a jet mass fraction $X_\mathrm{jet}>10^{-3}$ and checked that the
conclusions of this paper are invariant with respect to variations of this
choice.

The discrepancy between the inferred pressure of observed bubbles (via minimum
energy arguments of radio observations) and the ambient ICM pressure (as
inferred from X-ray observations) argues for a significant pressure contribution
of CR protons that cannot directly be inferred by any interaction process due to
the very low bubble densities
\citep{Birzan2008,Croston2008a,Laing2014,Croston2014,Heesen2018}.  CRs could be
accelerated at internal shocks of jets \citep{Perucho2007}. The strong
non-thermal emission in knots observed in jets supports this mechanism
\citep{Worrall2009,Laing2013,BarniolDuran2017}. As we do not resolve internal
shocks explicitly, we treat CR injection in a subgrid model. As most of the
injected kinetic energy would immediately be thermalized, we instead ensure a
minimum CR-to-thermal energy fraction $X_\mathrm{cr,acc}=
E_\rmn{cr}/E_\rmn{th}=1$ in every computational cell inside the jet/lobe for a
time $\tau_\mathrm{acc}=2\tau_\mathrm{jet}$, where $\tau_\mathrm{jet}$ is the
jet lifetime. When varying $\tau_\mathrm{acc}$ we find that the resulting
dynamical effects of late-time accelerated CRs ($t>\tau_{\rmn{jet}}$) are
negligible.  This is an important difference in comparison to
\cite{Weinberger2017} where the bubble CR pressure is specified at jet launch
and successive CR acceleration in the jet is not accounted for, which yields a
sub-dominant CR population.

\subsection{CR transport}
\label{sec:CRs}

CRs as charged particles are bound to stay on flux-frozen magnetic field
lines. As the magnetic field is transported alongside the gas, so are CRs.  In
addition to advection, CRs scatter on magnetic fluctuations which leads to
transport by diffusion and streaming relative to the rest frame of the gas,
mainly along the direction of the local mean magnetic field, which coincides
with the large-scale field \citep{Pfrommer2017}.

Anisotropically moving CRs in the frame of propagating Alfv\'en waves are
unstable to the streaming instability \citep{Kulsrud1969}. These CRs resonantly
excite Alfv\'en waves, which in turn causes the CRs' pitch angles to scatter and
eventually to isotropize in the Alfv\'en frame (that moves with speed
$v_\mathrm{A}$). Hence, in galaxy clusters, the streaming velocity $v_\mathrm{s}$
relative to the thermal plasma corresponds approximately to the Alfv\'en
velocity, i.e., $v_\mathrm{s}-v_\mathrm{A}\sim{}c^2/(3\nu{}l_\mathrm{cr})$,
where $\nu$ is the pitch angle scattering rate and
$l_\mathrm{cr}=P_\mathrm{cr}/\nabla P_\mathrm{cr}$ is the CR gradient length
scale \citep{Kulsrud2005}. In addition, CRs diffuse along field lines with a
diffusion coefficient
$\tilde{\kappa}\equiv{}c^2/(3\nu)\sim(v_\mathrm{s}-v_\mathrm{A})l_\mathrm{cr}$,
which makes diffusion negligible compared to streaming in the strong scattering
limit. Consequently, we introduce an effective CR diffusion coefficient
$\kappa_\mathrm{cr,A}\sim{}l_\mathrm{cr}v_\mathrm{A}$ that emulates the combined effects of
streaming and spatial diffusion \citep{Sharma2009,Wiener2017}.

Different damping mechanisms dissipate Alfv\'en waves, causing the streaming CRs
in steady state to continuously transfer part of their energy into heat
via {\it Alfv\'en wave heating} with a rate
\begin{equation}
\mathcal{H}_\mathrm{cr}=\left|\vect{v}_\mathrm{A}\bcdot \bnabla {P}_\mathrm{cr}\right|.
\end{equation} 
Note the dependence of the Alfv\'en heating rate on the CR pressure
gradient. This directly relates $\mathcal{H}_\mathrm{cr}$ to the CR diffusion
coefficient $\kappa_\mathrm{cr,A}$ in our effective model. 

Following this self-confined picture, CRs are treated as a secondary fluid with
adiabatic index $\gamma_\text{cr}=4/3$ \citep{Pfrommer2017}. We emulate
streaming through a combination of anisotropic diffusion and Alfv\'en losses
\citep{Wiener2017}. Our CRs are advected with the gas and anisotropically
diffuse with a constant diffusion coefficient
$\kappa_\parallel=10^{29}\ \mathrm{cm}^2\mathrm{s}^{-1}$ along $\vecbf{B}$ and
$\kappa_\perp=0$ perpendicular to it \citep{Pakmor2016a}. The particular value
of $\kappa_\parallel$ has been choosen so that it produces self-consistent
results in our simulations (as we will see later in Section
\ref{sec:cosmicrays_characteristics}).

The CR distribution in the lobes quickly becomes uniform as the magnetic field
confines the CRs initially to stay within the lobes. This leads to a negligible
CR pressure gradient and hence an insignificant Alfv\'en wave cooling rate
inside lobes. In our simulations, insufficiently resolved steep gradients at the
lobe interface would lead to artificially large (numerical) cooling rates.  We
prevent this by imposing a maximum jet tracer threshold $X_\mathrm{jet,cool}\leq10^{-3}$ for Alfv\'en wave
cooling to be active (for a discussion, see
Appendix \ref{sec:BubbleCooling}). Our simulations also include the cooling
  of CRs through Coulomb and hadronic interactions \citep{Pfrommer2017}, which
are however much smaller here in comparison to Alfv\'enic losses.

Physically, the jet fluid is magnetically unconnected to the ICM. In order
to study the evolution of the diffusing CR distribution accurately, the jet
launching region needs to be magnetically isolated to prevent spurious diffusion
at the resolution limit along field lines that connect the jet launch region to
the exterior ICM. Many individual injection events are present in our
simulation, which require accurate magnetic isolation before each individual
event. To this end, we project out the radial magnetic field components:
\begin{equation}
\label{eq:magnetic_isolation_bubble}
\vecbf{B} \to \left[1-g(r)\hat{\vecbf{r}}\hat{\vecbf{r}}\right]\vecbf{B},
\end{equation}
where $\hat{\vecbf{r}}$ is the radial unit vector measured
  from the centre of our spherical jet launching region and $g(r)=1-\left|
\cos\left(0.5\pi\left(x+\Delta x-1\right)/{\Delta x}\right)\right|$ for
$1-\Delta x<x<1+\Delta x$, where $x=r/r_\textrm{jet}$, $\Delta x=0.25$, and
$r_\mathrm{jet}$ is the radius of the jet launching region. Jets are launched from two regions on opposite sides of the centre. During the first
  injection event of a jet, the two bimodal jet regions are isotropically
  isolated, i.e., across the entire spherical surface (see
  equation~\ref{eq:magnetic_isolation_bubble}).  At all subsequent injection
  events while the jet is active, only the bubble hemisphere closer to the SMBH
is isolated in order to retain the velocity structure in the direction of motion
of the jet.

\subsection{Simulation models}
\label{sec:simulations}

\begin{table}
\begin{center}
\begin{tabular}{ c  c c c c c}
	\hline
   $P_\text{jet}$ & $X_{{B},\mathrm{ICM}}$ & $X_{{B},\mathrm{jet}}$ & $\tau_\text{jet}$ & $\kappa_\parallel$ & $E_\mathrm{jet}$\\
        $[\text{erg}\,\text{s}^{-1}]$ & &  & $[\text{Myr}]$ & $[\text{cm}^{2}\,\text{s}^{-1}]$ & $[10^{59}\,\mathrm{erg}]$\\
  \hline			
    \multicolumn{6}{l}{simulations with CRs}   \\
    \hline
  $4\times10^{43}$ & $0.05$ & $0.1$ & 10 & $10^{29}$ & $~~0.13$  \\
  $4\times10^{43}$ & $0.05$ & $0.1$ & 25 & $10^{29}$ & \textcolor{black}{$~~0.31$} \\
  $4\times10^{43}$ & $0.05$ & $0.1$ & 50 & $10^{29}$ & \textcolor{black}{$~~0.63$} \\
  $1\times10^{44}$ & $0.05$ & $0.1$ & 10 & $10^{29}$ & \textcolor{black}{$~~0.32$}  \\
  $1\times10^{44}$ & $0.05$ & $0.1$ & 25 & $10^{29}$ & \textcolor{black}{$~~0.79$}\\
  $1\times10^{44}$ & $0.05$ & $0.1$ & 50 & $10^{29}$ & \textcolor{black}{$~~1.58$} \\
  $2\times10^{44}$ & $0.05$ & $0.1$ & 10 & $10^{29}$ & \textcolor{black}{$~~0.63$} \\
  ${\bf2\times10^{44}}$ & ${\bf0.05}$ & ${\bf 0.1}$ & {\bf25} & ${\bf 10^{29}}$ & ${\bf ~~1.58}$ \\
  $2\times10^{44}$ & $0.05$ & $0.1$ & 50 & $10^{29}$ & \textcolor{black}{$~~3.15$} \\
  $4\times10^{44}$ & $0.05$ & $0.1$ & 10 & $10^{29}$ & \textcolor{black}{$~~1.26$} \\
  $4\times10^{44}$ & $0.05$ & $0.1$ & 25 & $10^{29}$ & \textcolor{black}{$~~3.15$} \\
  $4\times10^{44}$ & $0.05$ & $0.1$ & 50 & $10^{29}$ & \textcolor{black}{$~~6.31$} \\
  $1\times10^{45}$ & $0.05$ & $0.1$ & 10 & $10^{29}$ & \textcolor{black}{$~~3.15$}\\
  $1\times10^{45}$ & $0.05$ & $0.1$ & 25 & $10^{29}$ & \textcolor{black}{$~~7.88$} \\
  $1\times10^{45}$ & $0.05$ & $0.1$ & 50 & $10^{29}$ & $15.77$\\
  \hline
    \multicolumn{6}{l}{simulations without CRs}   \\
    \hline
  $2\times10^{44}$ & $0.05$ & $0.1$ & 25 & $0$ & $~~1.58$  \\
  $2\times10^{44}$ & $0~~~~$ & $0.1$ & 25 & $0$ & $~~1.58$ \\
  $2\times10^{44}$ & $0.05$ & $0~~$ & 25 & $0$ & $~~1.58$ \\
  $2\times10^{44}$ & $0~~~~$ & $0~~$ & 25 & $0$ & $~~1.58$ 
\end{tabular}
\end{center}
\caption{Jet parameters of the different models with combined jet power
  $P_\text{jet}$ of the bipolar outflow, external magnetic-to-thermal pressure
  ratio $X_{{B},\mathrm{ICM}}$, jet magnetic-to-thermal pressure ratio
  $X_{{B},\text{jet}}$, time of activity of the jet $\tau_\mathrm{jet}$, CR
  diffusion coefficient along the magnetic field $\kappa_\parallel$ and jet
  energy $E_\mathrm{jet}=P_\mathrm{jet}\tau_\mathrm{jet}$. The lower part of the
  table corresponds to our control runs without CRs, which are used for the
  analysis of the bubble stability. The fiducial run is marked in boldface.}
\label{Tab:JetPara}
\end{table}

Our MHD simulations are performed with the moving-mesh code \textsc{arepo}
\citep{Springel2010}, using an improved second-order hydrodynamic scheme with
least-squares-fit gradient estimates and a Runge-Kutta time integration
\citep{Pakmor2016}. The MHD fluxes across cell interfaces are computed with an
HLLD Riemann solver \citep{Pakmor2011,Pakmor2013} adopting the Powell scheme for
divergence control \citep{Powell1999}.

The bubble evolution is studied in a set of simulations listed in Table
\ref{Tab:JetPara}. The jets are active for a prescribed time $\tau_\mathrm{jet}$
with a certain jet power $P_\mathrm{jet}$. Our fiducial model
corresponds to $P_\mathrm{jet}=2~\times~10^{44}\ \mathrm{erg}\ \mathrm{s}^{-1}$,
$\tau_\mathrm{jet}=25\ \mathrm{Myr}$, $X_{B,\mathrm{ICM}}=0.05$,
$X_{B,\mathrm{jet}}=0.1$ and
$\kappa_\parallel=10^{29}\ \mathrm{cm}^2\ \mathrm{s}^{-1}$. We use this model in the
analysis unless stated otherwise.

\begin{table}
\begin{center}
\begin{tabular}{ l  l  l }
   \multicolumn{3}{l}{\textbf{Jet parameters}}  \\
  \hline			
  Jet density & $\rho_\text{target}$ & $10^{-28}\ \text{g}\ \text{cm}^{-3}$ \\
  Jet launching region & $r_\mathrm{j}$ & $5\ \text{kpc}$ \\
  CR acceleration & $X_\text{cr,acc}$ & $1$ \\
   & & \\
   \multicolumn{3}{l}{\textbf{Magnetic field parameters}}  \\
  \hline
  Injection scale & $k_\text{inj}$ & $37.5^{-1}\ \text{kpc}^{-1}$ \\     
   & & \\
   \multicolumn{3}{l}{\textbf{Resolution}}  \\
  \hline
  Target mass & $m_{\text{target},0}$ &  lower res.: $1.5\times10^{6}\ \text{M}_\odot$ \\
  &  &  high res.: $1.5\times10^{5}\ \text{M}_\odot$ \\
  Target volume & $V_{\text{target}}^{1/3}$ & lower res.: $405\ \text{pc}$ \\
  &  & high res.: $188\ \text{pc}$ \\
  Minimum volume & $V_\text{min}$ & $V_\text{target}/2$ 
\end{tabular}
\end{center}
\caption{A summary of our adopted parameters of the simulation.}
\label{Tab:JetParaUnchanged}
\end{table}

Sub-grid parameters and those responsible for the resolution of our simulation
are given in Table \ref{Tab:JetParaUnchanged}. In \citet{Weinberger2017}, we
showed that the distance travelled by bubbles depends on the numerical
resolution of the jet. Here, we focus on the previously dubbed high-resolution
simulation. The numerical convergence of our results in comparison to lower
resolution simulations is discussed in Appendix \ref{sec:resolution_study}. When
we compare several simulation models with varying parameters, we use the
lower resolution simulations as this recovers qualitatively the same
evolution of the jet at a significantly lower computational cost.

If included in the lobes ($X_\mathrm{jet}>10^{-3}$), CRs are generally
accelerated for time $\tau_\mathrm{acc}=2\tau_\mathrm{jet}$ and we use a
conversion fraction from thermal-to-CR energy of $X_\mathrm{cr,acc}=1$. Reducing
the fraction $X_\mathrm{cr,acc}$ only changes the normalisation of our CRs but
has no significant impact on the studied features. Also, a much larger fraction
of CRs is disfavoured as FRI jets are slowed down through the entrainment of
ambient material. A remaining population of shocks seems unlikely to accelerate
significant portions of the entrained protons. The conclusions of this paper
also remain robust against variations of $\tau_\mathrm{acc}$ as long as
$\tau_\mathrm{acc}\geq\tau_\mathrm{jet}$.

While magnetic tension initially generate significant levels of
turbulence, this decays over the run time of our simulation since we do not
model large scale flows due to substructure \citep{Bourne2017} and neglect gas
cooling, which also induces motions.  In reality, the cooling gas accretes onto
the central SMBH, powering jets in this process, leading to highly variable jet
powers and lifetimes.  Here, we instead decided to impose predefined jet
energies to gain insight into the parametric dependencies of mixing,
morphologies and CR distributions more easily. We postpone modelling of self-consistent jet
injection through accreted cooling gas, which will enable us to study the
long-term stability of the CC cluster due to CR heating.

\begin{figure}
\centering
\includegraphics[trim=0.75cm 0.7cm 7.95cm .45cm,clip=true,width=\evolutionfac\columnwidth]{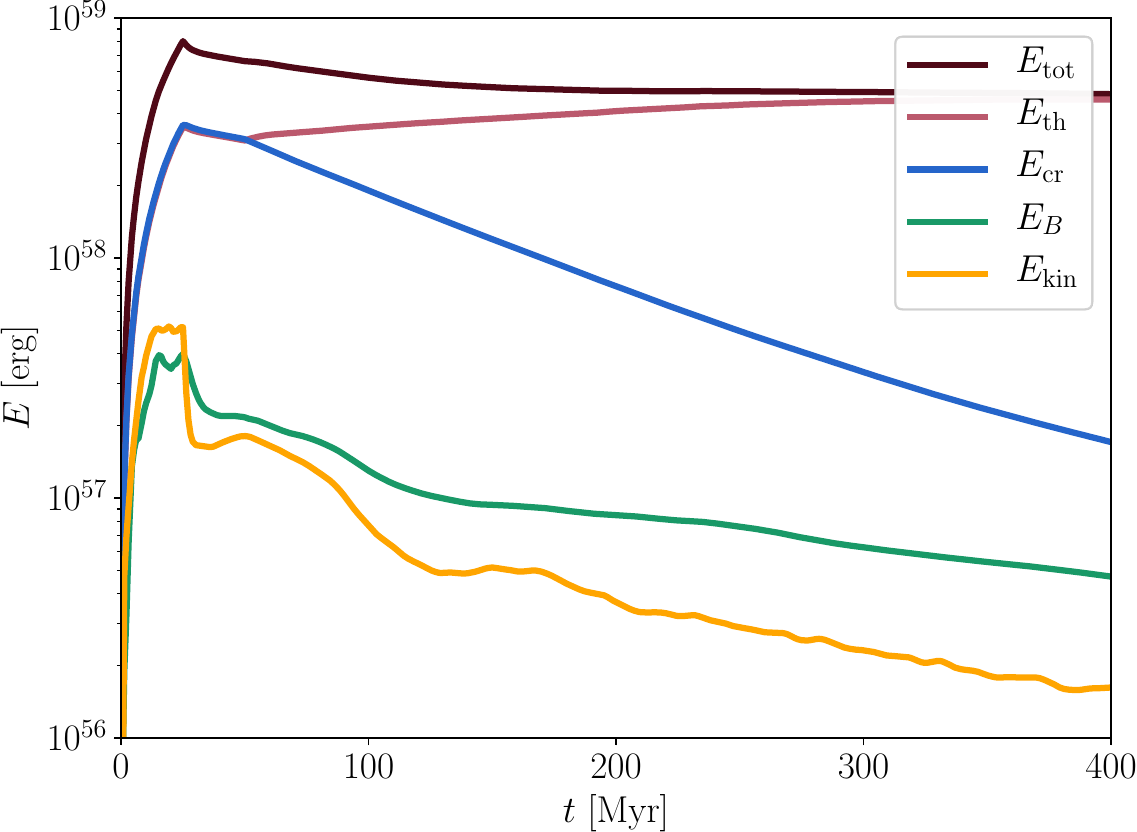}
    \caption{The time evolution of the energy components of the lobe. The lobe
      is defined as cells with jet mass fraction $X_\mathrm{jet}>10^{-3}$. The
      thermal and CR energy are initially in equipartition. Later, CRs diffuse
      out of the lobe and thermal gas is entrained, which leads to an overall
      increase in thermal energy after the jet becomes inactive.}
    \label{fig:energy_evolution}
\end{figure}

\begin{figure*}
\centering
\includegraphics[trim=0.25cm 0.6cm 1.3cm 1.45cm,clip=true,width=\evolutionfac\textwidth]{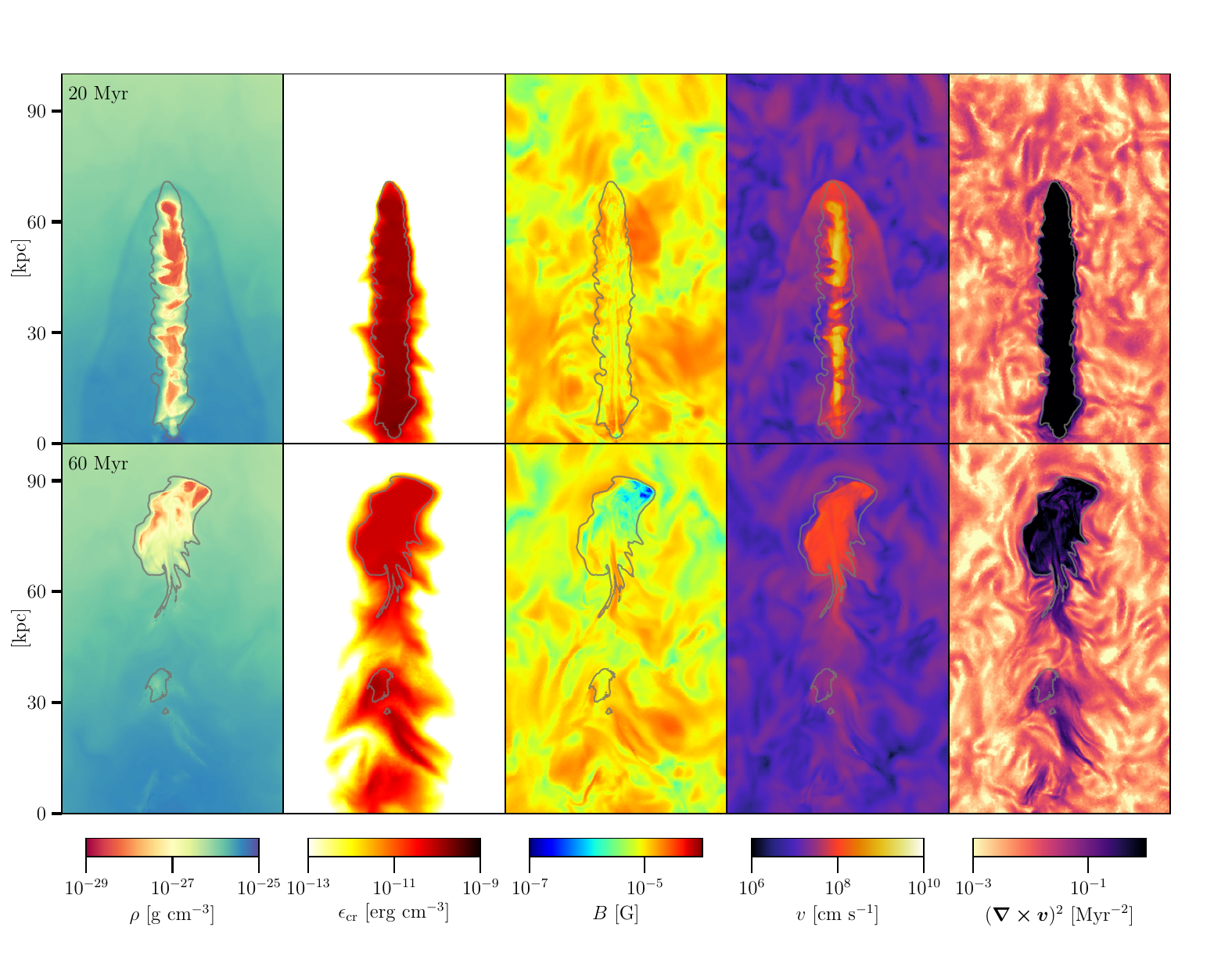}
    \caption{The mass density $\rho$, CR energy density $\epsilon_\mathrm{cr}$,
      magnetic field strength $B$, velocity $v$ and vorticity squared
      $(\nabla\times v)^2$ of the fiducial run are portrayed at times $20$ Myr
      and $60$ Myr. The images correspond to projections of thin layers
      ($100\ \mathrm{kpc}\times60\ \mathrm{kpc}\times4\ \mathrm{kpc}$) centred at
      $(50,0,0)$ and weighted by cell volume except for the velocity, which is
      weighted by the density. The grey contour corresponds to the jet tracer
      value $X_\mathrm{jet}=10^{-3}$. The characteristic transition from jet to
      lobe becomes evident.}
    \label{fig:evolution_absolute}
\end{figure*}

\begin{figure*}
\centering
\includegraphics[trim=0.25cm 0.65cm 1.4cm 1.45cm,clip=true,width=\evolutionfac\textwidth]{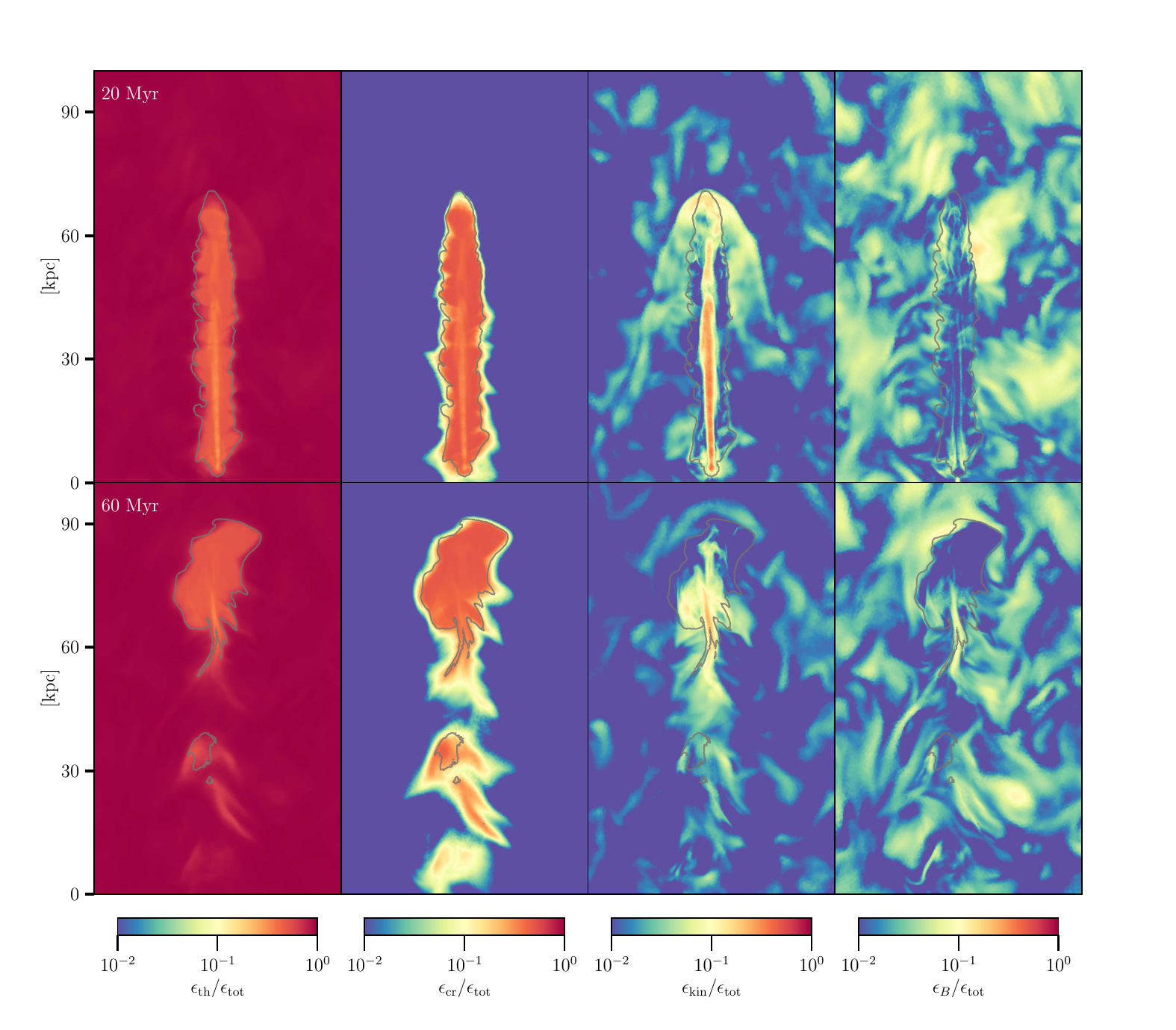}
    \caption{The thermal $\epsilon_\rmn{th}$, CR $\epsilon_\mathrm{cr}$, kinetic
      $\epsilon_\mathrm{kin}$ and magnetic $\epsilon_B$ energy density, which
      are normalised to the total energy density $\epsilon_\mathrm{tot}=
      \epsilon_\rmn{th}+\epsilon_\mathrm{cr}+\epsilon_\mathrm{kin}+\epsilon_B$
      of the fiducial run, are portrayed at times $20$ Myr and $60$ Myr. The
      images correspond to projections of thin layers
      ($100\ \mathrm{kpc}\times60\ \mathrm{kpc}\times4\ \mathrm{kpc}$), which are volume
      weighted and centred at $(50,0,0)$. The grey contour corresponds to the
      jet tracer value $X_\mathrm{jet}=10^{-3}$.}
    \label{fig:evolution_relative}
\end{figure*}

\begin{figure*}
\centering
\includegraphics[trim=.7cm -.2cm 1cm 1.cm,clip=true, width=0.51\textwidth]{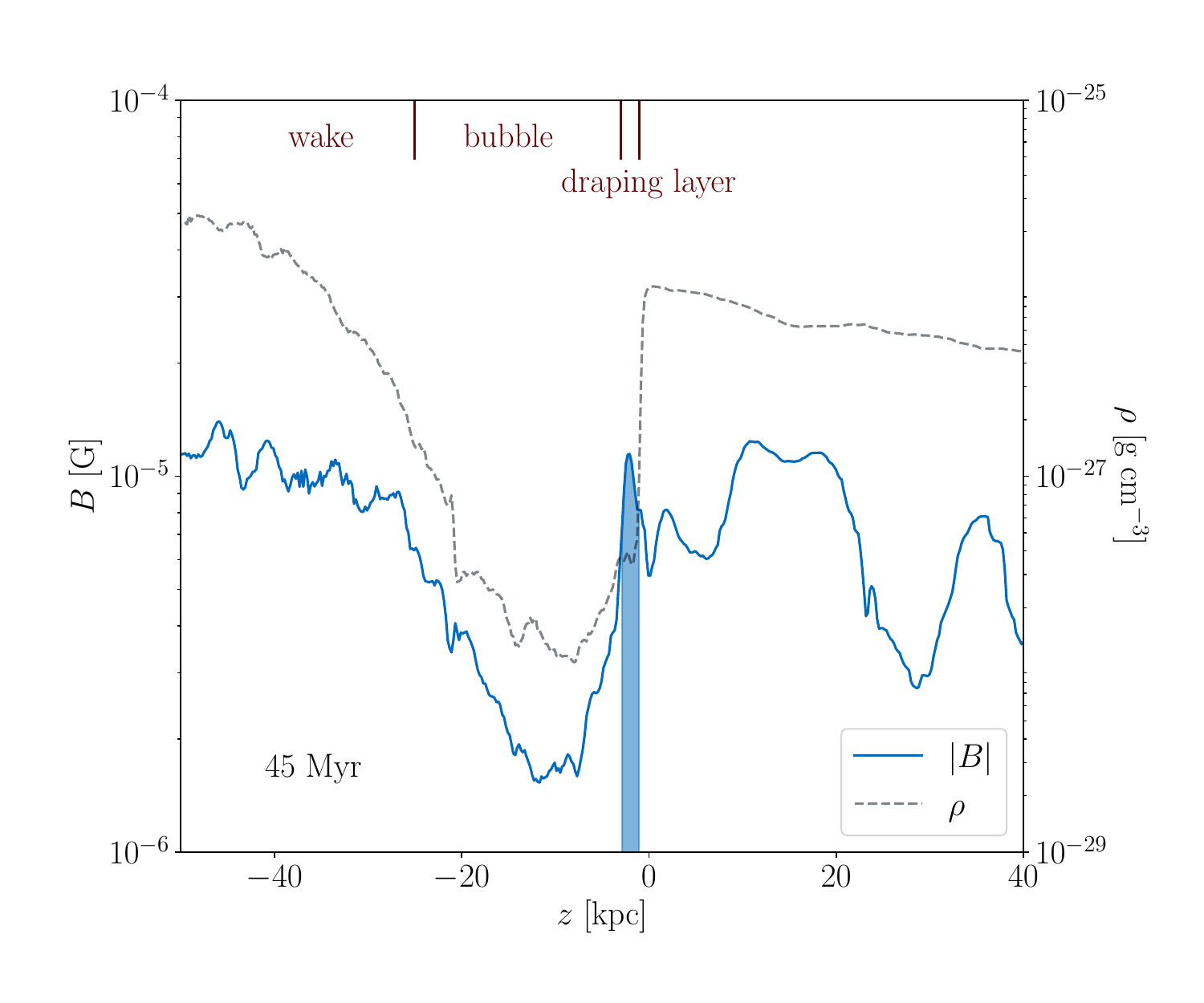}
\hspace{0.01\textwidth}
\includegraphics[trim=.7cm 0.1cm 3.2cm 2.cm,clip=true, width=0.46\textwidth]{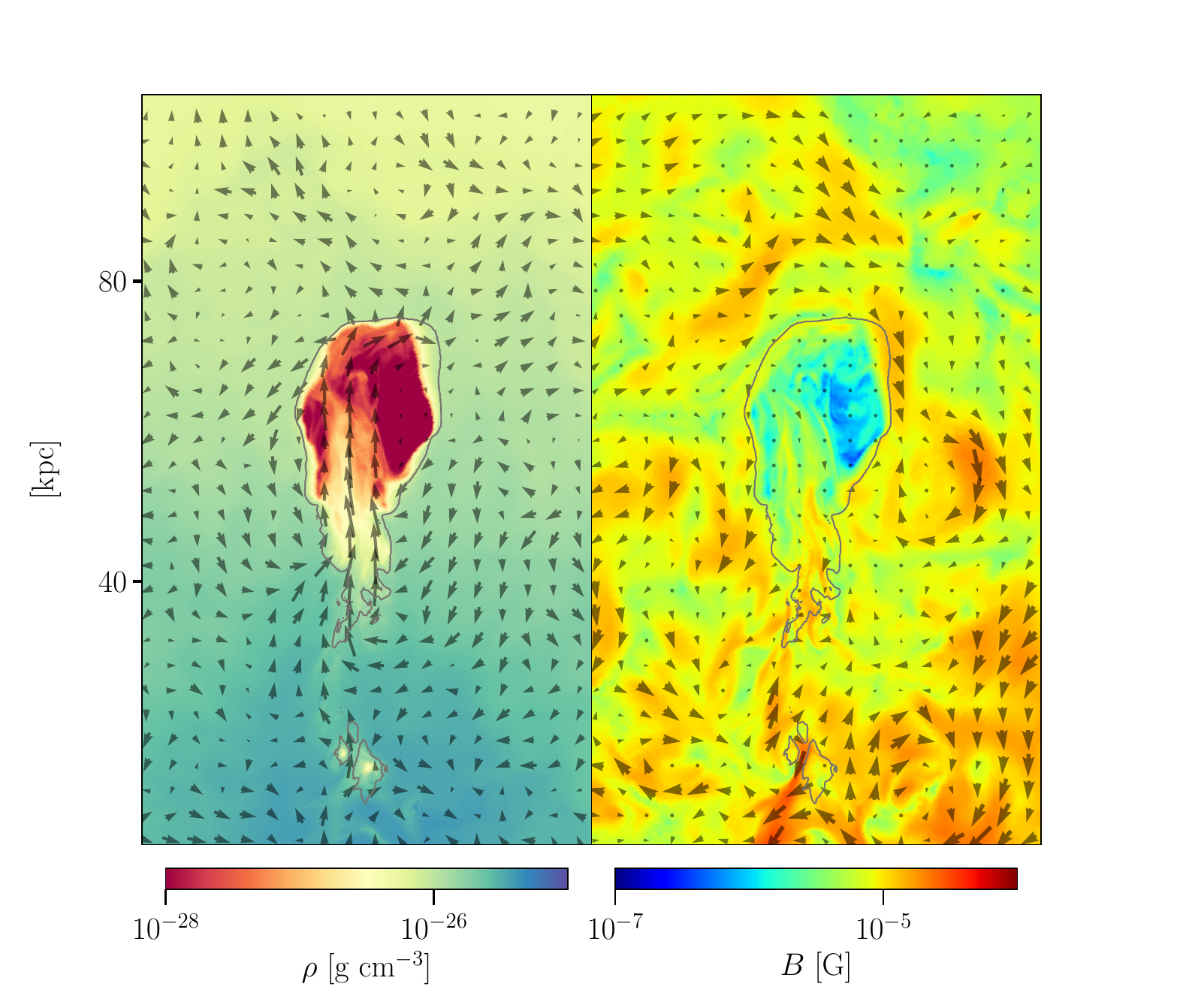}
    \caption{On the left, we show the mean magnetic field strength (blue) along
      a thin radially-aligned cylinder ($r=1\ \mathrm{kpc}$) through the bubble
      as a function of distance $z$ to the stagnation point at 45 Myr. We mark the
      thin magnetic draping layer with a filled blue contour and locate the
      extent of the bubble and its wake. To identify the extent of the bubble,
      the density is overlaid (black dashed). On the right, we show the density
      and magnetic field strength. The arrows correspond to the kinetic
      energy flux and the magnetic field directions, respectively, and the arrow
      length encodes the magnitude of these variables. The images correspond to
      projections of thin layers ($100\ \mathrm{kpc}\times60\ \mathrm{kpc}\times4\ \mathrm{kpc}$)
      centred on the wake of the bubble. The grey contour corresponds to the
      jet tracer value $X_\mathrm{jet}=10^{-3}$. It becomes apparent that the
      magnetic field amplification at the bubble surface is not due to the
      compression of gas but rather due to draping in the direction of
      movement. The width of the draping layer corresponds to the theoretically
      predicted value of $\approx2\ \mathrm{kpc}$.}
    \label{fig:draping_evolution}
\end{figure*}

\begin{figure*}
\centering
\includegraphics[trim=1.7cm 0.1cm 2.4cm 0.1cm,clip=true, width=0.49\textwidth]{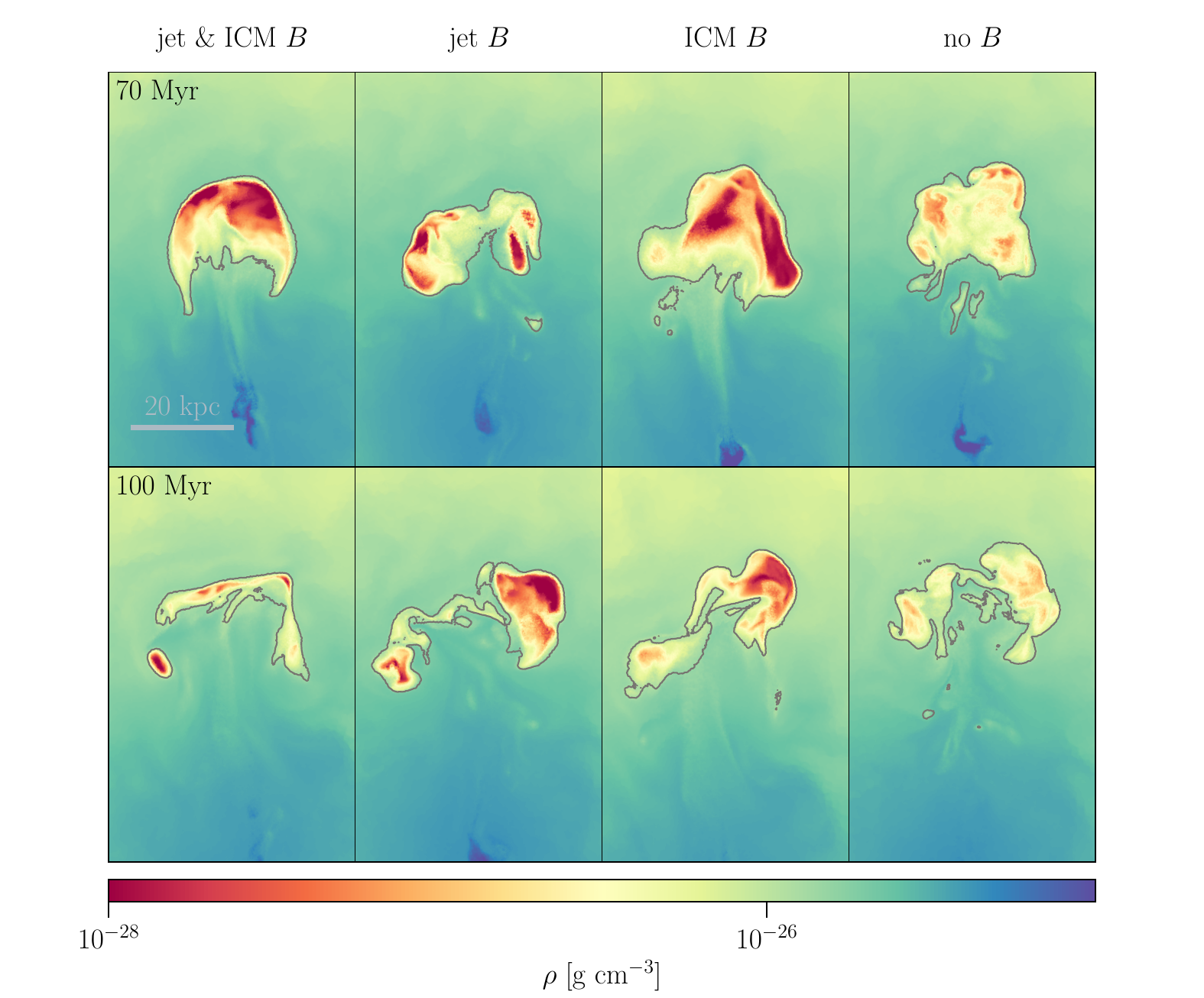}
\hspace{0.005\textwidth}
\includegraphics[trim=1.4cm -0.2cm 2.7cm 0.25cm,clip=true, width=0.49\textwidth]{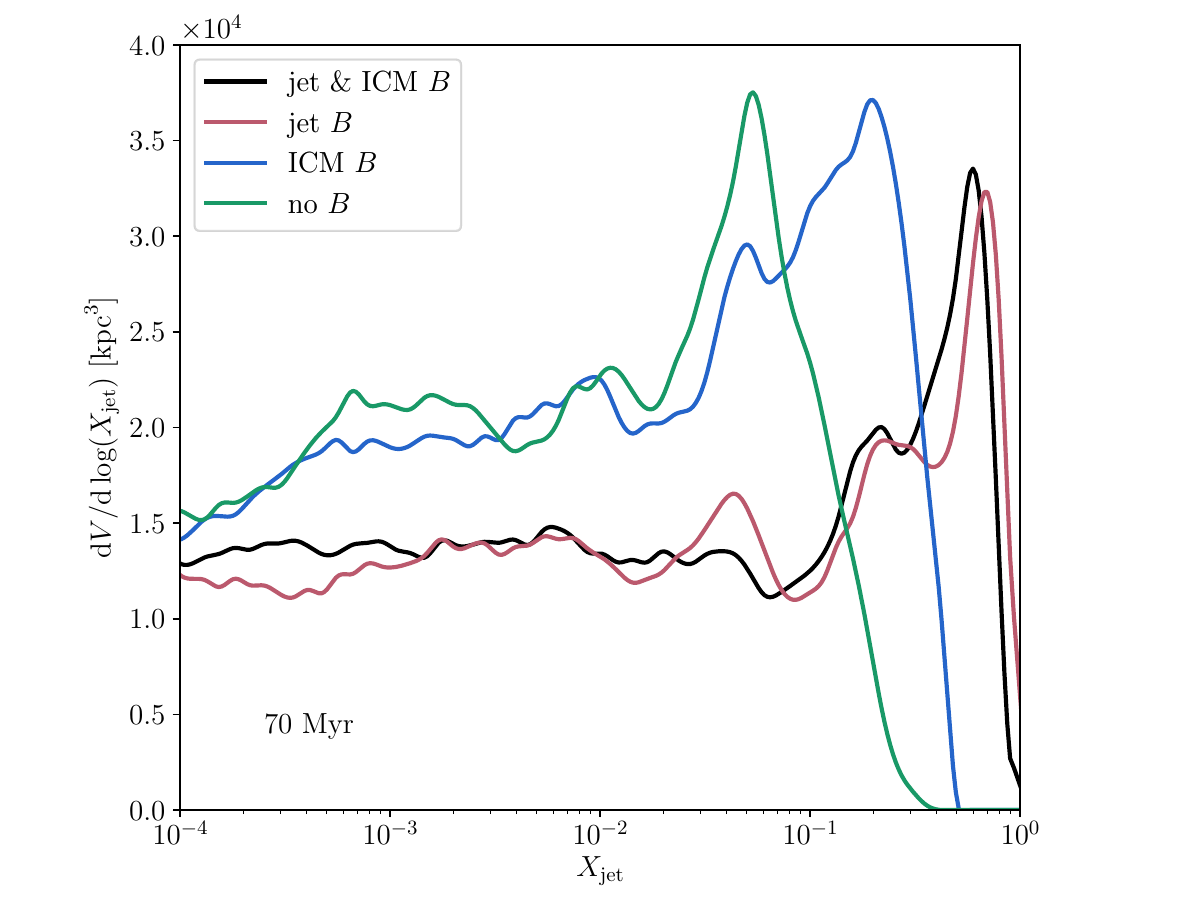}
\includegraphics[trim=1.4cm .2cm 2.7cm 0.55cm,clip=true, width=0.49\textwidth]{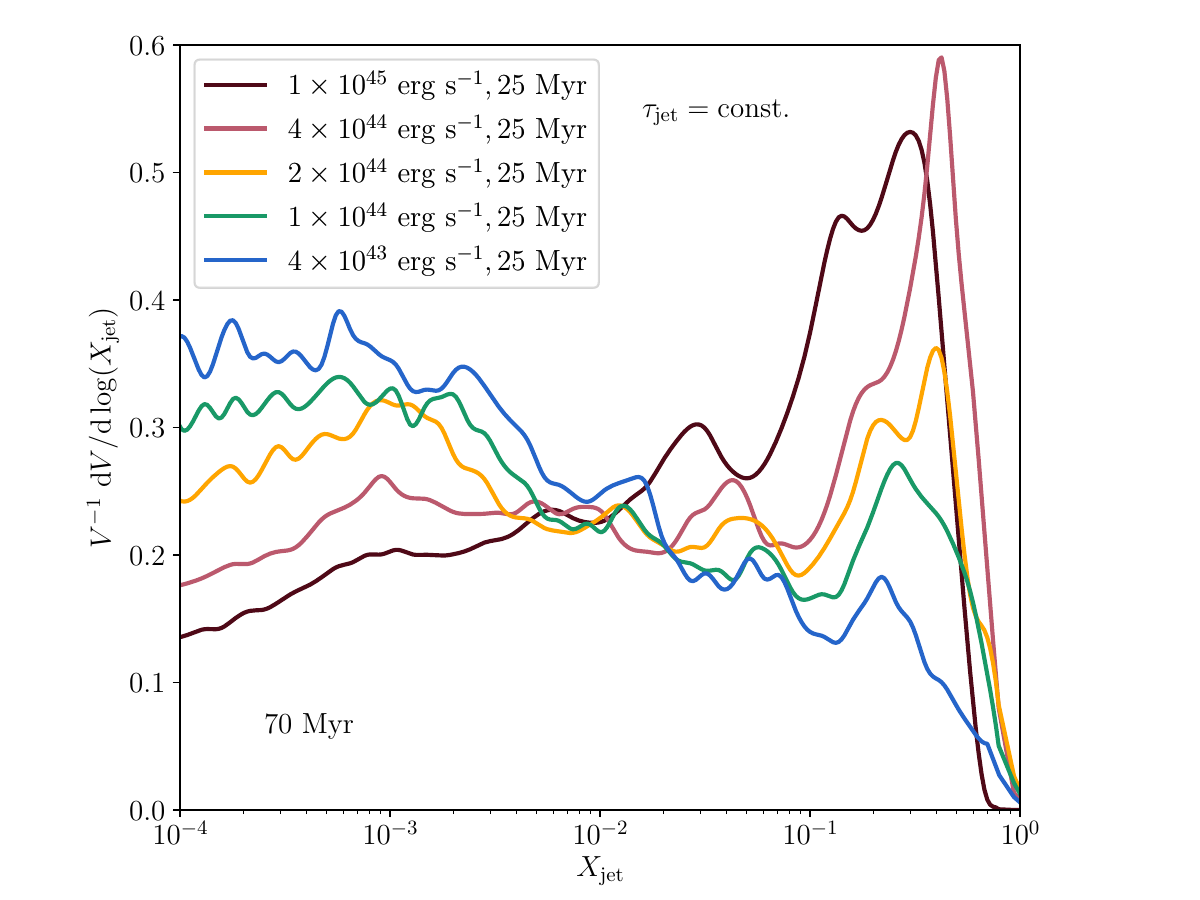}
\hspace{0.005\textwidth}
\includegraphics[trim=1.4cm .2cm 2.7cm 0.55cm,clip=true, width=0.49\textwidth]{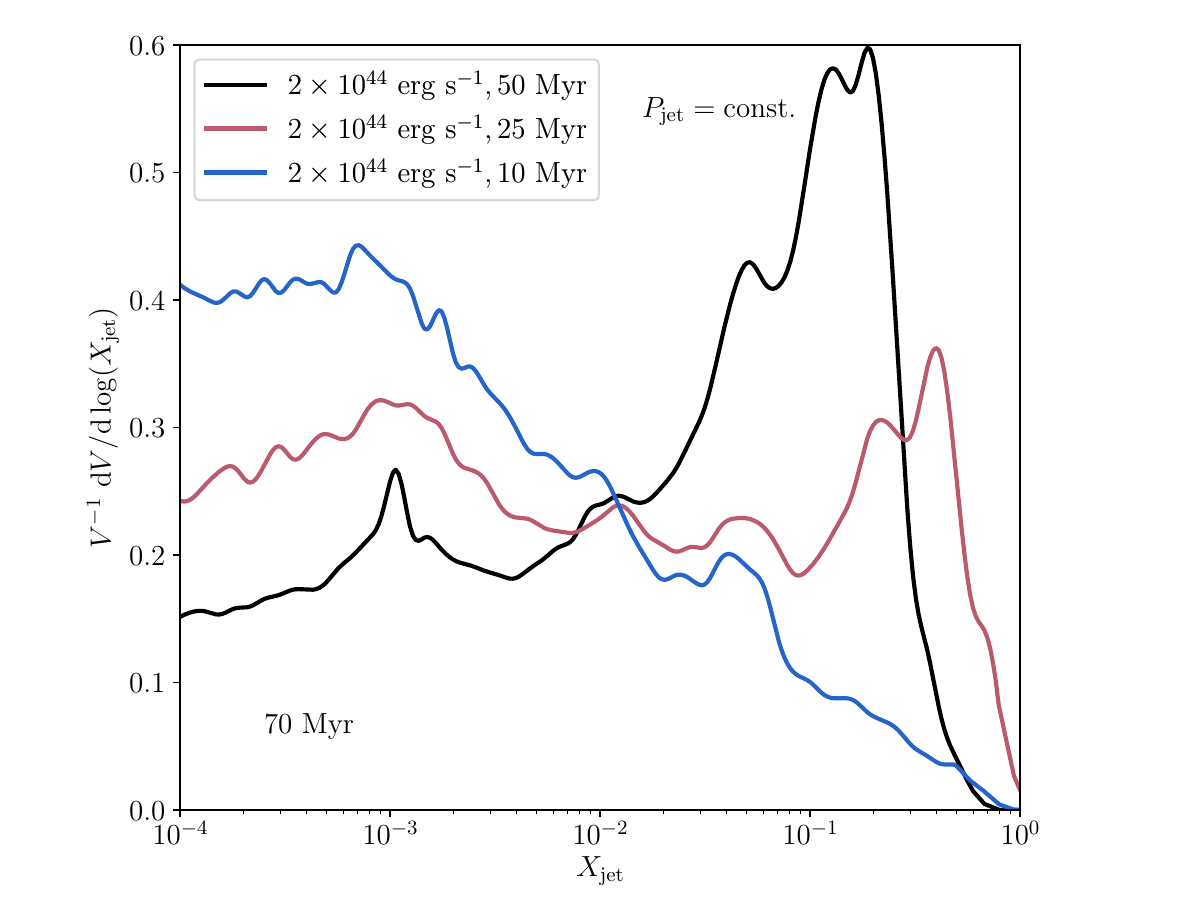}
\caption{The mixing efficiency of the bubble for different magnetic field
  parameters of our lower resolution simulations. On the top left, we show lobe-centred projections of thin layers ($80\ \mathrm{kpc}\times50\ \mathrm{kpc}\times4\ \mathrm{kpc}$)
  of the density with jet mass fractions overlaid ($X_\mathrm{jet}=10^{-3}$) for
  four simulations without CRs indicated above the panels. On the top right, the
  volume covered by gas, $\mathrm{d} V/\mathrm{d}\log(X_\mathrm{jet})$, with a
  given jet mass fraction $X_\mathrm{jet}$ is shown at 70 Myr for the same four
  simulations. The presence of magnetic fields significantly suppresses mixing
  in the bubble. The bottom panels portray the normalised volume covering
  fraction of a given jet mass fraction,
  $V^{-1}\ \mathrm{d}V/\mathrm{d}\log(X_\mathrm{jet})$. On the bottom left, we
  vary the jet power at constant time of jet activity
  $\tau_\mathrm{jet}=25\ \mathrm{Myr}$. On the bottom right, we show bubbles
  inflated by jets of constant power
  $P_\mathrm{jet}=2\times10^{44}\ \mathrm{erg/s}$. The peak of the volume
  fraction moves to the left for less energetic jets (decreased $P_\mathrm{jet}$
  or $\tau_\mathrm{jet}$) indicating increased mixing efficiencies for these
  jets. The lines are smoothed for clarity.}
    \label{fig:mixing_magneticField}
\end{figure*}

\section{Jet and bubble evolution}
\label{sec:evolution}
First, we discuss the evolution of our jets and bubbles in terms of global
quantities and with detailed maps of thermodynamical quantities.

\subsection{Global evolution}
The jets are initialised with a dominant kinetic energy component. Through
dissipation much of this kinetic energy is quickly thermalized. This would lead
to thermally dominated lobes, in conflict with X-ray observations. The subgrid acceleration scheme employed for CR maintains them in equipartition with thermal energy until $2\tau_\mathrm{acc}$. Thus, CRs share a significant
fraction of the energy and pressure in these simulations as shown in
Fig.~\ref{fig:energy_evolution} which shows the time evolution of different
energy components in the lobes. The strong dissipation of the kinetic energy of the jets effectively transfers kinetic energy into
CR and thermal energy. Compared to the initial magnetic-to-thermal pressure ratio in the jet $X_{B,\mathrm{jet}}=0.1$, the ratio drops by more than an order of magnitude due to the strong increase of thermal energy.

The thermal energy increases at later times whereas the CR energy decreases in
Fig.~\ref{fig:energy_evolution}. We adopt a morphology-based definition of lobes
with a jet mass fraction exceeding $10^{-3}$, which includes by definition
heavily mixed cells. Thereby, mixing leads to an increase of mass and thermal
energy in the lobes at later times. On the other hand, CRs continuously diffuse
out of the bubbles, lowering their total energy. Note that CR energy
losses due to cooling in the bubbles is solely restricted to the (almost
negligible) hadronic and Coulomb losses as we suppress Alfv\'enic cooling in the
bubble region (see Section~\ref{sec:CRs}).

The evolution of the thermodynamic cluster profiles
(Fig.~\ref{fig:radialProfile_general}) shows a decrease in density and an increase
in temperature at early times, which corresponds to the high thermal energy of
the propagating jet. Because this hot gas has such a low-density it
remains invisible in X-ray maps, which is in agreement with observations
\citep[e.g.,][]{Leccardi2008}. After the jet terminates and the bubbles rise
buoyantly, the mean thermodynamic profiles nearly recover the initial
conditions.

\subsection{Morphology}

The jet driving phase and subsequent phase of buoyant rise of the bubble are
portrayed in Figs.~\ref{fig:evolution_absolute} and \ref{fig:evolution_relative}
at 20~Myr and 60~Myr for the fiducial run. Figure~\ref{fig:evolution_absolute}
shows various thermodynamic variables projected in thin slices and
Fig.~\ref{fig:evolution_relative} displays various energy densities, normalised
to the total energy density. This enables direct comparisons of the dynamical
impact of the different components.

Initially, the jet penetrates into the ambient medium forming a bow shock at its
tip. The shock is visible in the first row of Fig.~\ref{fig:evolution_absolute}
as a discontinuity with increased density and velocity. The continuous mass flux
is deflected at the tangential discontinuity between the jet and shock. The
strong inflow expands horizontally and mixes with the shocked ICM gas
streaming around the jet. This leads to the creation of an extended, highly
turbulent region with high vorticity, see Fig.~\ref{fig:evolution_absolute}.

After the jet is switched off at 25 Myr, the remaining directed kinetic energy
flux thermalises and catches up with the previously injected material forming a
bubble, whose evolution is now solely driven by buoyancy. The high contrast in
density between the lobes and the ambient medium makes the setup susceptible to
the Rayleigh-Taylor instability. Thus, strong inflows develop in the wake of
the bubble. In the second row of Fig.~\ref{fig:evolution_absolute}, the
developing strong vorticity and high velocities are visible in the wake. They
cause strong mixing of the bubble material with the ambient medium, which is
accompanied by an increase in density.

The low magnetic field strength renders it dynamically irrelevant in the bulk of
the ICM (Fig.~\ref{fig:evolution_relative}). However, there are two cases where
the dynamics amplifies the magnetic field to the point that it plays a
significant role for the evolution of the system: first, there is significant
magnetic field amplification in the wake of the bubble
(Figs.~\ref{fig:evolution_absolute} and \ref{fig:evolution_relative}). This is
due to bubble-scale eddies that converge in the wake and adiabatically compress
the field and stretch it by the strong shear flows that become evident in the
vorticity map.

Second, the upwards motion of the bubbles in the ICM causes magnetic field lines
of the IGM to accumulate at the leading surface of the rising bubble, which is a
tangential discontinuity \citep{Pfrommer2010b}. These {\it draped} magnetic
field lines lead to the suppression of Kelvin-Helmholtz and Rayleigh-Taylor
instabilities (see also Section \ref{sec:draping}, for a detailed
discussion). The bubble is eventually disrupted by strong uplifts of wake
  material, which are able to penetrate through the centre of the bubble. If
  turbulence is absent the resulting torus structure remains stable on
  timescales of $>100\ \mathrm{Myr}$ \citep{Weinberger2017}. In our simulations,
  the onset of a developing (deformed) toroidal structure is visible after
  $\sim70-100\ \mathrm{Myr}$ (see Fig.~\ref{fig:mixing_magneticField}). However,
  the turbulent asymmetrical flow pattern of the uplift causes the break up of
  the lobe into two main pieces. If enough low density material remains, the
two cavities can continue to rise as two independent bubbles until the uplift
becomes too strong again and causes them to also break apart.

In addition to advection, CRs are allowed to anisotropically diffuse along
magnetic field lines. However, the draped magnetic layer on top of
the bubble confines the CRs inside and prevents escape ahead of the
bubble. Instead, CRs are able to diffuse out of the lower part of the bubble
along the strongly amplified magnetic filaments, which are bent by the strong
uplift and align with the jet axis (see Fig.~\ref{fig:evolution_absolute}).
The jet evolution and subsequent creation of the bubble is in excellent
agreement with previous work \citep[e.g.,][]{Lind1989,Reynolds2002}, suggesting
that the addition of internal and external magnetic fields as well as CRs does
not change the overall evolution of the system.

\section{Magnetic field evolution}
\label{sec:magneticfield}

In this section, we study the effect of magnetic draping at the bubble
interface, amplification of magnetic fields in the bubble's wake, and the effect
of magnetic fields on the mixing efficiency of the bubble fluid with the ICM.

\subsection{Magnetic draping and amplification}
\label{sec:draping}
Objects moving at super-Alfv\'enic speed through a magnetised medium accumulate
magnetic field lines at their interface. This magnetic draping effect occurs
only if the magnetic coherence scale is sufficiently large, i.e., $\lambda_B
\gtrsim R/\mathcal{M}_{\rmn{A}}$ where $R$ is the curvature radius at the
stagnation point and $\mathcal{M}_{\rmn{A}}$ is the Alfv\'enic Mach number
\citep{Pfrommer2010b}.  In a steady state, the rate at which new magnetic field
lines enter this strongly magnetised sheath balances the rate at which magnetic
field lines are advected over the bubble's surface to eventually leave the
draping layer. The magnetic tension exerted by this draping layer slows
down the object as the magnetic field is anchored in the ICM in ideal MHD and
thus acquires a large inertia \citep{Dursi2008}. Moreover, draping makes the
object more resilient against interface instabilities of the Kelvin-Helmholtz or
Rayleigh-Taylor type \citep{Dursi2007}. This draping layer inhibits any particle
transport across the bubble surface such as CR diffusion \citep{Ruszkowski2008},
heat conduction by thermal electrons and momentum transport or viscosity by
thermal protons, which stabilises sharp temperature and density
transitions observed in the ICM (i.e., cold fronts) against disruption
\citep{Vikhlinin2001,Lyutikov2006,Asai2007}. Previous simulations acknowledge
the stabilising effect of magnetic fields
\citep{Jones2005,ONeill2009,Bambic2018}. Here, for the first time, we present
results for the case of self-consistently inflated bubbles in a realistic
turbulent magnetised environment.

Magnetic draping creates a layer of thickness $l$ that is smaller than the
curvature radius $R$ of the bubble at the stagnation point by \citep{Dursi2008}:
\begin{equation}
  \label{eq:draping}
  \frac{l}{R}=\frac{1}{6\alpha\mathcal{M}^2_\mathrm{A}}
  =\frac{1}{3\alpha\beta\gamma\mathcal{M}^2}
  =\frac{1}{3}\frac{B_{\mathrm{ICM}}^2}{B_\mathrm{max}^2},
\end{equation} 
where $\alpha$ describes the magnetic-to-ram pressure ratio at the stagnation
point ($B_\mathrm{max}^2/8\pi=\alpha\rho v^2$), $\mathcal{M}_\mathrm{A}$
corresponds to the Alfv\'enic Mach number
$\mathcal{M}_\mathrm{A}=v/v_\mathrm{A}$, $\mathcal{M}$ is the sonic Mach number,
$\gamma=5/3$ is the adiabatic index and $X_{B,\mathrm{ICM}}=\beta^{-1}$ is the
thermal-to-magnetic pressure ratio in the upstream ICM. In our simulations,
$X_{B,\mathrm{ICM}}=0.05$, we determine $\mathcal{M}\approx0.2$ and take
$\alpha\approx2$ from \cite{Dursi2008}. Thus, we expect draping layer thickness
$l\approx2\ \mathrm{kpc}$ for a curvature radius $R=20\ \mathrm{kpc}$ at our
simulated bubble. This implies that we are able to numerically resolve the
draping layer due to our refinement prescriptions.

Figure~\ref{fig:draping_evolution} (left-hand panel) shows the magnetic field
strength (blue line) along the stagnation line of the bubble. At the bubble
edge, which is clearly identified by the sharp increase in density (black
dashed), the magnetic field shows a pronounced narrow peak with width
$l\approx2\ \mathrm{kpc}$. In agreement with our theoretical predictions, the
magnetic field rises from its mean value of $B_{\mathrm{ICM}}\approx5~\umu$G to
$B_\mathrm{max}\approx10~\umu$G in the draping layer (see
equation~\ref{eq:draping}).  In the magnetic field map (right-hand panel in
Fig.~\ref{fig:draping_evolution}), a thin enhanced magnetic field layer is also
visible that corresponds to the draping layer surrounding the
bubble. Consequently, we confirm that magnetic draping is an active process in
bubble dynamics in agreement with previous studies
\citep{Ruszkowski2007,Dursi2008}.

In the wake of the jet, magnetic field lines are stretched by differential
motions and compressed by converging downdrafts that compensate the upwards
motion of the bubble. This process amplifies magnetic fields in the wake and
aligns field lines with the jet axis (Fig.~\ref{fig:draping_evolution}), in
agreement with previous findings \citep{ONeill2010,Mendygral2012}. This has
important consequences for the CRs, which are confined in the bubble by the
draped magnetic field at the leading interface. Hence, escape from the bubble
only becomes possible in its wake as the converging eddies connect the bubble
interior magnetically with the ICM via the amplified magnetic filaments (see
Figs.~\ref{fig:evolution_absolute} and \ref{fig:draping_evolution}). As a
result, CRs are conducted diffusively along the magnetic filaments towards the
cluster centre in the opposite direction to the CR pressure gradient
(Fig.~\ref{fig:evolution_absolute}).

\begin{figure*}
\centering
\includegraphics[trim=0.25cm 0.65cm 1.4cm 1.45cm,clip=true,width=\evolutionfac\textwidth]{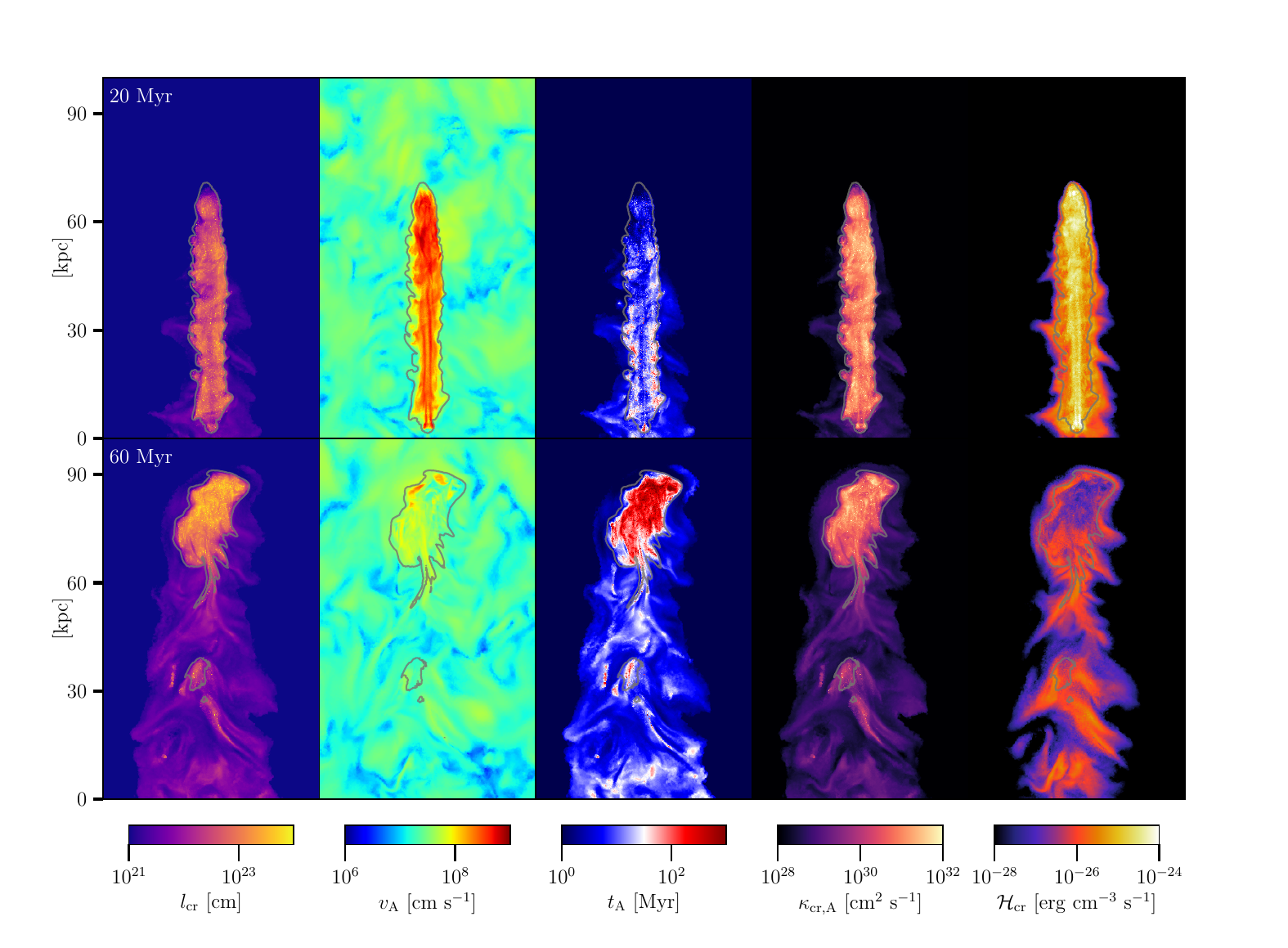}
\caption{Quantities related to CR transport in jet and lobe. We show the CR
  gradient length scale $l_\mathrm{cr}=P_\mathrm{cr}/\bnabla{}P_\mathrm{cr}$,
  the Alfv\'en velocity $v_\mathrm{A}$, the Alfv\'en cooling time scale of CRs,
  $t_\mathrm{A}=l_\mathrm{cr}/v_\mathrm{A}$, the instantaneous CR diffusion
  coefficient $\kappa_\mathrm{cr,A}=v_\mathrm{A}l_\mathrm{cr}$ and the Alfv\'en
  heating rate
  $\mathcal{H}_\mathrm{cr}=\left|\vect{v}_\mathrm{A}\cdot\bnabla{P}_\mathrm{cr}\right|$
  of the fiducial run at $20$ Myr and $60$~Myr. Here, we also display the
  measured heating rate inside the lobes to illustrate the strong increase of
  Alfv\'en heating at the lobe edges due to the large numerical CR pressure
  gradients there, which is for this reason suppressed in our simulations.  The
  images correspond to projections of thin layers
  ($100\ \mathrm{kpc}\times60\ \mathrm{kpc}\times4\ \mathrm{kpc}$) weighted with the
  volume and centred at $(50,0,0)$. We only show $l_\mathrm{cr}$,
  $t_\mathrm{A}$, $\kappa_\mathrm{cr,A}$ and $\mathcal{H}_\mathrm{cr}$ in
  regions with $\epsilon_\mathrm{cr}>10^{-14}\ \mathrm{erg}\ \mathrm{cm}^{-3}$
  to exclude an energetically subdominant CR population, which has partially
  experienced numerical diffusion. The CR gradient is unavailable due to
  numerical reasons for individual saturated cells which are shown with
  yellow. The grey contour corresponds to the jet tracer value
  $X_\mathrm{jet}=10^{-3}$.}
    \label{fig:evolution_CRparameters}
\end{figure*}

\begin{figure*}
\centering
\includegraphics[trim=1.75cm .85cm 5.9cm 0.25cm,clip=true, width=0.48\textwidth]{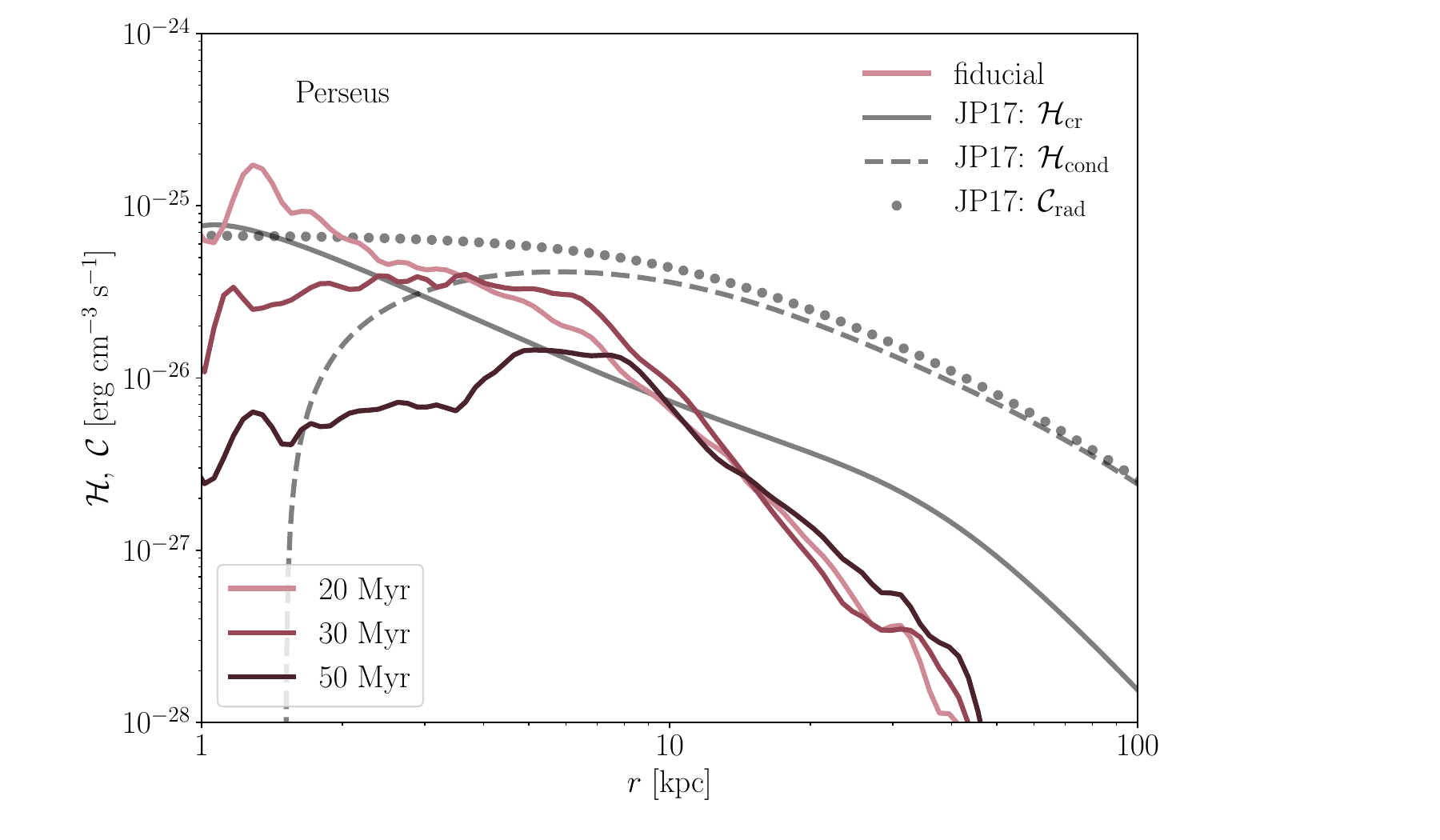}
\hspace{0.01\textwidth}
\includegraphics[trim=1.75cm .85cm 5.9cm 0.25cm,clip=true, width=0.48\textwidth]{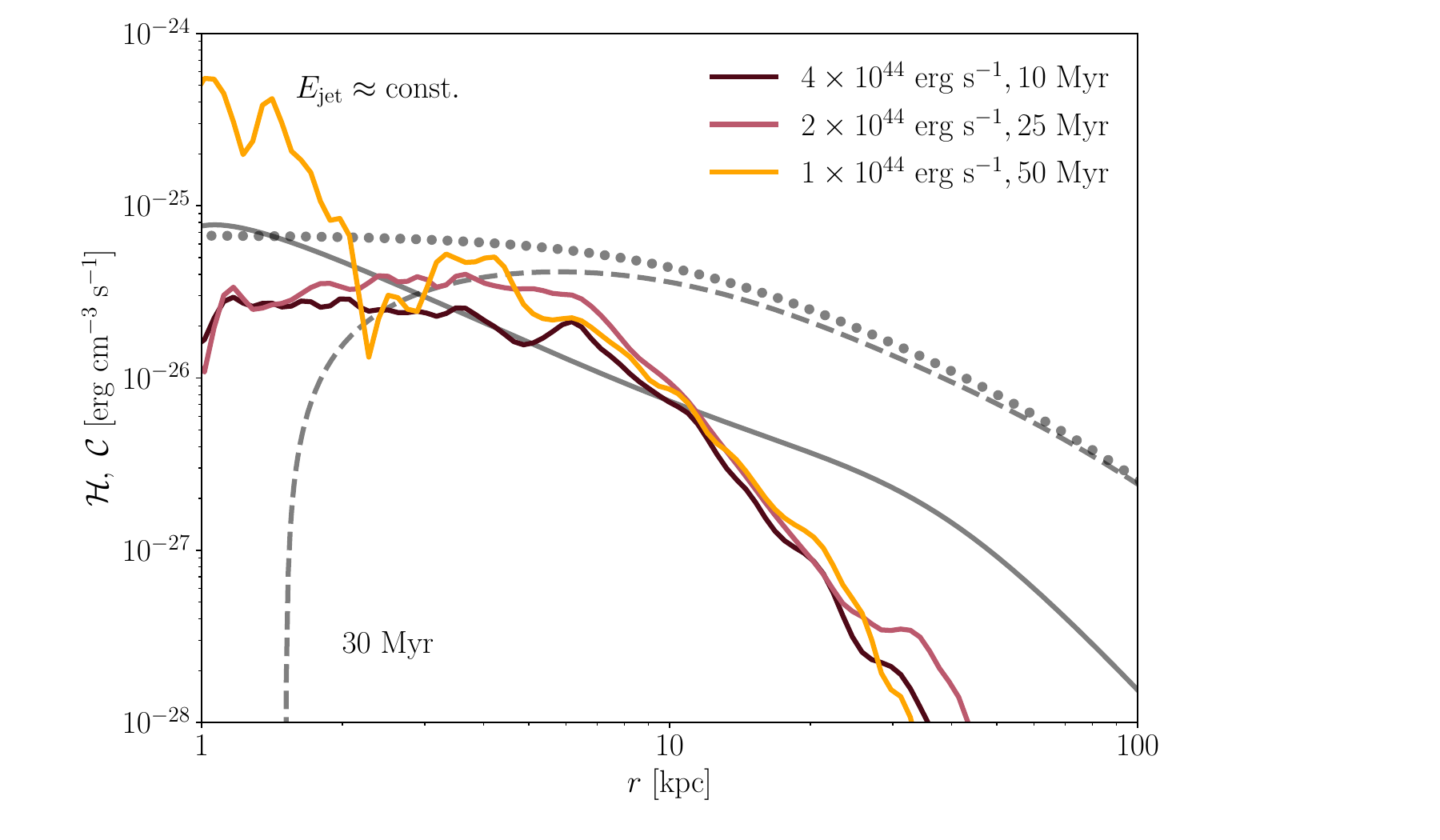}
\includegraphics[trim=1.75cm .85cm 5.9cm 0.25cm,clip=true, width=0.48\textwidth]{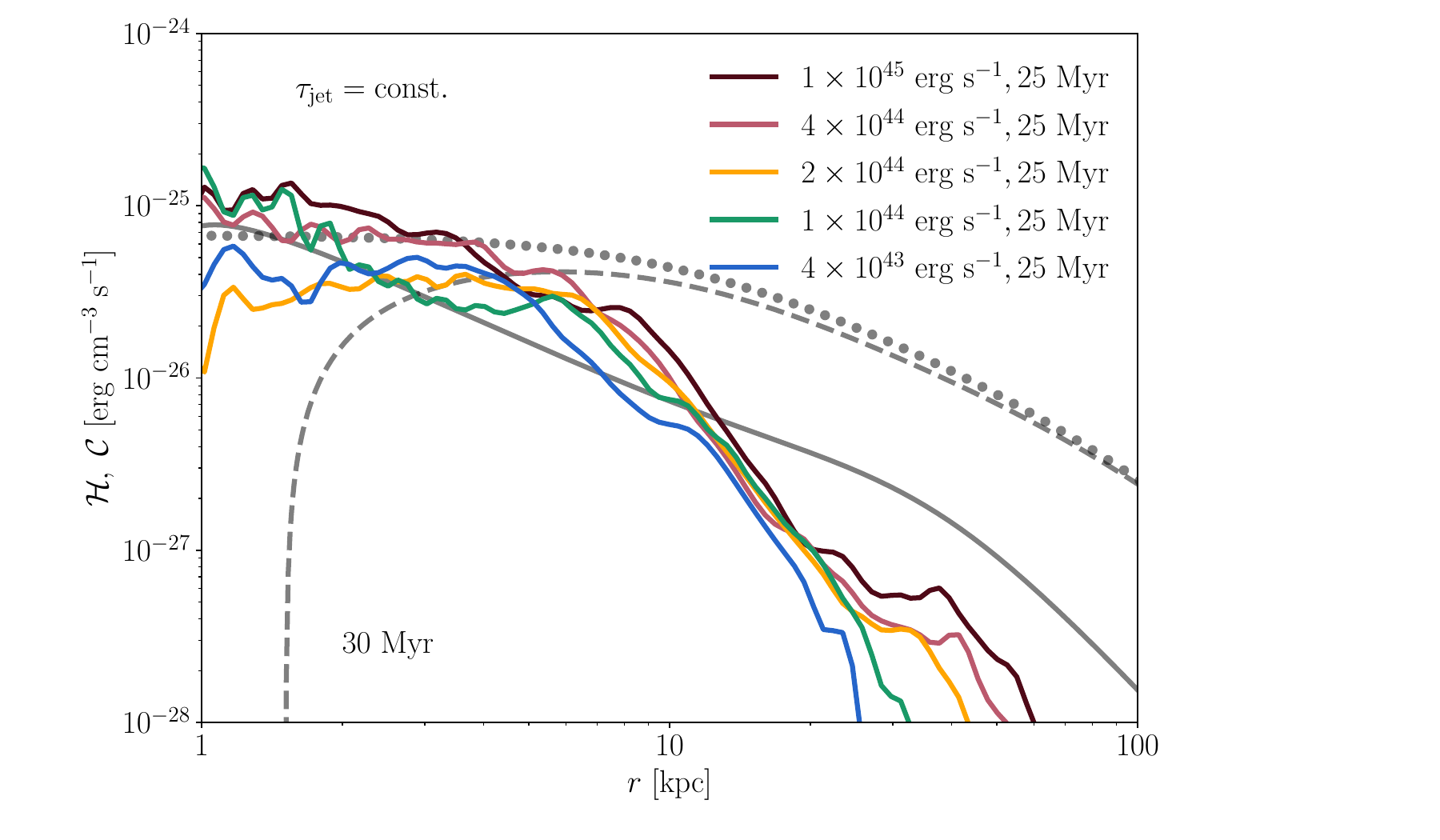}
\hspace{0.01\textwidth}
\includegraphics[trim=1.75cm .85cm 5.9cm 0.25cm,clip=true, width=0.48\textwidth]{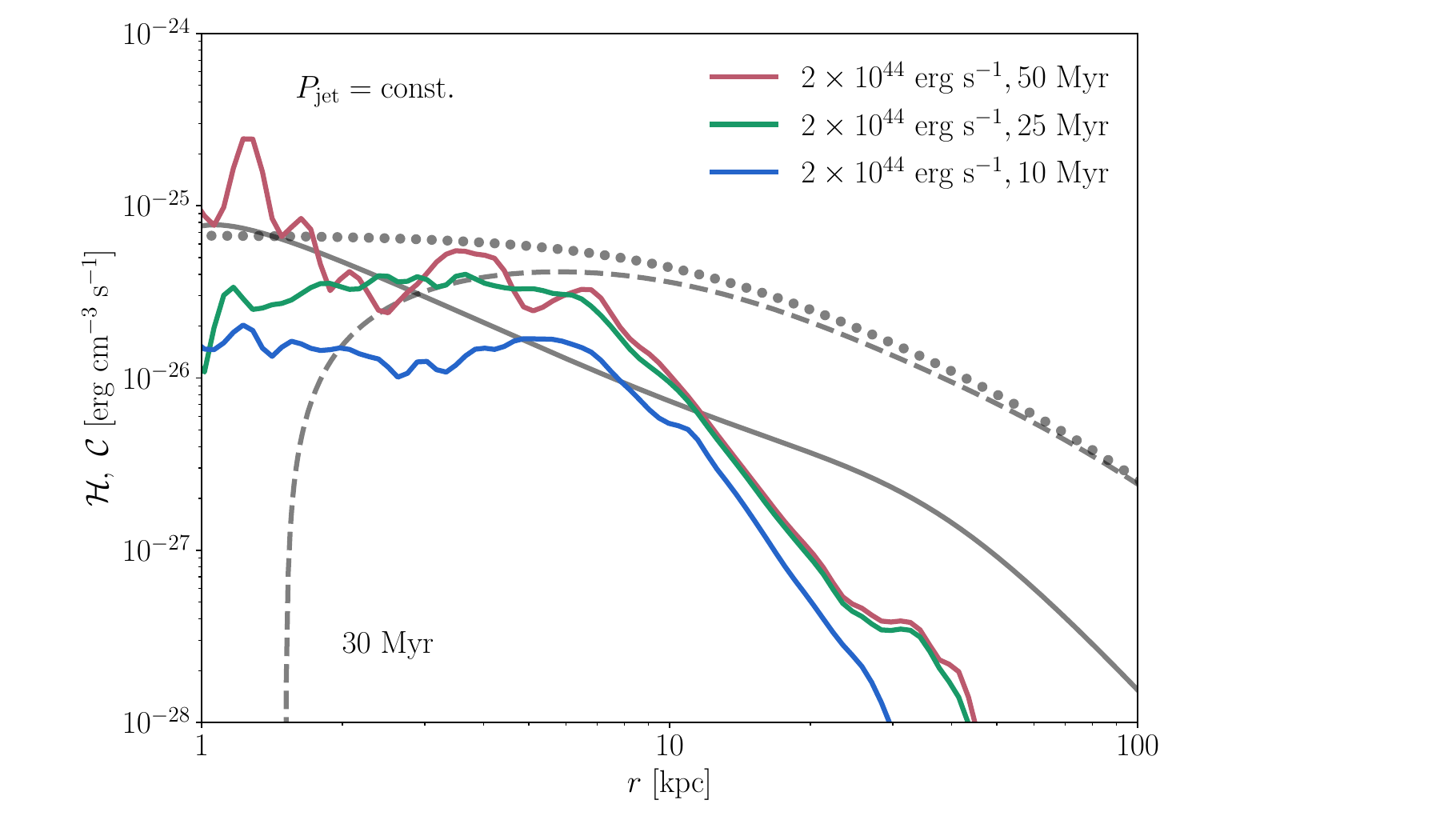}
\caption{Radial profiles of the Alfv\'en heating rate in the ICM due to CR
  streaming, $\mathcal{H}_\mathrm{cr}$ (restricting ourselves to ICM simulation
  cells with $X_\mathrm{jet}<10^{-3}$ and using our lower resolution
  simulations).  We compare our simulation to a spherically symmetric, steady
  state solution in which CR heating (solid) and conductive heating
  ($\mathcal{H}_\mathrm{cond}$, dashed) balances radiative cooling
  ($\mathcal{C}_\mathrm{rad}$, dotted) in the Perseus cluster
  \citep{Jacob2016a}.  On the upper left, we show different radial profiles at
  times 20, 30 and 50 Myr for our fiducial model. On the upper right, we compare
  profiles of jets with similar energy
  ($E_\mathrm{jet}\approx1.58\times10^{59}\ \mathrm{erg}$) but different
  $P_\mathrm{jet}$ and $\tau_\mathrm{jet}$: the heating rate profile is steeper
  for low-luminosity jets with longer activity times. On the bottom left, we
  compare jets with varying luminosity but constant duration
  $\tau_\mathrm{jet}=25$: while the maximum heating radius scales with jet
  luminosity, there is no clear trend of $\mathcal{H}_\mathrm{cr}$ with
  luminosity at smaller radius. On the bottom right, we compare jets with
  constant jet power
  $P_\mathrm{jet}=2\times10^{44}\ \mathrm{erg}\ \mathrm{s}^{-1}$ but varying jet
  activity time $\tau_\mathrm{jet}$: the prolonged CR production and ability to
  diffuse back to the centre for longer times compensates for the greater extent of the more energetic jets. The radial profiles generally
  correspond to volume-weighted averages.}
    \label{fig:radialProfile_CR}
\end{figure*}

\begin{figure*}
\centering
\includegraphics[trim=1.4cm 0cm 3.2cm 0.7cm,clip=true, width=0.49\textwidth]{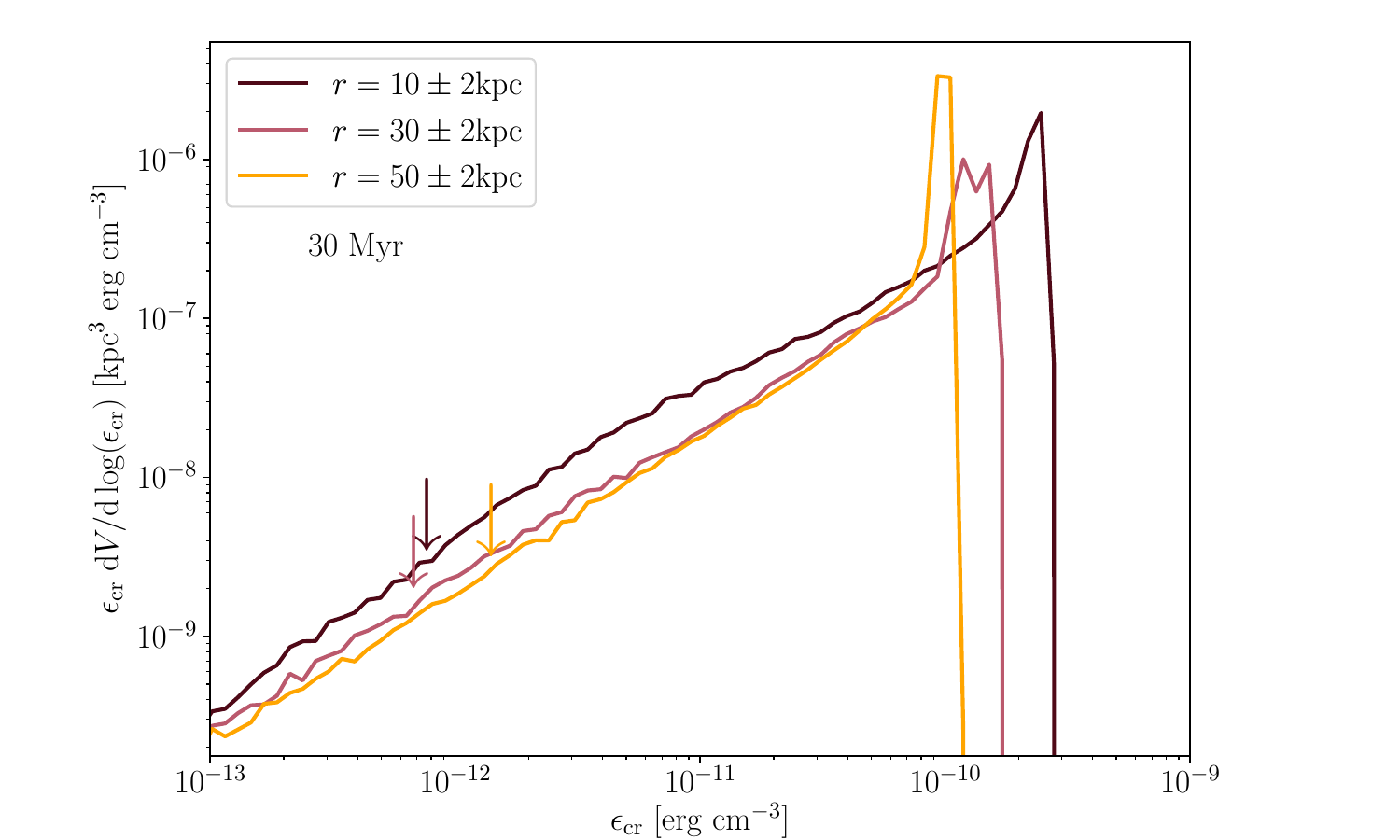}
\hspace{0.005\textwidth}
\includegraphics[trim=1.4cm 0cm 3.2cm 0.7cm,clip=true, width=0.49\textwidth]{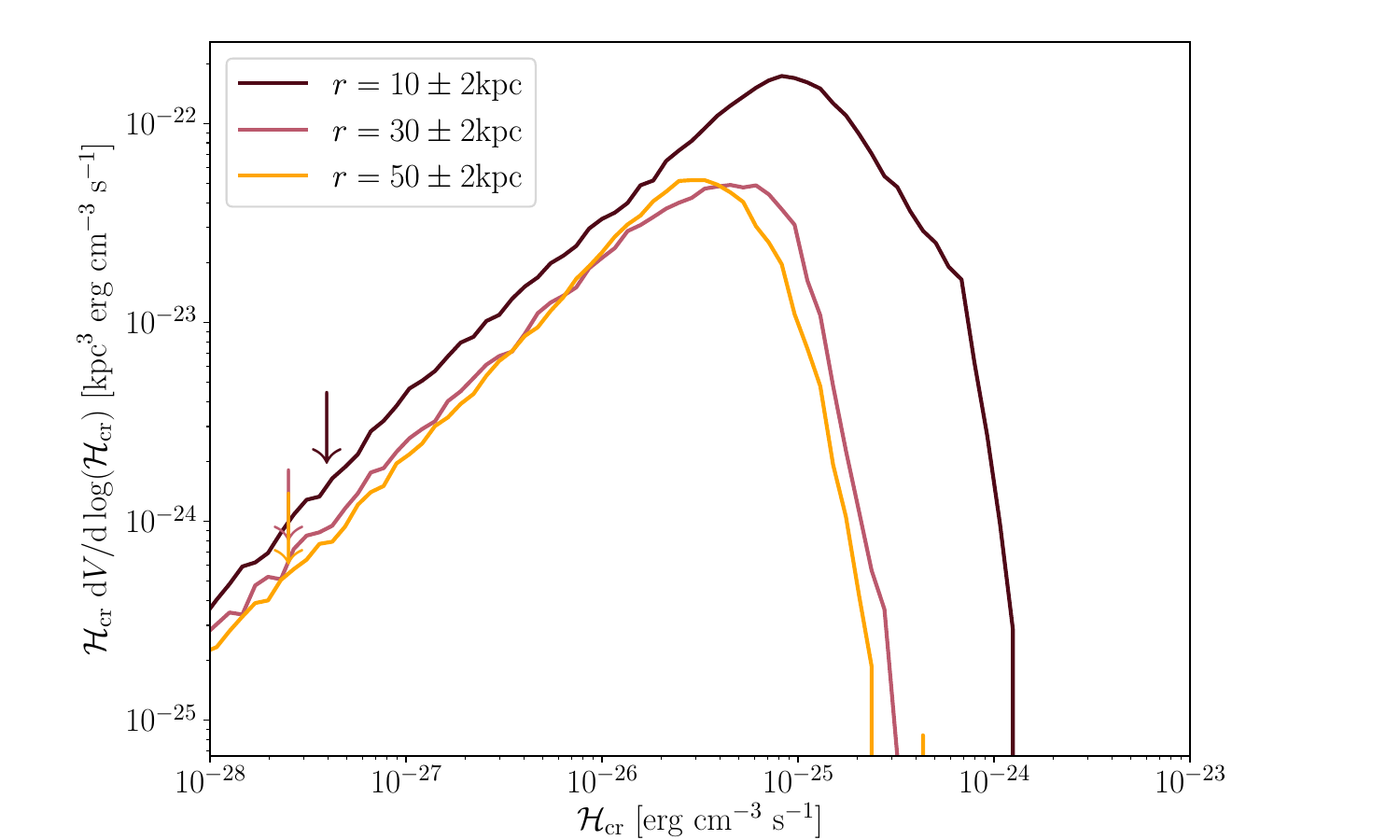}
\includegraphics[trim=1.55cm 0cm 3.2cm 0.55cm,clip=true, width=0.49\textwidth]{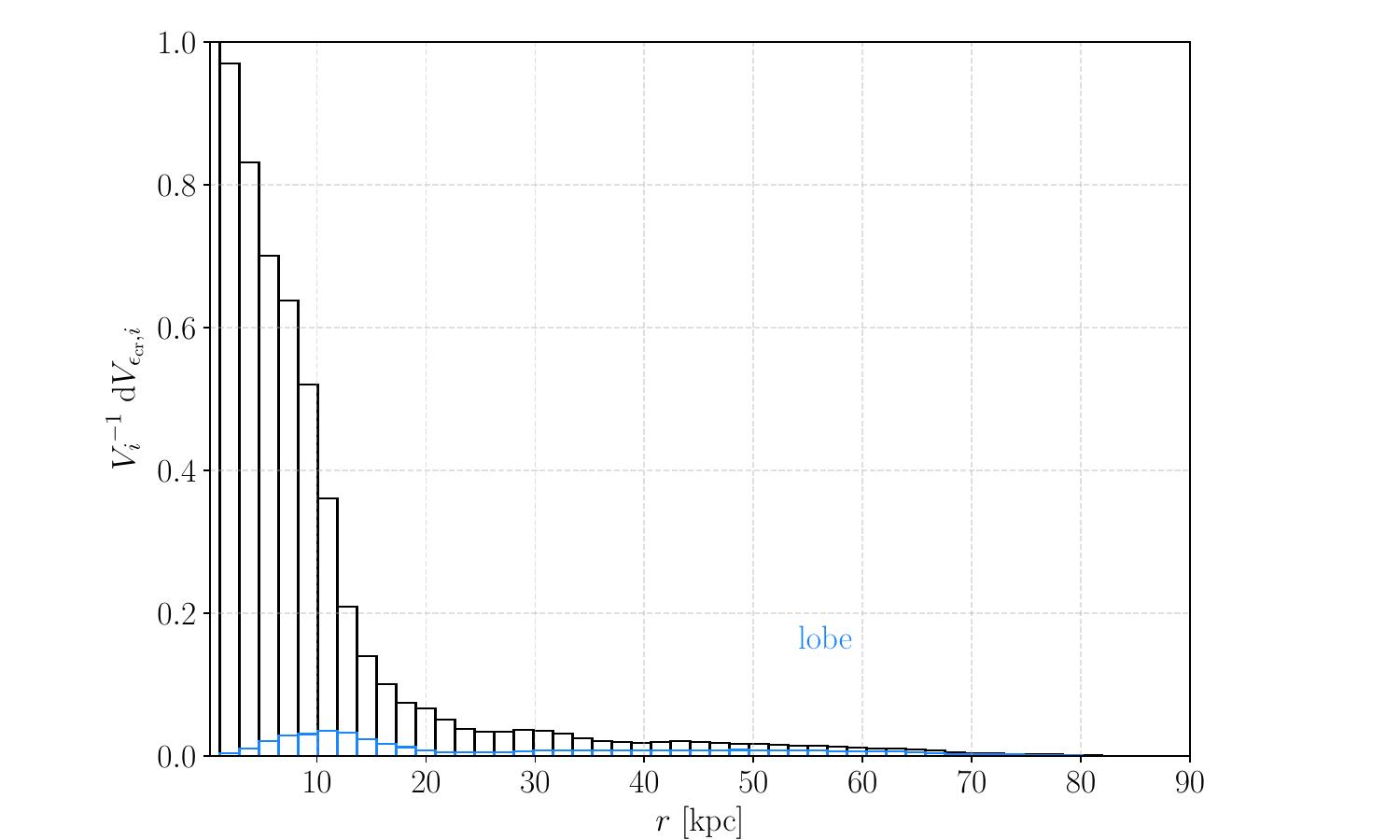}
\hspace{0.005\textwidth}
\includegraphics[trim=1.55cm 0cm 3.2cm 0.55cm,clip=true, width=0.49\textwidth]{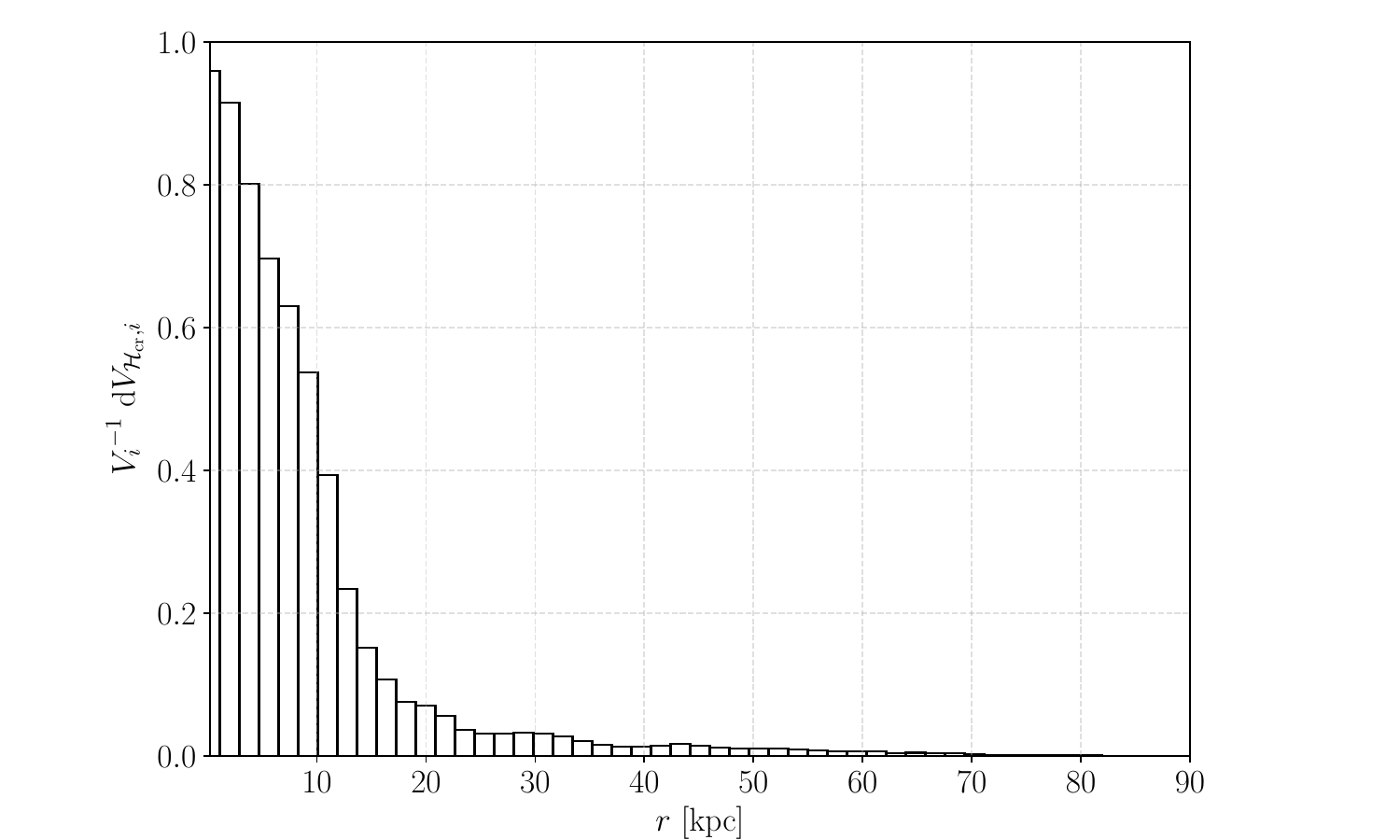}
\caption{Weighted volume distributions and filling factors of the CR
  energy density $\epsilon_\mathrm{cr}$ and the Alfv\'en heating rate
  $\mathcal{H}_\mathrm{cr}$ of our fiducial simulation at $30\ \mathrm{Myr}$. We
  show the weighted volume distribution of CR energy density
  $\epsilon_\mathrm{cr}\mathrm{d}V/\mathrm{d}\log(\epsilon_\mathrm{cr})$ (top
  left) and CR heating rate density
  $\mathcal{H}_\mathrm{cr}\mathrm{d}V/\mathrm{d}\log(\mathcal{H}_\mathrm{cr})$
  (top right), which characterize the isotropy of our CR distribution. The
  arrows mark the minimum CR energy density and heating rate density necessary
  to cover $3\sigma$ (99.8\%) of the CR energy and heating power at that radius,
  respectively. Due to their similarity, we adopt a common value for the floor
  values $\epsilon_\mathrm{cr,min}=10^{-12}\ \mathrm{erg}\ \mathrm{cm}^{-3}$ and
  $\mathcal{H}_\mathrm{cr,min}=3\times10^{28}\ \mathrm{erg}\ \mathrm{cm}^{-3}\ \mathrm{s}^{-1}$
  at all radii.  In the bottom panels, we show the filling factor of
  $\epsilon_\mathrm{cr}$ and $\mathcal{H}_\mathrm{cr}$ for cells above these
  thresholds.  Note that we suppress heating rates inside bubbles. The filling factor of the lobes ($X_\mathrm{jet}>10^{-3}$) is overplotted in the bottom
  left panel. The small filling factor of the lobes highlights the importance
  of CR diffusion to isotropize CRs. The large volume filling factor of
  $\epsilon_\mathrm{cr}$ and $\mathcal{H}_\mathrm{cr}$ becomes evident for
  $r\lesssim15\ \mathrm{kpc}$, suggesting that CR heating is isotropic at small
  radii.}
    \label{fig:CR_isotropy}
\end{figure*}

\subsection{Mixing}
\label{sec:mixing}
In order to analyse the mixing efficiency for different magnetic field
configurations, we use a suite of simulations without CRs (bottom in Table
\ref{Tab:JetPara}) to solely focus on magnetic effects. As discussed in the
previous section, the effect of draping stabilises the bubble against early
disruption from interface instabilities. However, the draping layer only
suppresses the growth of wave modes along the direction of the mean field but
not perpendicular to it. As the bubble rises in the cluster potential, its
surface is constantly warped and twisted by the turbulence, which causes
incomplete alignments of the external turbulent magnetic field with respect to
the bubble surface. Possibly, magnetic fields also cancel out through numerical
magnetic reconnection. These effects likely compromise the effects of draping
temporarily for our complex simulations in comparison to idealised
setups. In addition, we model the effect of helical fields, which develop in
  the turbulent bubble from the initially toroidal fields in the jet and also
  show stabilising effects \citep{Ruszkowski2007}.

The simulations in the top left panel of Fig.~\ref{fig:mixing_magneticField}
show a case with internal helical and external turbulent fields, two cases with
either one of the two field configurations and one without any magnetic
fields. Visually the morphology of the simulations including magnetic fields
appear similar to the ones without (accounting for projection effects). However,
the density contrast within the lobes is smaller in the simulation without
magnetic fields, indicating that mixing is suppressed by either the helical lobe
field and/or draped external fields. This can also be seen in the top right panel of
Fig.~\ref{fig:mixing_magneticField}, where we show the volume covered by a given
jet mass fraction $X_\mathrm{jet}$. Initially, the distribution peaks at
$X_\mathrm{jet}=1$ and if the bubble was perfectly insulated the distribution
would stay there. The faster the distribution moves to lower values of
$X_\mathrm{jet}$, the more efficiently does the bubble material mix with the
ICM. Conversely, a slow evolution indicates a significant suppression of
mixing. Accounting for magnetic fields in the ICM, which enables magnetic
draping, shifts the filling factor to higher values of
$X_\mathrm{jet}$ indicating an insulating effect of draping and preventing fast
mixing. In reality, this suppression of mixing should be even higher as the
driven external turbulence is weaker in the case without magnetic fields in
comparison to the case of ICM magnetic fields (see Section~\ref{sec:methods})
since turbulence amplifies mixing \citep[e.g.,][]{Ogiya2018}.

Similarly, our simulation with purely internal helical magnetic fields
suppresses mixing in comparison to the case without $\vecbf{B}$. This suggests
that our case with internal and external fields should show an even smaller
degree of mixing in comparison to either of the two individual cases with a
single magnetic component. However, there is only a slightly smaller jet mass
fraction retained in this double-magnetic case in comparison to the internal
field case (Fig.~\ref{fig:mixing_magneticField}). The loss of stability in the
case of the additional draping layer can be explained with the increase in
driven external turbulence for a magnetized ICM.

In the lower left panel of Fig.~\ref{fig:mixing_magneticField}, we compare
runs with constant jet lifetime ($\tau_\mathrm{jet}$) but varying jet power
($P_\mathrm{jet}$). We find a decrease in the mixing efficiency for jets with
higher power in agreement with \citet{Bruggen2002a}. Our normalisation ensures
comparability across differently sized lobes. The high-power jets penetrate the
inner region of the ICM as highly collimated outflows. Their disruption occurs
further out in the cluster atmosphere where the magnetic fluctuations and
thereby the level of turbulence is lower. This environment impedes mixing in
comparison to low-power jets which get disrupted inside the highly turbulent
cluster centre. In addition, Rayleigh Taylor instabilities should arise later
for larger cavities (higher power jets at constant $\tau_\mathrm{jet}$) as the
growth time scale of the instability increases for larger scales. This argument
is confirmed when we compare simulations with varying $\tau_\mathrm{jet}$, but
constant jet power (lower-right panel
Fig.~\ref{fig:mixing_magneticField}). Here, the larger cavities (larger
$\tau_\mathrm{jet}$) remain more stable. We conclude that less energetic jets
(decreased $P_\mathrm{jet}$ or $\tau_\mathrm{jet}$) show increased mixing
efficiencies of lobes with the ICM.

\section{Cosmic ray evolution}
\label{sec:cosmicrays}

After discussing the magnetic structure at the rising bubbles, we now turn our
attention to the distribution of CRs. First we examine the diffusive transport
of CRs and then detail Alfv\'en wave heating by CRs.

\subsection{CR diffusion and streaming}
\label{sec:cosmicrays_characteristics}

As discussed in Section~\ref{sec:CRs}, we model active CR transport via the
  anisotropic diffusion approximation to emulate CR streaming \citep[see
    also][]{Sharma2009}. To this end, we include CR energy losses through
Alfv\'en cooling and adopt a constant parallel diffusion coefficient
($\kappa_\parallel=10^{29}\ \mathrm{cm}^2\ \mathrm{s}^{-1}$) so that it
approximately matches the instantaneous CR diffusion coefficient
$\kappa_\mathrm{cr,A}\equiv{}l_\mathrm{cr}v_\mathrm{A}$ in the ICM.

This choice for $\kappa_\parallel$ is examined in
Fig.~\ref{fig:evolution_CRparameters}, which shows projections of thin layers of different
quantities related to CR transport. As expected, the CR population has a large
CR gradient length $l_\mathrm{cr}$ in the bubble (except for the boundary) as
the CR population quickly reaches a homogeneous distribution for our choice of
$\kappa_\parallel$. Outside the bubble, $l_\mathrm{cr}$ drops quickly. The
behaviour is echoed by the Alfv\'en velocity $v_\mathrm{A}$: inside the jet (and
bubble) it attains values up to $10^4\ \rmn{km}\ \rmn{s}^{-1}$ owing to the low
density and comparably large magnetic field whereas it fluctuates around a value
of $10^2\ \rmn{km}\ \rmn{s}^{-1}$ in the ICM. The distribution of Alfv\'en
cooling times, $t_\mathrm{A}=l_\mathrm{cr}/v_\mathrm{A}$, is a
direct consequence of this: $t_\mathrm{A}$ drops from values of $\approx1$~Gyr
inside the bubbles to values ranging from 10$-$30~Myrs, comparable to typical
jet duty time scales \citep[e.g.,][]{Vantyghem2014a,Turner2018}.

The combination of high $v_\mathrm{A}$ and $l_\mathrm{cr}$ inside the bubble
leads to a large value of the instantaneous CR diffusion coefficient
$\kappa_\mathrm{cr,A}\approx10^{32}\ \mathrm{cm}^2\mathrm{s}^{-1}$. While this
is much larger than our adopted constant diffusion coefficient of
$10^{29}\ \mathrm{cm}^{2}\mathrm{s}^{-1}$, the CR distribution in the bubble is
already homogeneous as can be seen in the CR energy density in
Fig.~\ref{fig:evolution_absolute}; increasing $\kappa_\parallel$ further
would not alter our results. Outside the bubble, the diffusion coefficient
 drops by three orders of magnitude to
$\kappa_\mathrm{cr,A}\approx10^{29}\ \mathrm{cm}^2\mathrm{s}^{-1}$. There, the
CRs are magnetically unconfined and the value of the diffusion coefficient
becomes crucial for accurately capturing the dynamics, justifying our choice of
$\kappa_\parallel$.

The small CR length scale $l_\mathrm{cr}$ at the bubble interface combined with
a high Alfv\'en velocity decreases the Alfv\'en cooling time $t_\mathrm{A}$
significantly to values of order 1~Myr. This would increase the Alfv\'en heating
rate at the edges and drain a significant amount of CR energy from the
bubble. In reality, the CR gradient would instead be smoothed out on the short
Alfv\'enic crossing time across the jet and stay flat during inflation of the
bubble and its evolution thereafter. This explains our initial choice of
limiting Alfv\'en cooling to regions outside the bubble
($X_\mathrm{jet}<10^{-3}$) as a numerical safeguard to prevent
numerically-induced CR cooling.

We discuss in Section~\ref{sec:magneticfield} that CRs can only escape through
the lower part of the bubble. As they are conducted out of the bubble they
remain confined to the magnetic field. Consequently, the vertically oriented,
magnified magnetic field lines are traced by the Alfv\'en heating rate. This
demonstrates the importance of simulating the exact structure of the magnetic
field in the vicinity of the bubble. It will be interesting to see how
substructure induced motions and radiative cooling influence this result.

\begin{figure*}
\centering
\includegraphics[trim=1.5cm .6cm 1.3cm .9cm,clip=true, width=\matrixfac\textwidth]{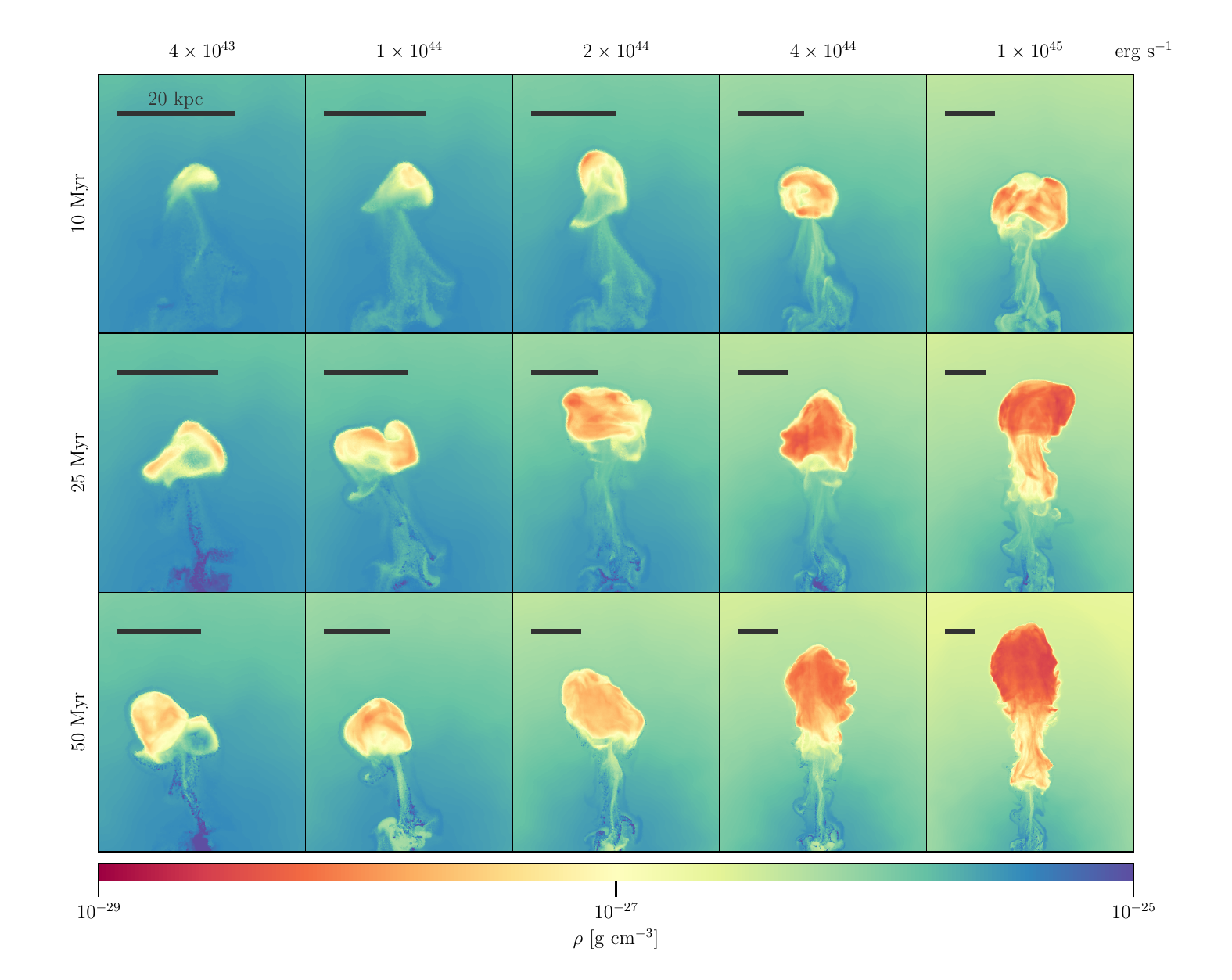}
\caption{Density projections for our lower resolution simulations with varying
  jet parameters. We show a full projection of the jet tracer-weighted density
  at 70~Myr. The jet activity time increases from top to bottom and the jet
  power increases from left to right as indicated. Jets with similar energy are
  ordered along diagonals from the bottom left to the top right. The scale bar
  corresponds to $20\ \mathrm{kpc}$ in all panels and decreases from left to
  right but stays constant along diagonals with
  $E_\rmn{jet}\approx\rmn{const}$.}
    \label{fig:matrix_density}
\end{figure*}
\begin{figure*}
\centering
\includegraphics[trim=1.5cm .65cm 1.35cm .85cm,clip=true, width=\matrixfac\textwidth]{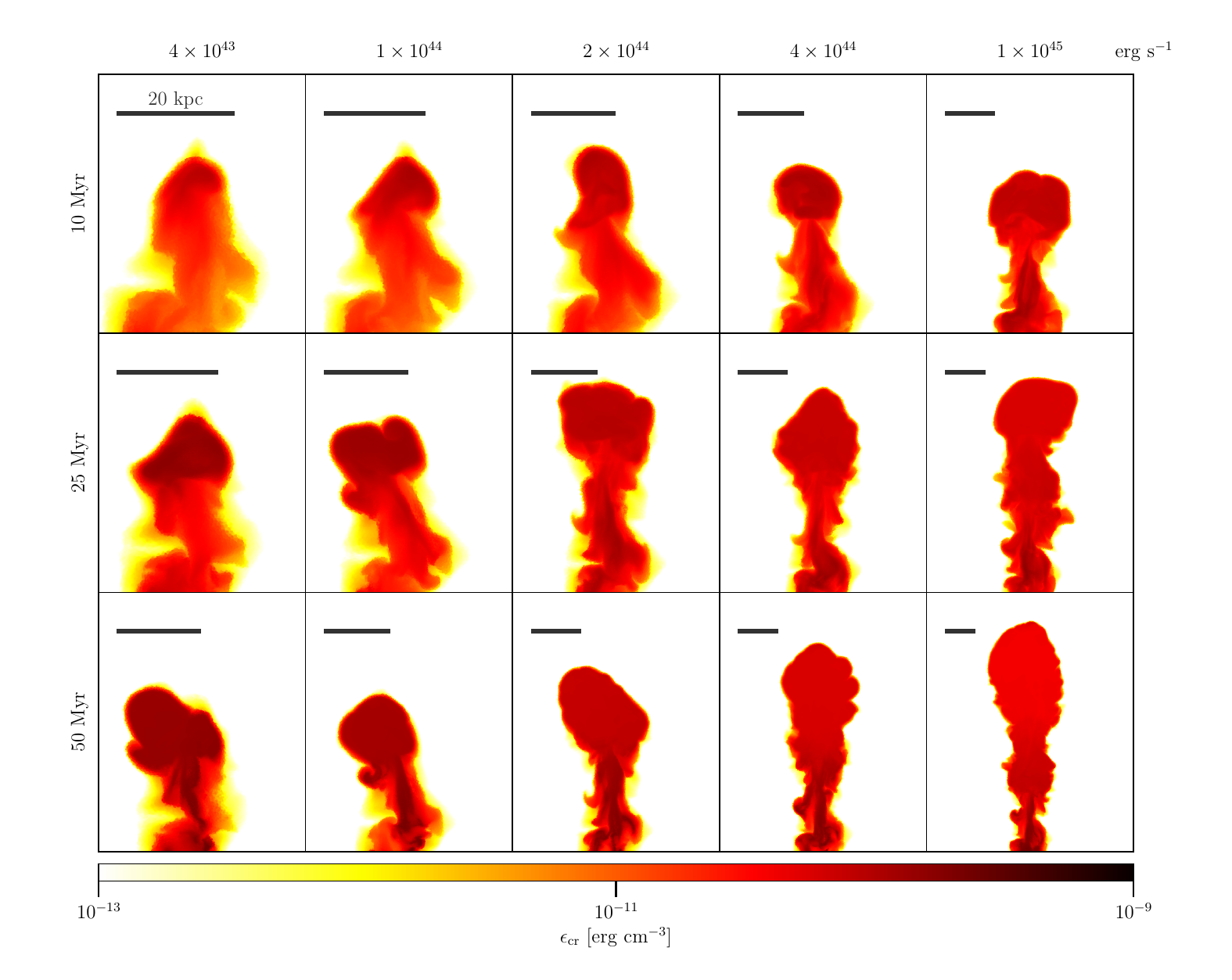}
\caption{Same as in Fig.~\ref{fig:matrix_density} but for the jet tracer-weighted CR
  energy density $\epsilon_\mathrm{cr}$. Even for bubbles with low density
    contrast (top left) there is significant CR energy density released into
    the ICM. }
    \label{fig:matrix_CRenergydensity}
\end{figure*}
\begin{figure*}
\centering
\includegraphics[trim=1.5cm .65cm 1.35cm .85cm,clip=true, width=\matrixfac\textwidth]{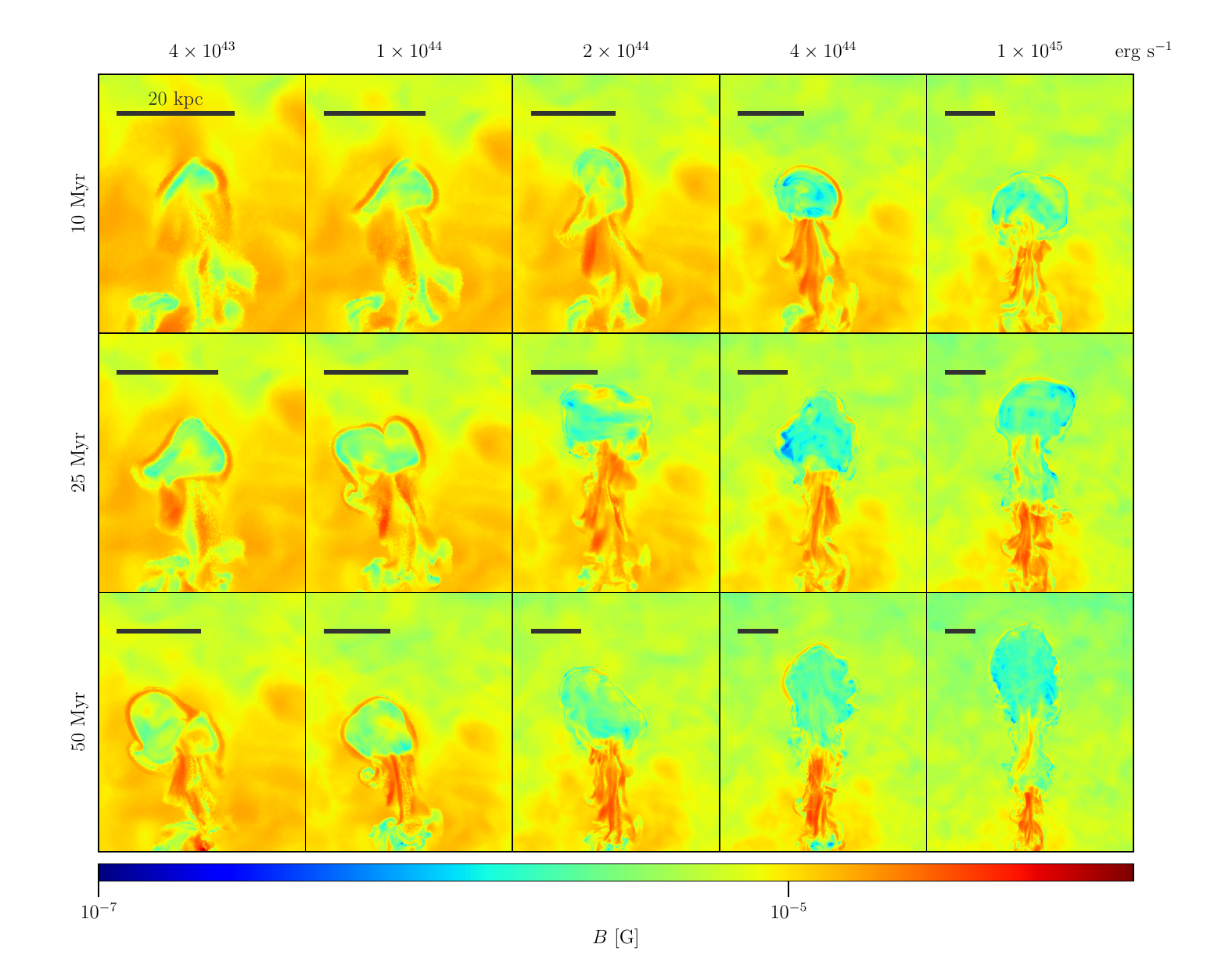}
\caption{Same as in Fig.~\ref{fig:matrix_density} but for the jet
  tracer-weighted magnetic field strength $B$. The red envelope surrounding
    the bubbles clearly show magnetic draping and magnetic fields in the wake of
    bubbles are strongly amplified. Note that the weighting procedure enhances
  the magnetic field in the wake of the bubble.}
    \label{fig:matrix_magneticfield}
\end{figure*}
\begin{figure*}
\centering
\includegraphics[trim=1.5cm .65cm 1.35cm .85cm,clip=true, width=\matrixfac\textwidth]{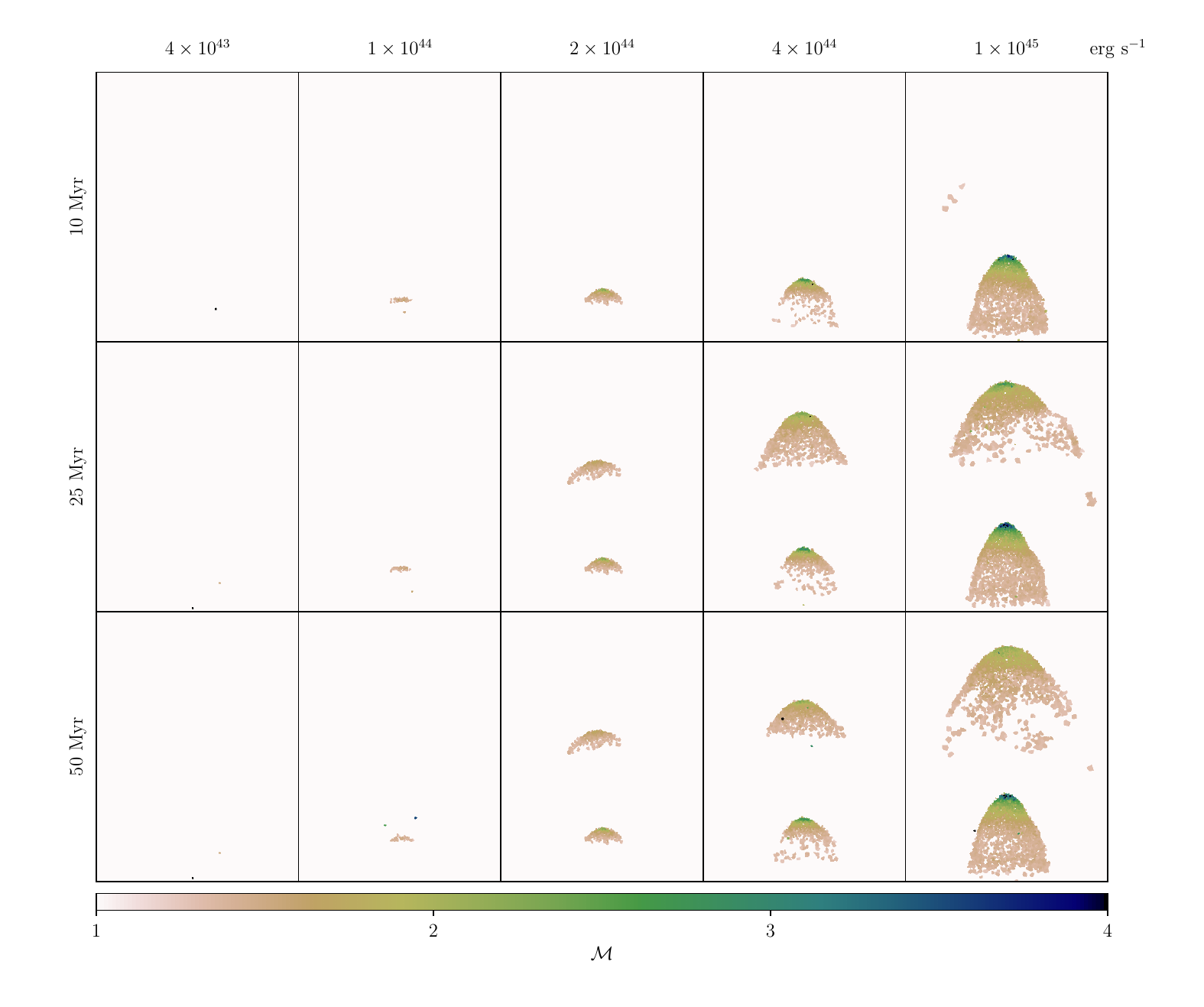}
\caption{Same as in Fig.~\ref{fig:matrix_density} but for the Mach number
  $\mathcal{M}$ that was weighted with the energy dissipation rate at the
  shocks. We overplot $\mathcal{M}$ projections of two snapshots, at
  $5\ \mathrm{Myr}$ and $20\ \mathrm{Myr}$. Here, projections have constant
  dimensions of $80\ \mathrm{kpc}\times60\ \mathrm{kpc}\times60\ \mathrm{kpc}$ centred at
  (40-0-0). Low-power jets show at best small Mach numbers whereas the Mach
  numbers of high power-jets decrease quickly with time. These characteristics
  are in agreement with observations of FRI and FRII, respectively.}
    \label{fig:matrix_machnumbers}
\end{figure*}

\subsection{CR distribution and Alfv\'en wave heating}

In Fig.~\ref{fig:radialProfile_CR} (upper left panel) we show the radial profile
of the Alfv\'en heating rate ($\mathcal{H}_\rmn{cr}$) of our fiducial model at
different times during and after the jet lifetime $\tau_{\rmn{jet}}=25$~Myr. In
agreement with our sub-grid model of CR Alfv\'en cooling in bubbles and in order
to focus solely on the heating of the ICM, we impose a jet tracer threshold of
$X_\rmn{jet}<10^{-3}$ to exclude artificially high cooling rates within the
bubble. The result is robust up to factors of two when we vary the jet tracer
threshold by an order of magnitude. We compare our simulated Alfv\'en heating
rates to theoretical predictions by \citet{Jacob2016a} for the Perseus cluster
who found steady-state solutions in which the heating rates due to Alfv\'en
heating (at small radii) and thermal conduction (at larger radii) balance
radiative cooling. Our simulated Alfv\'en heating rates are in good agreement
with the theoretical predictions up to $30\ \mathrm{Myr}$ after jet launch with
details depending on parameter choices as we will now discuss. At later times,
newly launched jets are expected to replenish the CR energy reservoir, which has
then significantly cooled via Alfv\'en wave losses.

The overall shape of the radial profile of $\mathcal{H}_\rmn{cr}$ is determined
by the jet energy, power and lifetime. For efficient CR heating in the centre of clusters, the exact value of the jet energy $E_\rmn{jet}$ proves to be crucial: if it is too small there is not enough CR energy injected and
the induced heating rate cannot balance radiative losses of the gas. On the
other hand, if the jets are too energetic they pierce out of the cluster
centre and reach the outskirts of the core, which makes it difficult for CRs to
diffuse back to the origin and to maintain a large heating rate.

For jets with $E_\mathrm{jet}=\rmn{const.}$ but varying luminosity and
lifetime, the profiles differ slightly (Fig.~\ref{fig:radialProfile_CR}, top
right panel). The heating rate profile is steeper for low-luminosity jets with
longer activity times. This is because low-luminosity jets are more quickly
decelerated by the inertia of the ambient ICM and CRs have more time to
diffuse back towards the cluster centre where they sustain a larger central
heating rate.

Jets with constant $\tau_\mathrm{jet}$ exhibit an increasing heating radius with
increasing $P_\mathrm{jet}$ (Fig.~\ref{fig:radialProfile_CR}, bottom left
panel). A larger jet luminosity corresponds to enhanced CR production (at
$\tau_\mathrm{jet}=\rmn{const.}$) while the jet also pushes to larger radii. At
small radii there is a larger variance of $\mathcal{H}_\mathrm{cr}$ because of
the fast CR transport to large radii in the jet which competes with the
backwards CR diffusion and advection towards the dense cluster centre.  Jets
with $P_\mathrm{jet}=\rmn{const.}$ produce almost self-similar
$\mathcal{H}_\mathrm{cr}$ profiles that scale with the amount of injected CR
energy (Fig.~\ref{fig:radialProfile_CR}, bottom right panel).

A successful heating mechanism in CC clusters is not required to act
isotropically throughout the {\it entire} core region. However, as cooling
material falls to the centre, eventually it should be heated at some inner
radius \citep{McNamara2012}, which poses requirements for the isotropy of the
proposed heating mechanism at {\it small} radii. To examine the volume-filling
of CR heating in our simulations, we first show the volume distributions of CR
energy density and Alfv\'en-heating rate in the top panels of
Fig.~\ref{fig:CR_isotropy}. Diffusion leads to a shallow CR floor in the cluster
centre and beyond at later times. To quantify the degree of CR isotropy, we
define a minimum amount of the CR energy density and Alfv\'en heating rate by
requiring that 3$\sigma$ (99.8\%) of CR energy and Alfv\'en heating power are
above these floor values, respectively.

For each concentric shell of radius $r$, we display the volume fraction covered
by cells with $\epsilon_\mathrm{cr}$ (or $\mathcal{H}_\mathrm{A}$) above these
floor values and normalise it to the volume of the shell at this radius (bottom
panels of Fig.~\ref{fig:CR_isotropy}). While the bipolar jets transport CRs
mainly along the jet axes, subsonic CR advection and diffusion strongly limit
lateral transport of CRs within a CR cooling time at large radii
($r\gtrsim15\ \mathrm{kpc}$), precluding isotropic heating there. In contrast,
Alfv\'en-wave heating is almost isotropic at small radii ($r<15\ \mathrm{kpc}$).

In fact, observations favour a smooth heating process with only minimal temporal
over-heating or -cooling \citep{Fabian2012}. Our simulations reproduce this
property as we deduce from the evolution of the pressure and temperature
profiles in Fig.~\ref{fig:radialProfile_general}. The profiles show little
variance after the initial perturbation in temperature and density due to the
propagating jet at $10\ \mathrm{Myr}$.

The match of $\mathcal{H}_\mathrm{cr}$ in our simulations to the steady-state
solutions \citep{Jacob2016a} validates that dynamically evolved jets with
plausible parameters can distribute CRs sufficiently well to successfully
balance the radiative cooling losses of the ICM via CR heating on timescales
$t\lesssim30$~Myr.  Future simulations including gas cooling and jet injection
coupled to accretion will scrutinise the feasibility of the model on long
timescales and whether it is self-regulating.

\begin{figure}
\centering
\includegraphics[trim=1.05cm .95cm 8.cm .75cm,clip=true, width=\columnwidth]{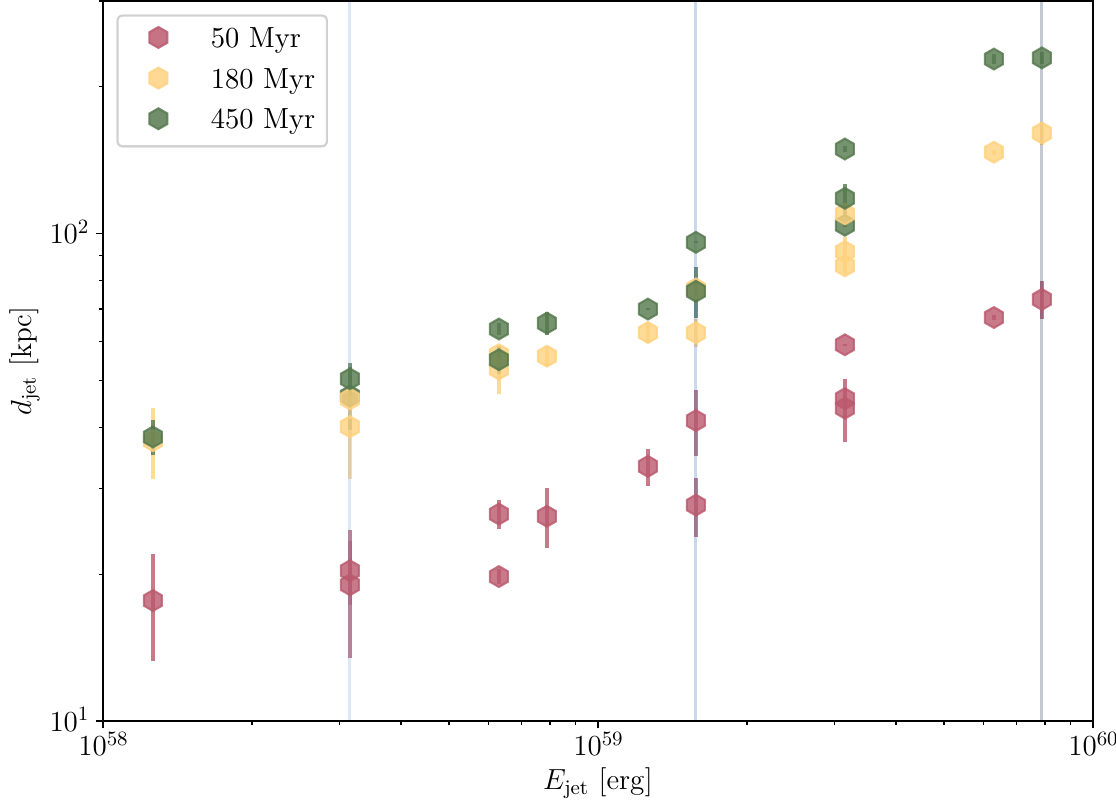}
\caption{Distance of jet travel vs. jet energy of our lower resolution
  simulations. The jet distance $d_\mathrm{jet}$ for individual jets is defined
  as the distance from the SMBH to the point to which half the jet mass
  $X_\mathrm{jet}m$ (of the upper hemisphere) has travelled (typically situated
  inside the lobe). We show the average distances of the upper and lower jets
  for every individual simulation (hexagons) and the error bar denotes the
  ranges for the two individual jets. There is a clear correlation of jet
  distance and energy at each of the three different simulation times (indicated
  by colour). The vertical (light blue) lines correspond to the three
  simulations shown in the left panel of Fig.~\ref{fig:CR_totalEnergy}. }
    \label{fig:distance_totalEnergy}
\end{figure}

\begin{figure*}
\centering
\includegraphics[trim=1.05cm .95cm 7.95cm .75cm,clip=true, width=0.47\textwidth]{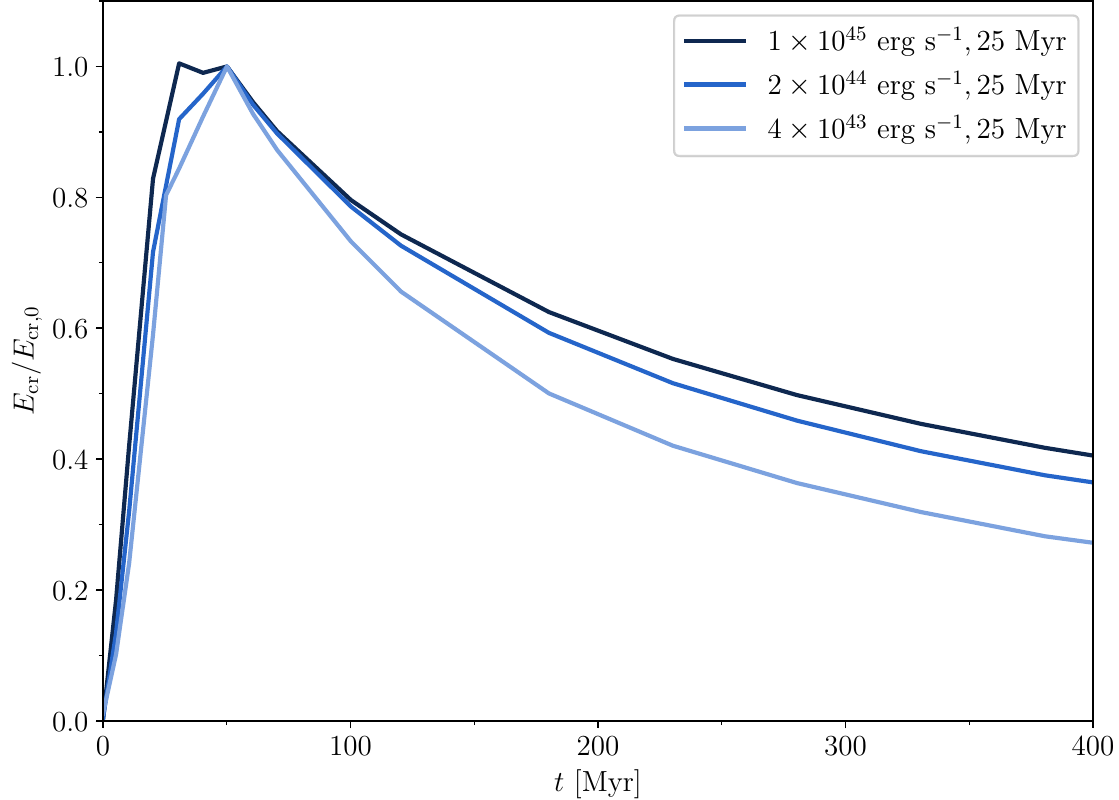}
\hspace{0.01\textwidth}
\includegraphics[trim=1.05cm .95cm 7.95cm .75cm,clip=true, width=0.47\textwidth]{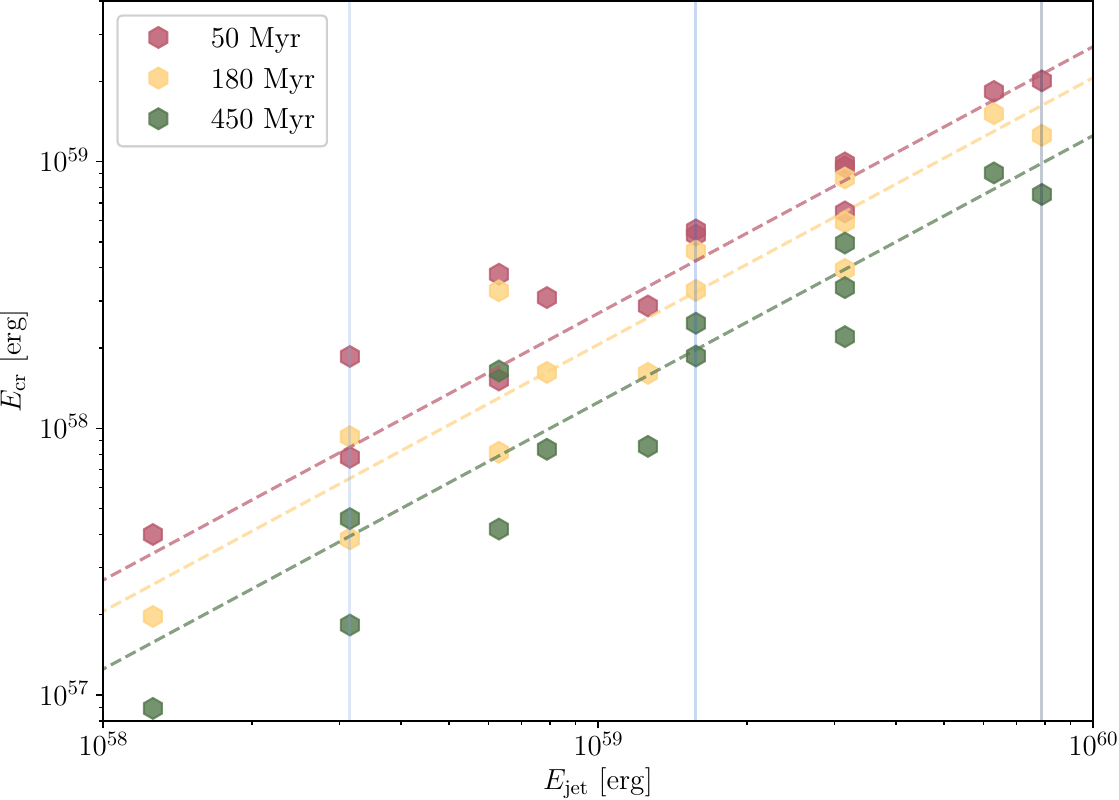}
\caption{The evolution of the total CR energy as a function of jet energy for
  three different jet luminosities of our lower resolution simulations. On the
  left, we show the CR energy $E_\mathrm{cr}$ as a function of time normalised
  to its maximum value $E_\mathrm{cr,0}$. During the jet stage, the CR energy
  increases steadily until $\tau_\rmn{acc}$ after which it decreases as a result
  of escaping CRs that suffer Alfv\'en wave losses in the ICM. On the right, we
  show the CR energy $E_\mathrm{cr}$ as a function of total jet energy
  $E_\mathrm{jet}$ for different simulations at three different times (colour
  coded). A linear fit to each of the three simulation times is shown with
  dashed lines and describes the simulations well. The significant scatter
  towards small jet energies corresponds to the increased mixing efficiency for
  these systems. The vertical (light blue) lines correspond to the simulations
  shown in the left panel. }
    \label{fig:CR_totalEnergy}
\end{figure*}

\section{Parameter study}
\label{sec:jetparameters}
The following parameter study focuses on jet power and lifetime. Here, we will
show that instead of these two parameters, jet energy appears to be the most
important parameter for determining jet morphology, CR distribution and magnetic
field structure, whereas the jet power determines the maximum attainable Mach
number.

\subsection{Bubble morphology}
In Fig.~\ref{fig:matrix_density}, we show how the bubble morphology changes with
varying jet parameters at 70~Myr. While the jet acceleration of CRs in our
  subgrid model is still ongoing for jets in the bottom row, the resulting
  dynamical effects of this late-time acceleration ($t>\tau_{\rmn{jet}}$) are
  negligible. Jet lifetime increases from top to bottom and jet power increases
from left to right. Jets with similar energy are ordered along diagonals from
the bottom left to the top right. In order to identify the highly anisotropic
features of the bubble, Figs.~\ref{fig:matrix_density},
\ref{fig:matrix_CRenergydensity}, and \ref{fig:matrix_magneticfield} show
projections weighted with the jet tracer mass fraction
$X_\rmn{jet}$. Additionally, we show in
Fig.~\ref{fig:matrix_machnumbers} the Mach number $\mathcal{M}$ weighted with
the energy dissipation rate at the shocks. Note that we assign a minimum value of
$X_\rmn{jet}=10^{-10}$ to every cell to also display the background. The
projection depth corresponds to the projection width.

We see that jet energy is responsible for setting the overall bubble
morphology. Bubbles inflated with a low-power jet with a long activity time
resemble bubbles originating from jets with high power but shorter lifetimes. In
Fig.~\ref{fig:distance_totalEnergy}, we show the mean distance travelled by the jet as a
function of the jet energy at three different times.  We find a power-law relation
$d_\rmn{jet}\propto E_\rmn{jet}^\alpha$ with $\alpha\approx0.4$. Jets with the
same energy reach similar heights, confirming the correlation. Because different
jets with $E_\rmn{jet}=\rmn{const.}$ produce bubbles of similar sizes, this
implies comparable Rayleigh-Taylor lifetimes (see Section~\ref{sec:mixing}).

Low-energy jets inflate smaller lobes (Fig.~\ref{fig:matrix_density}, to the
upper left), which terminate at lesser heights. They are deflected from
their original jet trajectories and show clear signs of ongoing mixing (as
indicated by the low density contrast with the ICM). These are the signatures of
FRI-type jets according to the Fanaroff-Riley (FR) classification
\citep{Fanaroff1974}. Increasing jet energy (top left to bottom right) results
in jets that penetrate the ICM to larger distances from the cluster centre. They
propagate mostly along the original jet direction and can sustain high-density
contrasts for longer times. These properties resemble jets of the FRII-type
category.

Producing realistic FRI jets in simulations requires resolving the radius of the jet with $\approx 10$ cells \citep{Anjiri2014}, which is difficult to achieve in our
large-scale simulations for our low-power jet models. In these idealised jet
simulations the occurrence of low-power FRI-type jets is then attributed to the
development of a turbulent structure rather then terminal shocks as in the case
of high-power FRII jets. The jet power that marks a threshold between these jet
categories is given by
$P_\mathrm{jet}\sim10^{43}\ \mathrm{erg}\ \mathrm{s}^{-1}$ according to
simulations \citep{Massaglia2016a}. Here, we can confirm that FRI-like jets are
obtained in our simulations, albeit at a somewhat higher threshold
luminosity. We address this point in Appendix~\ref{sec:resolution_study} and
find that while this threshold luminosity decreases with increasing numerical
resolution, general properties regarding distribution of magnetic fields, CRs
and heating rates remain qualitatively similar. Most importantly, we find that
jet energy appears to be the leading variable to distinguish between the main FR
jet features. Alternative scenarios for the origin of FRI jets include
de-focusing due to a magnetic kink instability in the jet
\citep{Tchekhovskoy2016}, mixing due to Kelvin-Helmholtz instabilities in
sheared relativistic flows \citep[e.g.,][]{Perucho2010} and mass entrainment
from stellar winds \citep[e.g.,][]{Wykes2015}.

We leave detailed morphological studies of bubbles in cooling clusters that are
generated through the interplay of accretion and jet launch for future
studies. Additionally, the interaction of subsequent generations of bubbles may
play a key role as they may merge and form large outflowing cavities
\citep{Cielo2018}. Finally, the jet lifetime cannot be arbitrarily increased to
form ever larger bubbles as these will inevitably fragment, generating multiple
disconnected bubbles in a turbulent environment \citep{Morsony2010}.

\subsection{CR distribution}

Looking at the distribution of CR energy density, $\epsilon_\rmn{cr}$, in
Fig.~\ref{fig:matrix_CRenergydensity}, it becomes apparent that
$\epsilon_\rmn{cr}$ is more homogeneous than the density across our models. This means that even if there is no visible cavity due to low
contrast in X-rays, there may still be a substantial CR population that is
responsible for efficient Alfv\'en wave heating.

The previously described trend that low-energy jets form bubbles that are easily
deflected and dispersed early-on translates to a very centrally localised CR
distribution with a high degree of isotropy
(Fig.~\ref{fig:matrix_CRenergydensity}). Conversely, high-energy jets develop
bubbles that stay intact out to large distances. CRs continuously diffuse out of
the bubble but the majority of the CR energy is transported to large radii. This
makes low-energy jets more efficient in heating the fast-cooling cluster
centres. These low-energy systems have a larger mixing efficiency and their
bubbles remain in the central regions of the cluster \citep{Mukherjee2016}.

The sequence of CR acceleration, bubble disruption, diffusive CR escape, and
successive Alfv\'en cooling is expected to reflect on the available CR energy
for individual jets (Fig.~\ref{fig:CR_totalEnergy}). Interestingly, we observe a
linear correlation between CR energy and the total jet energy. When considering
a sample of jets with the same energy, the jet with the longest lifetime
$\tau_\mathrm{jet}$ will accelerate CRs for longer periods of time. As CRs cool
over time, the low-power (high-lifetime) jet is expected to maintain a larger CR
energy, which explains the scatter at constant energy in
Fig.~\ref{fig:CR_totalEnergy}.

Another source of scatter in the linear relation is due to the difference in
mixing efficiency for jets with variable jet power. The left panel of
Fig.~\ref{fig:CR_totalEnergy} exemplifies this point as it shows the normalised
CR energy for jets with constant lifetime but varying jet power.  During the jet
stage, the CR energy increases as kinetic energy is dissipated and transferred
to CR energy until $\tau_\rmn{acc}$.  Afterwards, CR energy decreases as a
result of escaping CRs that suffer Alfv\'en wave losses in the ICM. A decrease
in jet power implies a decrease of CR energy in the entire cluster. As discussed
in Section~\ref{sec:mixing}, low-power jets mix more efficiently, which results
in earlier disruption times of the lobe and thereby an earlier onset of CR
cooling.

\subsection{Magnetic field structure}

The magnetic field structure shows the draping layer wrapped around the bubbles
in most of the different jet simulations
(Fig.~\ref{fig:matrix_magneticfield}). Even though the draping layer lies
outside the bubble interface with the ambient ICM, the numerically diffused
advective jet tracers still highlight this feature. Consequently, the jet
  tracer-weighted projection leads to apparent amplification ratios of
  $>2$. However, the actual ratios are on the order of $\sim2$ as discussed in
  Section \ref{sec:draping}. In some exceptional cases, the draping layer
remains absent. This is due to the turbulent nature of the ICM that causes
perturbations in the trajectory of the bubble and generates a corrugated bubble
interface as the bubble expands into a region of lower ambient pressure. The
local change of propagation direction forces the bubble to accumulate a new
draping layer. In addition, numerical reconnection of magnetic field lines in
the turbulent environment may temporarily erase the draping layer as fields of
different polarity accumulate in the layer.

We observe strongly amplified magnetic filaments in the wakes of every bubble,
which reach field strengths up to 30~$\umu$G. These elongated filaments align
approximately along the jet axis and point back to the cluster centre to which
they conduct the diffusing CRs. These strongly magnetised magnetic structures
resemble observed H$\alpha$ filaments in CC clusters. 

As we have seen, a large fraction of the jet tracers is mixed with the ICM in the
wake of the jet. This causes the magnetic field in the region to be
significantly enhanced compared to the surrounding magnetic field. Due to this
effect the contrast in magnetic field strength between wake region and ICM
increases for larger projection box sizes (bottom right of
Fig.~\ref{fig:matrix_magneticfield}).

\subsection{Shocks and Mach numbers}

Most observed Mach numbers of expanding lobes that reside in CC clusters are observed to be
at the order of $\mathcal{M}\sim 1$ \citep{McNamara2007}. Using a shock finder
in {\sc arepo} \citep{Schaal2015}, we detect and characterise the bow shock that
is driven into the ICM by the propagating lobes.  In
Fig.~\ref{fig:matrix_machnumbers}, we show projections of the Mach numbers of
this bow shock, weighted with the energy dissipation rate at two different
times, at 5 Myr and 20 Myr.

The shock strength scales with the jet power as expected. At $5\ \mathrm{Myr}$, the lobes of low-power jets ($\leq2\times10^{44}\ \mathrm{erg}\ \mathrm{s}^{-1}$)
predominantly exhibit Mach numbers $\mathcal{M}\lesssim 1.5$, similar to
observations. In contrast, the lobes of high-power jets
($>2\times10^{44}\ \mathrm{erg}\ \mathrm{s}^{-1}$) exhibit strong Mach numbers
$\mathcal{M}>4$. However after $20\ \mathrm{Myr}$ those have already decreased
to values of $\mathcal{M}\sim2-3$. After the jet is switched off, only
individual cells exhibit Mach numbers $\mathcal{M}>1$. Thus, even our lobes from high-power
jets only exhibit high Mach numbers for a short period of time and
low-to-intermediate Mach numbers for most of the time the jet is active.  We
conclude that our simulations successfully reproduce the observed low Mach
numbers for the lobes of low-power jets.

\section{Conclusions}
\label{sec:conclusions}
Using 3D MHD simulations with the moving-mesh code \textsc{arepo}, we study the
evolution of magnetised and CR-filled jets in an idealised Perseus galaxy
cluster. Following the jet-driven inflation of underdense bubbles, we study
their buoyant rise in the cluster atmosphere and how they interact with a
turbulent cluster magnetic field. The bubbles are exposed to interface
instabilities which finally disrupt the bubbles and enable initially confined
CRs to diffusively escape and to heat the ambient ICM. Here we summarise our
main findings:
\begin{itemize}
\item The accumulation of magnetic fields at the bubble interface as a result of
  buoyant bubble motion relative to the ambient ICM stabilises the bubble
  against the turbulent environment, suppresses Kelvin-Helmholtz instabilities,
  and reduces the mixing efficiency. Internal helical magnetic fields show a
  similar effect.
\item We find that a decrease in jet power and/or in total jet energy increases
  the mixing efficiency of the jet.
\item CRs inside the bubbles are confined by the draped magnetic field that
  inhibits diffusion across the bubble surface.
\item In the wake of the bubble, the magnetic field is strongly amplified and
  adiabatically compressed by converging inflows that are compensating the
  upwards motion of the bubble. Differential motions stretch the magnetic field
  so that it becomes filamentary and aligned along the jet axis. These strongly
  magnetised filaments acquire strengths of 30~$\umu$G and resemble observed
  H$\alpha$ filaments in clusters. We postpone a detailed study to future work.
\item These radial magnetic filaments connect the bubble interior to the ambient
  ICM, and allow CRs to diffusively escape into the ICM and heat the surrounding
  medium. Our simulated radial profiles of the CR-induced Alfv\'en wave heating
  rate match CR heating rates predicted by steady-state models of CC clusters
  extremely well \citep{Jacob2016a}. Inside a radius $r\lesssim15$~kpc, we find
  a volume-filling CR distribution that generates isotropic Alfv\'en wave
  heating, which is necessary for solving the cooling flow problem at the
  centres of clusters. The temporal evolution varies significantly such that
  time-dependent modeling becomes crucial.
\item A parameter study of different jet simulations with varying jet lifetime
  and jet power reveals that the jet energy is the critical parameter for
  determining the overall bubble morphology and CR distribution. Magnetic
  draping as well as the strong filamentary magnetic field amplification in the
  wakes is ubiquitous in our sample.
\item We find a high degree of coherence and decreasing mixing efficiency with
  increasing jet energies. This finding and the observed low Mach numbers show
  that we can reproduce the main features of both, FRI and FRII-like jets:
    FRII jets exhibit bipolar, lobe-brightened morphologies with high density
    contrasts that power high Mach numbers $\mathcal{M}\sim4$ in the ICM at
    early times. On the other hand, FRI jets are characterized by lower density
    contrasts, show more deflected and corrugated bubbles that generate a
    laterally more expanded CR distribution at the centre, and do not drive
    detectable shocks into the ICM.
\end{itemize}
These results encourage further studies of the impact of CR-filled AGN bubbles on
radiatively cooling CC cluster atmospheres. Accounting for accretion onto SMBHs
and successive jet formation will enable us to find out whether we can obtain a
self-regulated CR heating-radiative cooling cycle.

\section*{Acknowledgements}
We thank the anonymous referee for helpful suggestions and comments.
This work has been supported by the European Research Council under ERC-CoG grant CRAGSMAN-646955, ERC-StG grant EXAGAL-308037 and by the Klaus Tschira Foundation.



\bibliographystyle{mnras}

\bibliography{library}

\begin{thebibliography}{}
\makeatletter
\relax
\def\mn@urlcharsother{\let\do\@makeother \do\$\do\&\do\#\do\^\do\_\do\%\do\~}
\def\mn@doi{\begingroup\mn@urlcharsother \@ifnextchar [ {\mn@doi@}
  {\mn@doi@[]}}
\def\mn@doi@[#1]#2{\def\@tempa{#1}\ifx\@tempa\@empty \href
  {http://dx.doi.org/#2} {doi:#2}\else \href {http://dx.doi.org/#2} {#1}\fi
  \endgroup}
\def\mn@eprint#1#2{\mn@eprint@#1:#2::\@nil}
\def\mn@eprint@arXiv#1{\href {http://arxiv.org/abs/#1} {{\tt arXiv:#1}}}
\def\mn@eprint@dblp#1{\href {http://dblp.uni-trier.de/rec/bibtex/#1.xml}
  {dblp:#1}}
\def\mn@eprint@#1:#2:#3:#4\@nil{\def\@tempa {#1}\def\@tempb {#2}\def\@tempc
  {#3}\ifx \@tempc \@empty \let \@tempc \@tempb \let \@tempb \@tempa \fi \ifx
  \@tempb \@empty \def\@tempb {arXiv}\fi \@ifundefined
  {mn@eprint@\@tempb}{\@tempb:\@tempc}{\expandafter \expandafter \csname
  mn@eprint@\@tempb\endcsname \expandafter{\@tempc}}}

\bibitem[\protect\citeauthoryear{Anjiri, Mignone, Bodo  \& Rossi}{Anjiri
  et~al.}{2014}]{Anjiri2014}
Anjiri M.,  Mignone A.,  Bodo G.,   Rossi P.,  2014, \mn@doi [MNRAS]
  {10.1093/mnras/stu1004}, 442, 2228

\bibitem[\protect\citeauthoryear{Asai, Fukuda  \& Matsumoto}{Asai
  et~al.}{2007}]{Asai2007}
Asai N.,  Fukuda N.,   Matsumoto R.,  2007, \mn@doi [ApJ] {10.1086/518235},
  663, 816

\bibitem[\protect\citeauthoryear{Bambic, Morsony  \& Reynolds}{Bambic
  et~al.}{2018}]{Bambic2018}
Bambic C.~J.,  Morsony B.~J.,   Reynolds C.~S.,  2018, preprint
  (arXiv:1801.06233)

\bibitem[\protect\citeauthoryear{B{\^{i}}rzan, Rafferty, McNamara, Wise  \&
  Nulsen}{B{\^{i}}rzan et~al.}{2004}]{Birzan2004}
B{\^{i}}rzan L.,  Rafferty D.~A.,  McNamara B.~R.,  Wise M.~W.,   Nulsen P.
  E.~J.,  2004, \mn@doi [ApJ] {10.1086/383519}, 607, 800

\bibitem[\protect\citeauthoryear{B{\^{i}}rzan, McNamara, Nulsen, Carilli  \&
  Wise}{B{\^{i}}rzan et~al.}{2008}]{Birzan2008}
B{\^{i}}rzan L.,  McNamara B.~R.,  Nulsen P. E.~J.,  Carilli C.~L.,   Wise
  M.~W.,  2008, \mn@doi [ApJ] {10.1086/591416}, 686, 859

\bibitem[\protect\citeauthoryear{Bonafede, Feretti, Murgia, Govoni, Giovannini,
  Dallacasa, Dolag  \& Taylor}{Bonafede et~al.}{2010}]{Bonafede2010}
Bonafede A.,  Feretti L.,  Murgia M.,  Govoni F.,  Giovannini G.,  Dallacasa
  D.,  Dolag K.,   Taylor G.~B.,  2010, \mn@doi [A{\&}A]
  {10.1051/0004-6361/200913696}, 513, A30

\bibitem[\protect\citeauthoryear{Bourne \& Sijacki}{Bourne \&
  Sijacki}{2017}]{Bourne2017}
Bourne M.~A.,  Sijacki D.,  2017, \mn@doi [MNRAS] {10.1093/mnras/stx2269}, 472,
  4707

\bibitem[\protect\citeauthoryear{Br{\"{u}}ggen \& Kaiser}{Br{\"{u}}ggen \&
  Kaiser}{2001}]{Bruggen2001}
Br{\"{u}}ggen M.,  Kaiser C.~R.,  2001, \mn@doi [MNRAS]
  {10.1046/j.1365-8711.2001.04494.x}, 325, 676

\bibitem[\protect\citeauthoryear{Br{\"{u}}ggen \& Kaiser}{Br{\"{u}}ggen \&
  Kaiser}{2002}]{Bruggen2002}
Br{\"{u}}ggen M.,  Kaiser C.~R.,  2002, \mn@doi [Nature] {10.1038/nature00857},
  418, 301

\bibitem[\protect\citeauthoryear{Br{\"{u}}ggen, Kaiser, Churazov  \&
  En{\ss}lin}{Br{\"{u}}ggen et~al.}{2002}]{Bruggen2002a}
Br{\"{u}}ggen M.,  Kaiser C.~R.,  Churazov E.,   En{\ss}lin T.~A.,  2002,
  \mn@doi [MNRAS] {10.1046/j.1365-8711.2002.05233.x}, 331, 545

\bibitem[\protect\citeauthoryear{Churazov, Br{\"{u}}ggen, Kaiser,
  B{\"{o}}hringer  \& Forman}{Churazov et~al.}{2001}]{Churazov2001}
Churazov E.,  Br{\"{u}}ggen M.,  Kaiser C.~R.,  B{\"{o}}hringer H.,   Forman
  W.,  2001, \mn@doi [ApJ] {10.1086/321357}, 554, 261

\bibitem[\protect\citeauthoryear{Churazov, Forman, Jones  \&
  B{\"{o}}hringer}{Churazov et~al.}{2003}]{Churazov2003}
Churazov E.,  Forman W.,  Jones C.,   B{\"{o}}hringer H.,  2003, \mn@doi [ApJ]
  {10.1086/374923}, 590, 225

\bibitem[\protect\citeauthoryear{Cielo, Babul, Antonuccio-Delogu, Silk  \&
  Volonteri}{Cielo et~al.}{2018}]{Cielo2018}
Cielo S.,  Babul A.,  Antonuccio-Delogu V.,  Silk J.,   Volonteri M.,  2018,
  preprint (arXiv:1801.04276)

\bibitem[\protect\citeauthoryear{Croston \& Hardcastle}{Croston \&
  Hardcastle}{2014}]{Croston2014}
Croston J.~H.,  Hardcastle M.~J.,  2014, \mn@doi [MNRAS]
  {10.1093/mnras/stt2436}, 438, 3310

\bibitem[\protect\citeauthoryear{Croston, Hardcastle, Birkinshaw, Worrall  \&
  Laing}{Croston et~al.}{2008}]{Croston2008a}
Croston J.~H.,  Hardcastle M.~J.,  Birkinshaw M.,  Worrall D.~M.,   Laing
  R.~A.,  2008, \mn@doi [MNRAS] {10.1111/j.1365-2966.2008.13162.x}, 386, 1709

\bibitem[\protect\citeauthoryear{Croston, Ineson  \& Hardcastle}{Croston
  et~al.}{2018}]{Croston2018}
Croston J.~H.,  Ineson J.,   Hardcastle M.~J.,  2018, MNRAS, 476, 1614

\bibitem[\protect\citeauthoryear{Duran, Tchekhovskoy  \& Giannios}{Duran
  et~al.}{2016}]{BarniolDuran2017}
Duran R.~B.,  Tchekhovskoy A.,   Giannios D.,  2016, \mn@doi [MNRAS]
  {10.1093/mnras/stx1165}, 469, 4957

\bibitem[\protect\citeauthoryear{Dursi}{Dursi}{2007}]{Dursi2007}
Dursi L.~J.,  2007, \mn@doi [ApJ] {10.1086/521997}, 670, 221

\bibitem[\protect\citeauthoryear{Dursi \& Pfrommer}{Dursi \&
  Pfrommer}{2008}]{Dursi2008}
Dursi L.~J.,  Pfrommer C.,  2008, \mn@doi [ApJ] {10.1086/529371}, 677, 993

\bibitem[\protect\citeauthoryear{En{\ss}lin \& Br{\"{u}}ggen}{En{\ss}lin \&
  Br{\"{u}}ggen}{2002}]{Ensslin2002}
En{\ss}lin T.~A.,  Br{\"{u}}ggen M.,  2002, \mn@doi [MNRAS]
  {10.1046/j.1365-8711.2002.05261.x}, 331, 1011

\bibitem[\protect\citeauthoryear{Fabian}{Fabian}{2012}]{Fabian2012}
Fabian A.~C.,  2012, \mn@doi [ARA{\&}A] {10.1146/annurev-astro-081811-125521},
  50, 455

\bibitem[\protect\citeauthoryear{Fabian et~al.,}{Fabian
  et~al.}{2000}]{Fabian2000}
Fabian A.~C.,  et~al., 2000, \mn@doi [MNRAS]
  {10.1046/j.1365-8711.2000.03904.x}, 318, L65

\bibitem[\protect\citeauthoryear{Fabian, Sanders, Allen, Crawford, Iwasawa,
  Johnstone, Schmidt  \& Taylor}{Fabian et~al.}{2003}]{Fabian2003b}
Fabian A.~C.,  Sanders J.~S.,  Allen S.~W.,  Crawford C.~S.,  Iwasawa K.,
  Johnstone R.~M.,  Schmidt R.,   Taylor G.~B.,  2003, \mn@doi [MNRAS]
  {10.1038/1747}, 344, L43

\bibitem[\protect\citeauthoryear{Fabian et~al.,}{Fabian
  et~al.}{2011}]{Fabian2011}
Fabian A.~C.,  et~al., 2011, \mn@doi [MNRAS]
  {10.1111/j.1365-2966.2011.19402.x}, 418, 2154

\bibitem[\protect\citeauthoryear{Fabian, Walker, Russell, Pinto, Sanders  \&
  Reynolds}{Fabian et~al.}{2017}]{Fabian2017}
Fabian A.~C.,  Walker S.~A.,  Russell H.~R.,  Pinto C.,  Sanders J.~S.,
  Reynolds C.~S.,  2017, \mn@doi [MNRAS] {10.1093/mnrasl/slw170}, 464, L1

\bibitem[\protect\citeauthoryear{Fanaroff \& Riley}{Fanaroff \&
  Riley}{1974}]{Fanaroff1974}
Fanaroff B.~L.,  Riley J.~M.,  1974, \mn@doi [MNRAS] {10.1093/mnras/167.1.31P},
  167, 31P

\bibitem[\protect\citeauthoryear{Guo \& Mathews}{Guo \&
  Mathews}{2011}]{Guo2011}
Guo F.,  Mathews W.~G.,  2011, \mn@doi [ApJ] {10.1088/0004-637X/728/2/121},
  728, 9

\bibitem[\protect\citeauthoryear{Guo \& Oh}{Guo \& Oh}{2008}]{Guo2008}
Guo F.,  Oh S.~P.,  2008, \mn@doi [MNRAS] {10.1111/j.1365-2966.2007.12692.x},
  384, 251

\bibitem[\protect\citeauthoryear{Heesen et~al.,}{Heesen
  et~al.}{2018}]{Heesen2018}
Heesen V.,  et~al., 2018, \mn@doi [MNRAS] {10.1093/mnras/stx2869}, 474, 5049

\bibitem[\protect\citeauthoryear{Heinz, Br{\"{u}}ggen, Young  \&
  Levesque}{Heinz et~al.}{2006}]{Heinz2006}
Heinz S.,  Br{\"{u}}ggen M.,  Young A.,   Levesque E.,  2006, \mn@doi [MNRASL]
  {10.1111/j.1745-3933.2006.00243.x}, 373, 65

\bibitem[\protect\citeauthoryear{Hillel \& Soker}{Hillel \&
  Soker}{2016}]{Hillel2015}
Hillel S.,  Soker N.,  2016, \mn@doi [MNRAS] {10.1093/mnras/stv2483}, 455, 2139

\bibitem[\protect\citeauthoryear{{Hitomi Collaboration} et~al.,}{{Hitomi
  Collaboration} et~al.}{2016}]{HitomiCollaboration2016}
{Hitomi Collaboration} et~al., 2016, \mn@doi [Nature] {10.1038/nature18627},
  535, 117

\bibitem[\protect\citeauthoryear{Jacob \& Pfrommer}{Jacob \&
  Pfrommer}{2017a}]{Jacob2016a}
Jacob S.,  Pfrommer C.,  2017a, \mn@doi [MNRAS] {10.1093/mnras/stx132}, 467,
  1449

\bibitem[\protect\citeauthoryear{Jacob \& Pfrommer}{Jacob \&
  Pfrommer}{2017b}]{Jacob2016b}
Jacob S.,  Pfrommer C.,  2017b, \mn@doi [MNRAS] {10.1093/mnras/stx132}, 467,
  1478

\bibitem[\protect\citeauthoryear{Jones \& {De Young}}{Jones \& {De
  Young}}{2005}]{Jones2005}
Jones T.~W.,  {De Young} D.~S.,  2005, \mn@doi [ApJ] {10.1086/429157}, 624, 586

\bibitem[\protect\citeauthoryear{Kannan, Vogelsberger, Pfrommer, Weinberger,
  Springel, Hernquist, Puchwein  \& Pakmor}{Kannan et~al.}{2017}]{Kannan2017}
Kannan R.,  Vogelsberger M.,  Pfrommer C.,  Weinberger R.,  Springel V.,
  Hernquist L.,  Puchwein E.,   Pakmor R.,  2017, \mn@doi [The Astrophysical
  Journal Letters] {10.3847/2041-8213/aa624b}, 837, L18

\bibitem[\protect\citeauthoryear{Kim \& Narayan}{Kim \&
  Narayan}{2003}]{Kim2003}
Kim W.-T.,  Narayan R.,  2003, \mn@doi [ApJ] {10.1086/378153}, 596, 889

\bibitem[\protect\citeauthoryear{Kuchar \& En{\ss}lin}{Kuchar \&
  En{\ss}lin}{2011}]{Kuchar2011}
Kuchar P.,  En{\ss}lin T.~A.,  2011, \mn@doi [A{\&}A]
  {10.1051/0004-6361/200913918}, 529, 13

\bibitem[\protect\citeauthoryear{Kulsrud}{Kulsrud}{2005}]{Kulsrud2005}
Kulsrud R.~M.,  2005, {Plasma Physics for Astrophysics}.
Princeton University Press, Princeton, NJ

\bibitem[\protect\citeauthoryear{Kulsrud \& Pearce}{Kulsrud \&
  Pearce}{1969}]{Kulsrud1969}
Kulsrud R.,  Pearce W.~P.,  1969, \mn@doi [ApJ] {10.1086/149981}, 156, 445

\bibitem[\protect\citeauthoryear{Laing \& Bridle}{Laing \&
  Bridle}{2013}]{Laing2013}
Laing R.~A.,  Bridle A.~H.,  2013, \mn@doi [MNRAS] {10.1093/mnras/stt531}, 432,
  1114

\bibitem[\protect\citeauthoryear{Laing \& Bridle}{Laing \&
  Bridle}{2014}]{Laing2014}
Laing R.~A.,  Bridle A.~H.,  2014, \mn@doi [MNRAS] {10.1093/mnras/stt2138},
  437, 3405

\bibitem[\protect\citeauthoryear{Leccardi \& Molendi}{Leccardi \&
  Molendi}{2008}]{Leccardi2008}
Leccardi A.,  Molendi S.,  2008, \mn@doi [A{\&}A]
  {10.1051/0004-6361:200809538}, 486, 359

\bibitem[\protect\citeauthoryear{Li, Ruszkowski  \& Bryan}{Li
  et~al.}{2017}]{Li2016a}
Li Y.,  Ruszkowski M.,   Bryan G.~L.,  2017, \mn@doi [ApJ]
  {10.3847/1538-4357/aa88c1}, 847, 106

\bibitem[\protect\citeauthoryear{Lind, Payne, Meier  \& Roger}{Lind
  et~al.}{1989}]{Lind1989}
Lind K.~R.,  Payne D.~G.,  Meier D.~L.,   Roger D.~B.,  1989, ApJ, 344, 89

\bibitem[\protect\citeauthoryear{Loewenstein, Zweibel  \& Begelman}{Loewenstein
  et~al.}{1991}]{Loewenstein1991}
Loewenstein M.,  Zweibel E.~G.,   Begelman M.~C.,  1991, ApJ, 377, 392

\bibitem[\protect\citeauthoryear{Lyutikov}{Lyutikov}{2006}]{Lyutikov2006}
Lyutikov M.,  2006, \mn@doi [MNRAS] {10.1111/j.1365-2966.2006.10835.x}, 373, 73

\bibitem[\protect\citeauthoryear{Martizzi \& Quataert}{Martizzi \&
  Quataert}{2018}]{Martizzi2018}
Martizzi D.,  Quataert E.,  2018, preprint (arXiv:1805.06461)

\bibitem[\protect\citeauthoryear{Massaglia, Bodo, Rossi, Capetti  \&
  Mignone}{Massaglia et~al.}{2016}]{Massaglia2016a}
Massaglia S.,  Bodo G.,  Rossi P.,  Capetti S.,   Mignone A.,  2016, \mn@doi
  [A{\&}A] {10.1051/0004-6361/201629375}, 596, A12

\bibitem[\protect\citeauthoryear{McNamara \& Nulsen}{McNamara \&
  Nulsen}{2007}]{McNamara2007}
McNamara B.~R.,  Nulsen P. E.~J.,  2007, \mn@doi [ARA{\&}A]
  {10.1146/annurev.astro.45.051806.110625}, 45, 117

\bibitem[\protect\citeauthoryear{McNamara \& Nulsen}{McNamara \&
  Nulsen}{2012}]{McNamara2012}
McNamara B.~R.,  Nulsen P. E.~J.,  2012, \mn@doi [New J. Phys.]
  {10.1088/1367-2630/14/5/055023}, 14, 40

\bibitem[\protect\citeauthoryear{Mendygral, Jones  \& Dolag}{Mendygral
  et~al.}{2012}]{Mendygral2012}
Mendygral P.~J.,  Jones T.~W.,   Dolag K.,  2012, \mn@doi [ApJ]
  {10.1088/0004-637X/750/2/166}, 750, 17pp

\bibitem[\protect\citeauthoryear{Morganti, Fanti, Gioia, Harris, Parma  \& de
  Ruiter}{Morganti et~al.}{1988}]{Morganti1988}
Morganti R.,  Fanti R.,  Gioia I.~M.,  Harris D.~E.,  Parma P.,   de Ruiter H.,
   1988, A{\&}A, 189, 11

\bibitem[\protect\citeauthoryear{Morsony, Heinz, Br{\"{u}}ggen  \&
  Ruszkowski}{Morsony et~al.}{2010}]{Morsony2010}
Morsony B.~J.,  Heinz S.,  Br{\"{u}}ggen M.,   Ruszkowski M.,  2010, \mn@doi
  [MNRAS] {10.1111/j.1365-2966.2010.17059.x}, 407, 1277

\bibitem[\protect\citeauthoryear{Mukherjee, Bicknell, Sutherland  \&
  Wagner}{Mukherjee et~al.}{2016}]{Mukherjee2016}
Mukherjee D.,  Bicknell G.~V.,  Sutherland R.,   Wagner A.,  2016, \mn@doi
  [MNRAS] {10.1093/mnras/stw1368}, 461, 967

\bibitem[\protect\citeauthoryear{Navarro, Frenk  \& White}{Navarro
  et~al.}{1996}]{Navarro1996}
Navarro J.~F.,  Frenk C.~S.,   White S. D.~M.,  1996, \mn@doi [ApJ]
  {10.1086/177173}, 462, 563

\bibitem[\protect\citeauthoryear{Navarro, Frenk  \& White}{Navarro
  et~al.}{1997}]{Navarro1997}
Navarro J.~F.,  Frenk C.~S.,   White S. D.~M.,  1997, \mn@doi [ApJ]
  {10.1086/304888}, 490, 493

\bibitem[\protect\citeauthoryear{O'Neill \& Jones}{O'Neill \&
  Jones}{2010}]{ONeill2010}
O'Neill S.~M.,  Jones T.~W.,  2010, \mn@doi [ApJ]
  {10.1088/0004-637X/710/1/180}, 710, 180

\bibitem[\protect\citeauthoryear{O'Neill, {De Young}  \& Jones}{O'Neill
  et~al.}{2009}]{ONeill2009}
O'Neill S.~M.,  {De Young} D.~S.,   Jones T.~W.,  2009, \mn@doi [ApJ]
  {10.1063/1.3293072}, 694, 1317

\bibitem[\protect\citeauthoryear{Ogiya, Biernacki, Hahn  \& Teyssier}{Ogiya
  et~al.}{2018}]{Ogiya2018}
Ogiya G.,  Biernacki P.,  Hahn O.,   Teyssier R.,  2018, preprint
  (arXiv:1802.02177)

\bibitem[\protect\citeauthoryear{Pakmor \& Springel}{Pakmor \&
  Springel}{2013}]{Pakmor2013}
Pakmor R.,  Springel V.,  2013, \mn@doi [MNRAS] {10.1093/mnras/stt428}, 432,
  176

\bibitem[\protect\citeauthoryear{Pakmor, Bauer  \& Springel}{Pakmor
  et~al.}{2011}]{Pakmor2011}
Pakmor R.,  Bauer A.,   Springel V.,  2011, \mn@doi [MNRAS]
  {10.1111/j.1365-2966.2011.19591.x}, 418, 1392

\bibitem[\protect\citeauthoryear{Pakmor, Springel, Bauer, Mocz, Munoz, Ohlmann,
  Schaal  \& Zhu}{Pakmor et~al.}{2016a}]{Pakmor2016}
Pakmor R.,  Springel V.,  Bauer A.,  Mocz P.,  Munoz D.~J.,  Ohlmann S.~T.,
  Schaal K.,   Zhu C.,  2016a, \mn@doi [MNRAS] {10.1093/mnras/stv2380}, 455,
  1134

\bibitem[\protect\citeauthoryear{Pakmor, Pfrommer, Simpson, Kannan  \&
  Springel}{Pakmor et~al.}{2016b}]{Pakmor2016a}
Pakmor R.,  Pfrommer C.,  Simpson C.~M.,  Kannan R.,   Springel V.,  2016b,
  \mn@doi [MNRAS] {10.1093/mnras/stw1761}, 462, 2603

\bibitem[\protect\citeauthoryear{Perucho \& Mart{\'{i}}}{Perucho \&
  Mart{\'{i}}}{2007}]{Perucho2007}
Perucho M.,  Mart{\'{i}} J.~M.,  2007, \mn@doi [MNRAS]
  {10.1111/j.1365-2966.2007.12454.x}, 382, 526

\bibitem[\protect\citeauthoryear{Perucho, Mart{\'{i}}, Cela, Hanasz, Cruz  \&
  Rubio}{Perucho et~al.}{2010}]{Perucho2010}
Perucho M.,  Mart{\'{i}} J.~M.,  Cela J.~M.,  Hanasz M.,  Cruz R.~D.,   Rubio
  F.,  2010, A{\&}A, 519, A41

\bibitem[\protect\citeauthoryear{Perucho, Mart{\'{i}}, Quilis  \&
  Borja-Lloret}{Perucho et~al.}{2017}]{Perucho2017}
Perucho M.,  Mart{\'{i}} J.~M.,  Quilis V.,   Borja-Lloret M.,  2017, \mn@doi
  [MNRAS] {10.1093/mnrasl/slx115}, 471, L120

\bibitem[\protect\citeauthoryear{Peterson \& Fabian}{Peterson \&
  Fabian}{2006}]{Peterson2006}
Peterson J.~R.,  Fabian A.~C.,  2006, \mn@doi [Phys. Rep.]
  {10.1016/j.physrep.2005.12.007}, 427, 1

\bibitem[\protect\citeauthoryear{Pfrommer}{Pfrommer}{2013}]{Pfrommer2013}
Pfrommer C.,  2013, \mn@doi [ApJ] {10.1088/0004-637X/779/1/10}, 779, 10

\bibitem[\protect\citeauthoryear{Pfrommer \& Dursi}{Pfrommer \&
  Dursi}{2010}]{Pfrommer2010b}
Pfrommer C.,  Dursi J.,  2010, \mn@doi [Nat. Phys.] {10.1038/nphys1657}, 6, 520

\bibitem[\protect\citeauthoryear{Pfrommer, Pakmor, Schaal, Simpson  \&
  Springel}{Pfrommer et~al.}{2017}]{Pfrommer2017}
Pfrommer C.,  Pakmor R.,  Schaal K.,  Simpson C.~M.,   Springel V.,  2017,
  \mn@doi [MNRAS] {10.1093/mnras/stw2941}, 465, 4500

\bibitem[\protect\citeauthoryear{Powell, Roe, Linde, Gombosi  \& {De
  Zeeuw}}{Powell et~al.}{1999}]{Powell1999}
Powell K.~G.,  Roe P.~L.,  Linde T.~J.,  Gombosi T.~I.,   {De Zeeuw} D.~L.,
  1999, \mn@doi [Journal of Computational Physics] {10.1006/jcph.1999.6299},
  154, 284

\bibitem[\protect\citeauthoryear{Quataert}{Quataert}{2008}]{Quataert2008}
Quataert E.,  2008, \mn@doi [ApJ] {10.1086/525248}, 673, 758

\bibitem[\protect\citeauthoryear{Reynolds, Heinz  \& Begelman}{Reynolds
  et~al.}{2001}]{Reynolds2001}
Reynolds C.~S.,  Heinz S.,   Begelman M.~C.,  2001, ApJ, 549, 179

\bibitem[\protect\citeauthoryear{Reynolds, Heinz  \& Begelman}{Reynolds
  et~al.}{2002}]{Reynolds2002}
Reynolds C.~S.,  Heinz S.,   Begelman M.~C.,  2002, \mn@doi [MNRAS]
  {10.1046/j.1365-8711.2002.04724.x}, 332, 271

\bibitem[\protect\citeauthoryear{Reynolds, McKernan, Fabian, Stone  \&
  Vernaleo}{Reynolds et~al.}{2005}]{Reynolds2005}
Reynolds C.~S.,  McKernan B.,  Fabian A.~C.,  Stone J.~M.,   Vernaleo J.~C.,
  2005, \mn@doi [MNRAS] {10.1111/j.1365-2966.2005.08643.x}, 357, 242

\bibitem[\protect\citeauthoryear{Reynolds, Balbus  \& Schekochihin}{Reynolds
  et~al.}{2015}]{Reynolds2015}
Reynolds C.~S.,  Balbus S.~A.,   Schekochihin A.~A.,  2015, \mn@doi [ApJ]
  {10.1088/0004-637X/815/1/41}, 815, 41

\bibitem[\protect\citeauthoryear{Ruszkowski, En{\ss}lin, Br{\"{u}}ggen, Heinz
  \& Pfrommer}{Ruszkowski et~al.}{2007}]{Ruszkowski2007}
Ruszkowski M.,  En{\ss}lin T.~A.,  Br{\"{u}}ggen M.,  Heinz S.,   Pfrommer C.,
  2007, \mn@doi [MNRAS] {10.1111/j.1365-2966.2007.11801.x}, 378, 662

\bibitem[\protect\citeauthoryear{Ruszkowski, En{\ss}lin, Br{\"{u}}ggen,
  Begelman  \& Churazov}{Ruszkowski et~al.}{2008}]{Ruszkowski2008}
Ruszkowski M.,  En{\ss}lin T.~A.,  Br{\"{u}}ggen M.,  Begelman M.~C.,
  Churazov E.,  2008, \mn@doi [MNRAS] {10.1111/j.1365-2966.2007.12659.x}, 383,
  1359

\bibitem[\protect\citeauthoryear{Ruszkowski, Yang  \& Reynolds}{Ruszkowski
  et~al.}{2017}]{Ruszkowski2017a}
Ruszkowski M.,  Yang H. Y.~K.,   Reynolds C.~S.,  2017, \mn@doi [ApJ]
  {10.3847/1538-4357/aa79f8}, 844, 13

\bibitem[\protect\citeauthoryear{Sanders \& Fabian}{Sanders \&
  Fabian}{2008}]{Sanders2008}
Sanders J.~S.,  Fabian A.~C.,  2008, \mn@doi [MNRASL]
  {10.1111/j.1745-3933.2008.00549.x}, 390, L93

\bibitem[\protect\citeauthoryear{Schaal \& Springel}{Schaal \&
  Springel}{2015}]{Schaal2015}
Schaal K.,  Springel V.,  2015, \mn@doi [MNRAS] {10.1093/mnras/stu2386}, 446,
  3992

\bibitem[\protect\citeauthoryear{Sharma, Chandran, Quataert  \& Parrish}{Sharma
  et~al.}{2009}]{Sharma2009}
Sharma P.,  Chandran B. D.~G.,  Quataert E.,   Parrish I.~J.,  2009, \mn@doi
  [ApJ] {10.1088/0004-637X/699/1/348}, 699, 348

\bibitem[\protect\citeauthoryear{Sijacki \& Springel}{Sijacki \&
  Springel}{2006}]{Sijacki2006}
Sijacki D.,  Springel V.,  2006, \mn@doi [MNRAS]
  {10.1111/j.1365-2966.2006.10752.x}, 371, 1025

\bibitem[\protect\citeauthoryear{Sijacki, Pfrommer, Springel  \&
  En{\ss}lin}{Sijacki et~al.}{2008}]{Sijacki2008}
Sijacki D.,  Pfrommer C.,  Springel V.,   En{\ss}lin T.~A.,  2008, \mn@doi
  [MNRAS] {10.1111/j.1365-2966.2008.13310.x}, 387, 1403

\bibitem[\protect\citeauthoryear{Soker}{Soker}{2003}]{Soker2003}
Soker N.,  2003, \mn@doi [MNRAS] {10.1046/j.1365-8711.2003.06548.x}, 342, 463

\bibitem[\protect\citeauthoryear{Springel}{Springel}{2010}]{Springel2010}
Springel V.,  2010, \mn@doi [MNRAS] {10.1111/j.1365-2966.2009.15715.x}, 401,
  791

\bibitem[\protect\citeauthoryear{Sternberg \& Soker}{Sternberg \&
  Soker}{2008}]{Sternberg2008a}
Sternberg A.,  Soker N.,  2008, \mn@doi [MNRASL]
  {10.1111/j.1745-3933.2008.00512.x}, 389, 13

\bibitem[\protect\citeauthoryear{Tchekhovskoy \& Bromberg}{Tchekhovskoy \&
  Bromberg}{2016}]{Tchekhovskoy2016}
Tchekhovskoy A.,  Bromberg O.,  2016, \mn@doi [MNRAS] {10.1093/mnrasl/slw064},
  461, L46

\bibitem[\protect\citeauthoryear{Turner}{Turner}{2018}]{Turner2018}
Turner R.~J.,  2018, \mn@doi [MNRAS] {10.1093/mnras/sty433}, 476, 2522

\bibitem[\protect\citeauthoryear{Vantyghem, McNamara, Russell, Main, Nulsen,
  Wise, Hoekstra  \& Gitti}{Vantyghem et~al.}{2014}]{Vantyghem2014a}
Vantyghem A.~N.,  McNamara B.~R.,  Russell H.~R.,  Main R.~A.,  Nulsen P.
  E.~J.,  Wise M.~W.,  Hoekstra H.,   Gitti M.,  2014, \mn@doi [MNRAS]
  {10.1093/mnras/stu1030}, 442, 3192

\bibitem[\protect\citeauthoryear{Vikhlinin, Markevitch  \& Murray}{Vikhlinin
  et~al.}{2001}]{Vikhlinin2001}
Vikhlinin A.,  Markevitch M.,   Murray S.~S.,  2001, \mn@doi [ApJ]
  {10.1086/320078}, 551, 160

\bibitem[\protect\citeauthoryear{Weinberger, Ehlert, Pfrommer, Pakmor  \&
  Springel}{Weinberger et~al.}{2017}]{Weinberger2017}
Weinberger R.,  Ehlert K.,  Pfrommer C.,  Pakmor R.,   Springel V.,  2017,
  \mn@doi [MNRAS] {10.1093/mnras/stx1409}, 470, 4530

\bibitem[\protect\citeauthoryear{Wentzel}{Wentzel}{1971}]{Wentzel1971}
Wentzel G.~W.,  1971, ApJ, 163, 503

\bibitem[\protect\citeauthoryear{Wiener, Oh  \& Guo}{Wiener
  et~al.}{2013}]{Wiener2013}
Wiener J.,  Oh S.~P.,   Guo F.,  2013, \mn@doi [MNRAS] {10.1093/mnras/stt1163},
  434, 2209

\bibitem[\protect\citeauthoryear{Wiener, Pfrommer  \& Oh}{Wiener
  et~al.}{2017}]{Wiener2017}
Wiener J.,  Pfrommer C.,   Oh S.~P.,  2017, \mn@doi [MNRAS]
  {10.1093/mnras/stx127}, 467, 906

\bibitem[\protect\citeauthoryear{Worrall}{Worrall}{2009}]{Worrall2009}
Worrall D.~M.,  2009, \mn@doi [Astronomy and Astrophysics Review]
  {10.1007/s00159-008-0016-7}, 17, 1

\bibitem[\protect\citeauthoryear{Wykes, Hardcastle, Karakas  \& Vink}{Wykes
  et~al.}{2015}]{Wykes2015}
Wykes S.,  Hardcastle M.~J.,  Karakas A.~I.,   Vink J.~S.,  2015, \mn@doi
  [MNRAS] {10.1093/mnras/stu2440}, 447, 1001

\bibitem[\protect\citeauthoryear{Yang, Reynolds, Yang  \& Reynolds}{Yang
  et~al.}{2016}]{Yang2016a}
Yang H.-Y.~K.,  Reynolds C.~S.,  Yang H.-Y.~K.,   Reynolds C.~S.,  2016,
  \mn@doi [ApJ] {10.3847/0004-637X/818/2/181}, 818, 181

\bibitem[\protect\citeauthoryear{Zweibel}{Zweibel}{2013}]{Zweibel2013}
Zweibel E.~G.,  2013, \mn@doi [Phys. Plasmas] {10.1063/1.4807033}, 20, 055501

\makeatother
\end{thebibliography}



\appendix

\section{Magnetic Field Generation}
\label{sec:magneticfield_generation}
The external turbulent magnetic field follows a Kolmogorov spectrum in agreement
with observations \citep{Bonafede2010,Kuchar2011}. The generation of the initial
conditions for the magnetic field follows largely Appendix A in
\citet{Ruszkowski2007}. The Gaussian-distributed random magnetic field with
vanishing mean ($\langle\vecbf{B} \rangle =\mathbf{0}$ but $\sqrt{\langle\vecbf{B}^2\rangle} \neq 0$) is set up with a Kolmogorov power spectrum in
Fourier space. The magnetic field is scaled to ensure a shell-averaged constant
magnetic-to-thermal pressure ratio $X_{B,\mathrm{ICM}}=P_B/P_\rmn{th}$
throughout the cluster. The three components of the magnetic field $B_i$
($i\in\{1,2,3\}$) are treated independently to ensure that the final
distribution of $\vecbf{B}(\vecbf{x})$ has a random phase. The large discrepancy
between minimum cell size and computational box size necessitates the
interpolation of fields from multiple nested Cartesian grids with increasing
resolution onto our initial setup.

First, we compute Gaussian-distributed field components that obey a
one-dimensional power spectrum $P_i(k)$, given by $P_i(k) \propto k^2
|\tilde{B}_i (k)|^2 $, with
\begin{equation}
  |\tilde{B}_i (k)|^2 =
  \begin{cases} A, & k < k_{\mathrm{inj}}, \\
    A \left( \dfrac{k}{k_{\mathrm{inj}}}\right)^{-11/3}, & k_{\mathrm{inj}} \leq k,
  \end{cases}
 \end{equation} 
where $A$ is a normalisation constant, $k = |\vecbf{k}|$, and $k_{\mathrm{inj}}$
is the injection scale of the field. Modes on large scales
($k<k_{\mathrm{inj}}$) follow a random white-noise distribution while modes in
the inertial range ($k>k_{\mathrm{inj}}$) obey a Kolmogorov power spectrum. For
each magnetic field component, we set up a complex field such that
\begin{equation}
\label{reB}
      [\Re(\tilde{B}_i (\vecbf{k})), \Im(\tilde{B}_i (\vecbf{k}))]
      = [G_1(\Lambda_1, \Lambda_2, \sigma_k), G_2(\Lambda_1, \Lambda_2, \sigma_k)],
\end{equation}
where $\Lambda_1$ and $\Lambda_2$ are uniform random deviates so that the
function $G_i(\Lambda_1, \Lambda_2, \sigma_k)$ ($i\in\{1,2\}$) returns
Gaussian-distributed values with standard deviation $\sigma_k=\tilde{B}_i$ for
every value of $k$. We then normalise the spectrum to the desired variance of
the magnetic field components in real space, $\sigma_B$ using Parseval's
theorem,
\begin{equation}
\sigma_B^2 = \dfrac{1}{N^2} \sum_{i}\sum_{k_j} |\tilde{B}_i (k_j)|^2.
\end{equation}
To eliminate overlapping magnetic field lines between different nested meshes,
we (i) reorient radial field lines in the overlap region and (ii) remove the
inner part of our coarser mesh via a spherical tapering function and replace it
with the tapered high-resolution mesh. Since this process introduces divergences
in the magnetic field, we iteratively perform divergence cleaning steps while
accounting for the reorientation of radial magnetic field:
\begin{enumerate}
\item {\it Divergence cleaning in Fourier space.} We eliminate the field
  component in the direction of $\vect{k}$, via the projection operator:
  \begin{equation}
    \tilde{\vecbf{B}} \to \tilde{\vecbf{B}} - \hat{\vecbf{k}} (\hat{\vecbf{k}}
    \cdot \tilde{\vecbf{B}})
  \end{equation}
  in order to fulfil the constraint $\bnabla\bcdot\vecbf{B}=0$.
\item {\it Field rescaling to constant $X_{B,\mathrm{ICM}}$.} We rescale the
  magnetic field to obtain a constant average value of
  $\langle{}X_{B,\mathrm{ICM}}(r)\rangle$ in thin concentric shells of radius
  $r$ around the cluster centre.
\item {\it Cleaning and smoothing transition regions between meshes.}  In order to
  prevent interconnecting field lines between meshes, we force the radial
  magnetic field lines to bend in the overlap regions of different meshes. To
  this end, we remove the radial component of the magnetic field in real space
  via
  \begin{equation}
    \label{eq:magnetic_isolation}    
    {\vecbf{B}} \to
    \left[1-g(r)\hat{\vecbf{r}}\hat{\vecbf{r}}\right]{\vecbf{B}}
  \end{equation}
  where $\hat{\vecbf{r}}$ is the unit radial vector and
  \begin{equation}
    \label{eq:tapering}
    g(r)=1-\left| \cos
    \left(\frac{\pi}{2}\frac{x+\Delta x-1}{\Delta x}\right)\right|
  \end{equation}
  for $1-\Delta x<x<1+\Delta x$ with $x=r/r_\textrm{m}$ and $\Delta x=0.05$. The
  overlap radius of the two meshes is given by $r_\mathrm{m}$.
  
Afterwards, we taper the field strength in the overlapping regions of one mesh
with a spline $f_S(x)=0.5\cos\left(\pi(r-r_m)/d_m\right)$, where the taper width
$d_m$ is set to the larger (outer) cell size of the two adjacent meshes. Its
neighbouring mesh uses the spline $f_C(x)=1-f_S(x)$.

Steps (i)-(iii) are repeated until the divergence of the magnetic field has
sufficiently decreased. The resulting field is interpolated on our adaptive,
smoothly varying mesh in the initial conditions, which are setup in hydrostatic
equilibrium. To maintain this equilibrium, the temperature is varied according
to
\begin{equation}
n k_\mathrm{B}\delta T=-\frac{\delta \vect{B}^2}{8\pi}.
\end{equation} 
Finally, we relax the mesh with {\sc arepo} so that a remaining (small) divergence
is cleaned with the Powell algorithm. To reverse the decrease of the magnetic
field strength due to the conversion from magnetic to turbulent energy, we
rescale the magnetic field with a constant factor and obtain
$X_{B,\mathrm{ICM}}$.
\end{enumerate}

\section{Resolution study}
\label{sec:resolution_study}

\begin{figure*}
\centering
\includegraphics[trim=0.25cm 0.6cm 1.3cm 1.55cm,clip=true,width=\highresfac\textwidth]{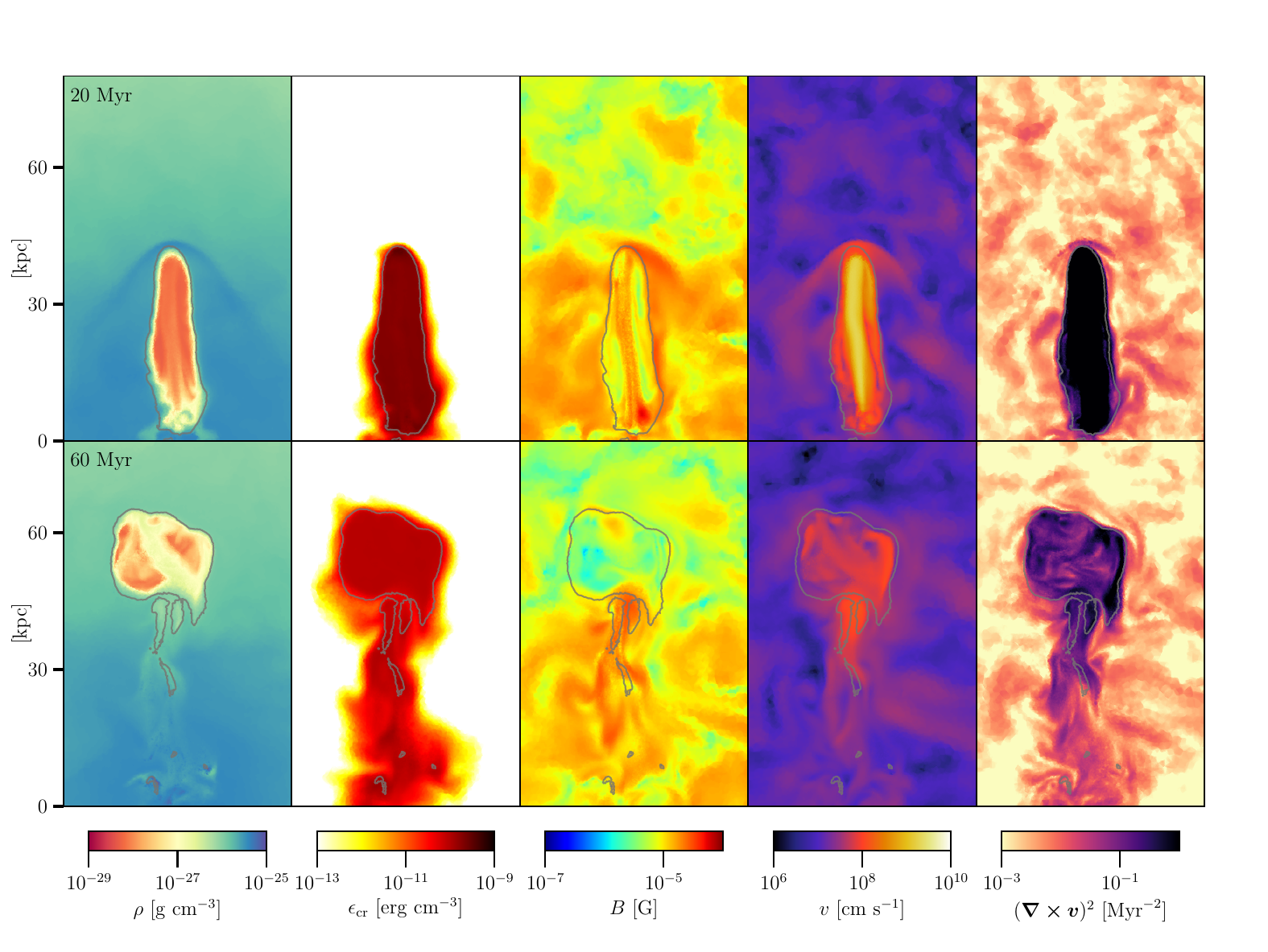}
\caption{Same model as in Fig.~\ref{fig:evolution_absolute} at lower resolution
  with projection dimensions
  $80\ \mathrm{kpc}\times50\ \mathrm{kpc}\times4\ \mathrm{kpc}$. Because magnetic tension
  is less well resolved and more affected by numerical diffusion, the external
  magnetic field strength and turbulent motions are smaller in
  amplitude. Moreover, at lower numerical resolution the velocity gradient at
  the jet foot is not sufficiently resolved, which implies that the jet does not
  travel as far as for the high-resolution simulation.  Thus, momentum transport
  is less efficient here.}
    \label{fig:highRes_evolution_absolute}
\end{figure*}

\begin{figure*}
\centering
\includegraphics[trim=0.25cm 0.6cm 1.3cm 1.55cm,clip=true,width=\highresfac\textwidth]{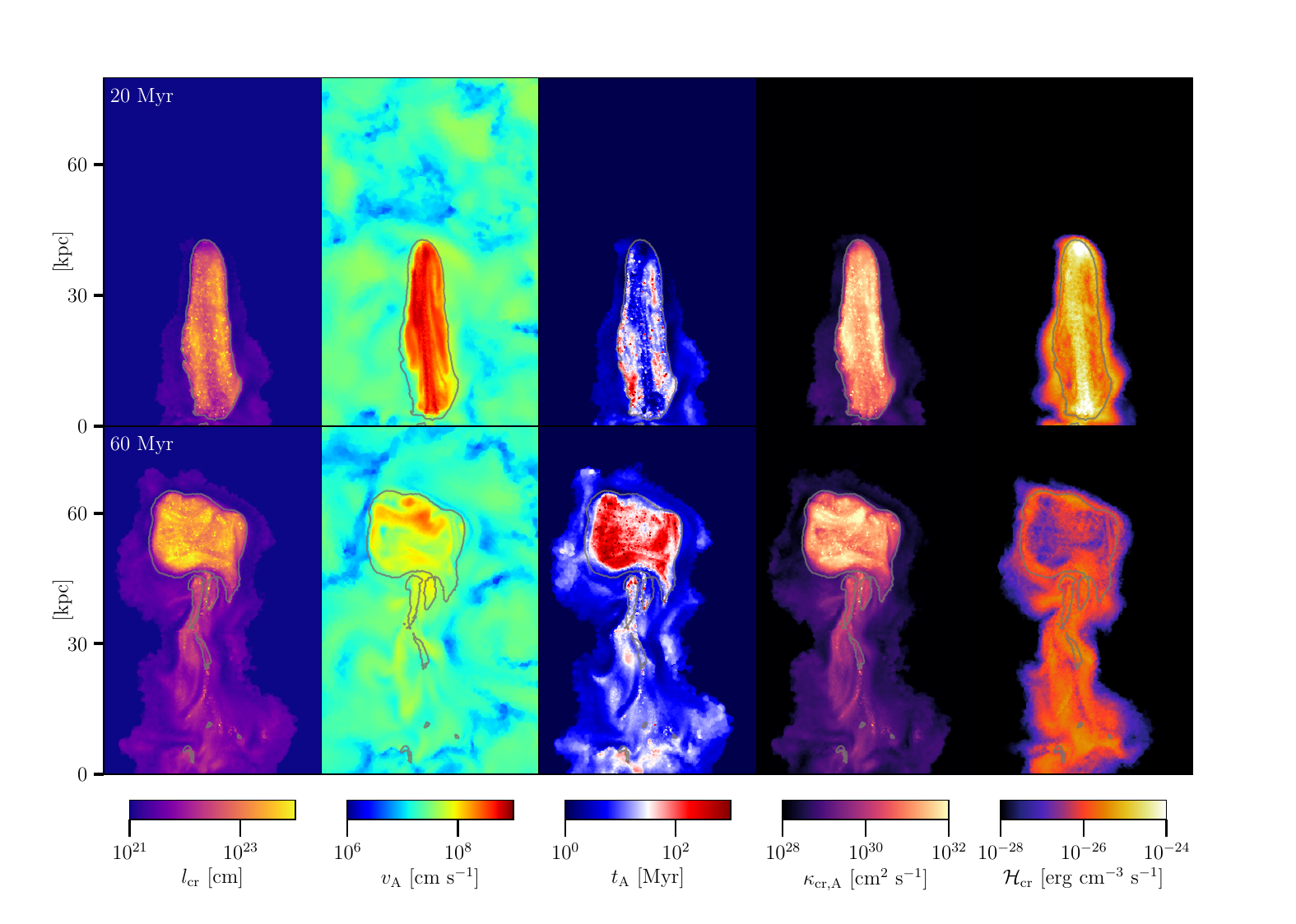}
\caption{Same model as in Fig.~\ref{fig:evolution_CRparameters} at lower
  resolution with projection dimensions $80\ \mathrm{kpc}\times50
  \mathrm{kpc}\times4\ \mathrm{kpc}$. The exact distribution of CRs depends somewhat
  on resolution because jet the distance of jet travel increases with numerical
  resolution. However, the quantities related to CR transport
  ($\kappa_{\rmn{cr,A}}$ and $\mathcal{H}_{\rmn{cr}}$) show general agreement
  with our high-resolution simulation.}
    \label{fig:highRes_evolution_CRparameters}
\end{figure*}

\begin{figure}
\centering
\includegraphics[trim=1.75cm .85cm 5.9cm 0.25cm,clip=true, width=\columnwidth]{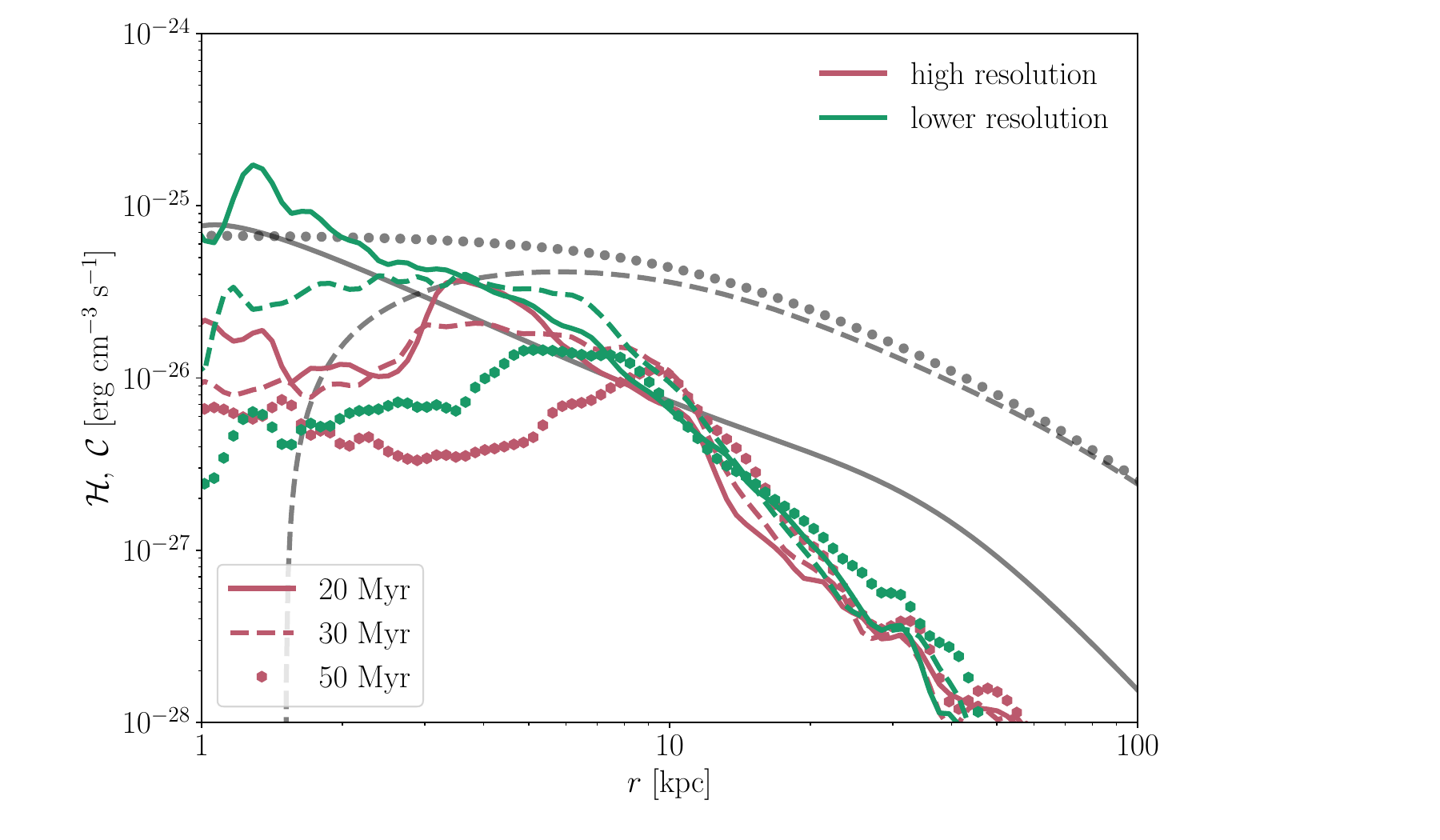}
\caption{Comparing the Alfv\'en-wave heating rate due to
  streaming CRs for our low- and high-resolution simulations at $t=20$, 30
  and 50~Myr (see also Fig.~\ref{fig:radialProfile_CR}). While the 
  jet in the high-resolution simulation propagates further, the CR heating rate drops 
 slightly in the cluster centre by a factor of less than two.}
    \label{fig:highRes_radialProfile_CR}
\end{figure}

\begin{figure}
\centering
\includegraphics[trim=1.4cm .3cm 2.7cm 0.6cm,clip=true, width=\columnwidth]{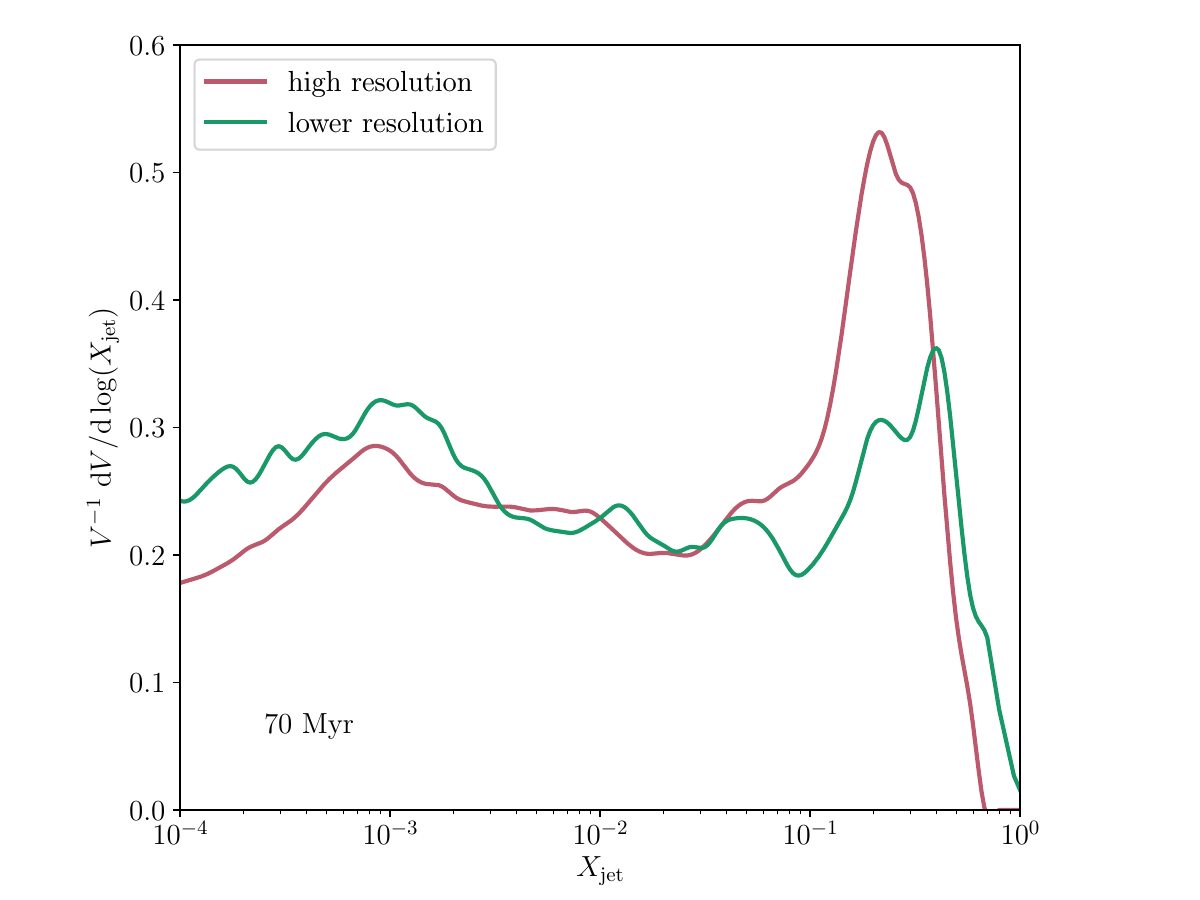}
\caption{Comparing the mixing
  efficiency of the bubble for CR simulations at different numerical resolutions
  (cf.\ Fig.~\ref{fig:mixing_magneticField}). To ensure an unbiased comparison
  we show the normalised filling factor of a given jet mass fraction
  $V^{-1}\ \mathrm{d}V/\mathrm{d}\log(X_\mathrm{jet})$. Mixing is suppressed at
  higher resolutions due to the more compact jet that propagates to larger
  distances. The lines are smoothed for clarity.}
    \label{fig:highRes_mixing_magneticField}
\end{figure}

To test numerical convergence, we compare simulation runs with the fiducial jet
parameters at high and lower resolution
(cf.\ Table~\ref{Tab:JetParaUnchanged}). The overall evolution
(Fig.~\ref{fig:highRes_evolution_absolute}) of the jet remains qualitatively
similar. However there are (smaller) quantitative differences such as the
emergence of Kelvin-Helmholtz instabilities on the jet surface in our
high-resolution simulation. That instability remain dynamically subdominant as
the more dominant Rayleigh-Taylor instability starts to develop. However, the
magnetic field strengths in the ICM remain larger for longer time scales in
comparison to the run at fiducial resolution since we better resolve magnetic
tension due to the lower level of numerical diffusion. This increased level of
magnetic turbulence causes the bubble to change direction more often at late
times.

As discussed in \citet{Weinberger2017}, the distance travelled by the jet is
resolution dependent. For this, resolving the velocity structure of the jet
proves crucial. The jet expands more laterally at lower resolution as the
prescribed jet width used for all simulations is insufficient to resolve the velocity gradient of the jet. The jet compensates by broadening. This spreads the area
of the effective momentum leading to a decrease of the jet velocity and thus a
smaller distance travelled at lower resolution.  However, the properties of CRs
remain robust against changes in resolution
(Fig.~\ref{fig:highRes_evolution_CRparameters}
vs. Fig.~\ref{fig:evolution_CRparameters} ).

Figure~\ref{fig:highRes_radialProfile_CR} compares radial profiles of the
Alfv\'en heating rate for our fiducial and high-resolution simulations.  Because
better resolved jets travel further, the Alfv\'en heating rate is somewhat
increased at larger distances for the high resolution run. This faster transport
to larger distance comes at the price of a somewhat reduced heating rate at
smaller radii. However, these differences stay below a factor of two, reinforcing
the robustness of our result on CR heating with respect to numerical resolution.

Even though the increased level of turbulence should be able to amplify the
mixing efficiency, the high-resolution run shows a significantly lower degree of
mixing in comparison to the fiducial (see
Fig.~\ref{fig:highRes_mixing_magneticField}). Possible causes of this are (i)
faster transport of CRs to larger distance which delays the onset of the
Rayleigh-Taylor instability and the successive disruption of the bubble and (ii)
better resolved magnetic field structures in the jet as well as in the draping
layer. Additionally, (iii) increased numerical diffusivity at lower resolution
facilitates mixing.

In conclusion, all discussed results are qualitatively robust to changes in
resolution. However, quantitative findings may be subject to (minor)
revision. This is in particular the case for the transition energy of FRI- and
FRII-like jets, which moves to lower values with increasing resolution, thus,
resolving the discrepancy of our runs with idealised jet simulations
\citep{Massaglia2016a}. Here, we choose fiducial (moderate) resolution since we
run a comprehensive parameter study to address the impact of CRs and turbulent
magnetic fields on a large variety of jet luminosities and activity
times. Moreover, this papers also serves as a first step towards studying
self-regulated CR-AGN feedback in cosmological cluster simulations in which we
have resolution requirements not too dissimilar from the adopted resolution.

\section{Bubble cooling}
\label{sec:BubbleCooling}

\begin{figure}
\centering
\includegraphics[trim=1.75cm .85cm 5.9cm 0.25cm,clip=true, width=\columnwidth]{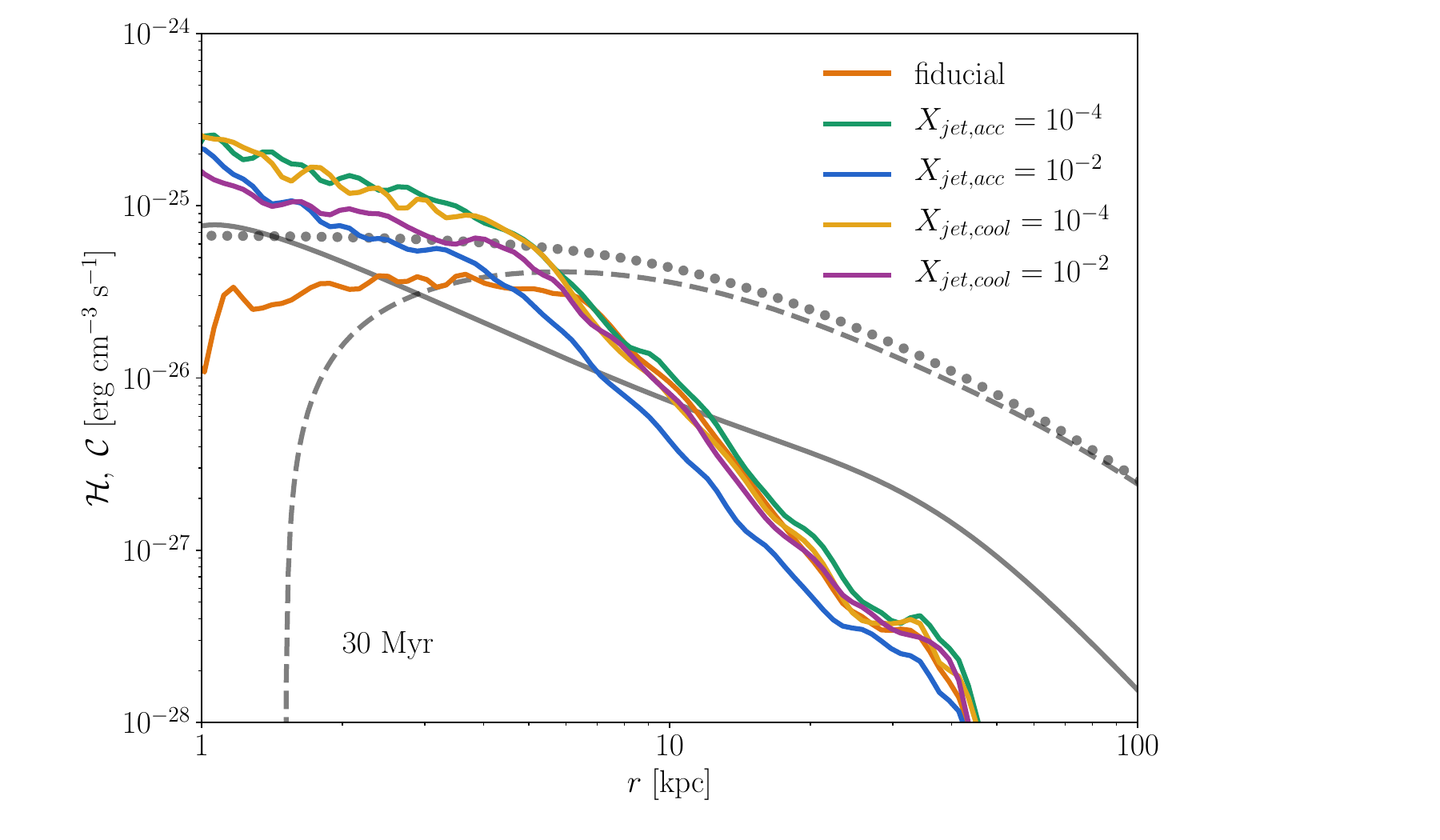}
\caption{Same as in Fig.~\ref{fig:radialProfile_CR}, here showing profiles of
  lower resolution simulations with varying jet tracer thresholds
  for subgrid CR acceleration $X_\mathrm{jet,acc}$ and CR Alfv\'en cooling
  $X_\mathrm{jet,cool}$ at $30\ \mathrm{Myr}$. Varying the thresholds by an
  order of magnitude in both directions, the resulting heating rate differs at
  most by a factor of a few at small radii and converges at large radii.}
    \label{fig:bubbleCooling_CRradprof}
\end{figure}

To remain consistent with our definition of the lobe, i.e.,
$X_\mathrm{jet}<10^{-3}$, we set the jet tracer threshold for CR acceleration
and cooling to $X_\mathrm{jet,acc}=X_\mathrm{jet,cool}=10^{-3}$. To test the
robustness of our choice, Fig.~\ref{fig:bubbleCooling_CRradprof} shows radial
profiles of the CR-induced Alfv\'en wave heating rate if we vary
$X_\mathrm{jet,acc}$ and $X_\mathrm{jet,cool}$ each by one order of
magnitude. In the inner 10 kpc, the CR distributions agree within a factor of a
few. Further out in the vicinity of the bubbles, the radial profiles agree
well. We generally find that changing the cooling threshold $X_\mathrm{jet,cool}$ has a
smaller impact than varying the acceleration threshold $X_\mathrm{jet,acc}$.


\bsp
\label{lastpage}
\end{document}